\documentclass[12pt]{article}
\usepackage{setspace}

\include{marginsfqx}

\usepackage{color}
\usepackage[pctex32]{graphics}
\newcommand{\ra}{\rangle }
\newcommand{\la}{\langle }
\newcommand{\ket}[1]{| #1 \rangle }
\newcommand{\bra}[1]{\langle #1 | }
\newcommand{ \ave}[2]{ \langle #1 | #2 | #1 \rangle }
\newcommand{\amp }[2]{\langle #1|#2 \rangle }
\newcommand{\weakv}[3]{ \frac{\langle #1|#2| #3\rangle}{\langle #1 | #3 \rangle }}
\newcommand{\beq}{\begin{equation}}
\newcommand{\eeq}{\end{equation}}

\newcommand{\upn}{^{(N)}}

\newcommand{\beqa}{\begin{eqnarray}}
\newcommand{\eeqa}{\end{eqnarray}}
\newcommand{\beqar}{\begin{eqnarray*}}
\newcommand{\eeqar}{\end{eqnarray*}}

\def \la {\langle}
\def \ra {\rangle}

\def \d {\delta}

\def \u {\uparrow}
\def \d {\downarrow}

\newcommand{\ltwid}{\mathrel{\raise.3ex\hbox{$<$\kern-.75em\lower1ex\hbox{$\sim$}}}}
\newcommand{\gtwid}{\mathrel{\raise.3ex\hbox{$>$\kern-.75em\lower1ex\hbox{$\sim$}}}}
\input colornam.sty
\input epsf.sty
\usepackage{graphicx}
\graphicspath{%
    {converted_graphics/}
    {D:/v5v6bugSamples/tollaksen/}
}

\begin{document}

\centerline{ \bf \large New Insights on Time-Symmetry in Quantum Mechanics}

\bigskip

\centerline{\small   Yakir Aharonov and Jeff Tollaksen}

\bigskip
\centerline{Center for Quantum Studies}
\centerline{Department of Physics and Department of Computational and Data Sciences}
\centerline{College of Science, George Mason University, Fairfax, VA 22030}

\date{\today}

\bigskip

The ``time-asymmetry" attributed to the standard formulation of Quantum Mechanics (QM) was inherited from a reasonable tendency learned from Classical Mechanics (CM)
to predict the future based on initial conditions: once the equations of motion are fixed in CM, then the initial and
final conditions are not independent, only one can be fixed arbitrarily.  
In contrast, as a result of the uncertainty principle, the relationship between
initial and final conditions within QM can be one-to-many: two ``identical" particles with identical environments 
can subsequently exhibit different properties under identical measurements.
These subsequent identical measurements provide fundamentally new information about the system which could not in principle be obtained from the initial conditions.
Although this lack of causal relations seemed to conflict with basic tenets of science, many justified it by arguing ``nature is capricious." This lead to Einstein's objection ``God doesn't play dice."  Nevertheless, after 100 years of experimental verification, QM has won over Einstein's objection.

QM's ``time-asymmetry" is the assumption that measurements only have consequences {\bf after} they are performed, i.e. towards the future.  Nevertheless, a positive spin was placed on QM's non-trivial relationship between initial and final conditions by Aharonov, Bergmann and Lebowitz (ABL)~\cite{abl} who showed that the new information obtained from measurements was also relevant for the {\bf past} of every quantum-system and not just the future.  This inspired ABL to re-formulate QM in terms of {\em Pre-and-Post-Selected-ensembles}.
The traditional paradigm for ensembles is to simply prepare systems in a particular state and thereafter subject them to a variety of experiments.  These are ``pre-selected-only-ensembles." 
For pre-{\bf and-post-selected}-ensembles, we add one more step, a subsequent measurement or post-selection.  
By collecting only a subset of the outcomes for this later measurement, we see that the ``pre-selected-only-ensemble"   can be divided into sub-ensembles according to the results of this subsequent ``post-selection-measurement."  
Because pre-and-post-selected-ensembles are the most refined quantum ensemble, they are of fundamental importance and subsequently led to the {\em two-vector} or {\em Time-Symmetric re-formulation of Quantum Mechanics} (TSQM)~\cite{av,jmav}.  TSQM  provides a complete description of a quantum-system at a given moment by using two-wavefunctions, one evolving from the past towards the future (the one utilized in the standard paradigm) and a second one, evolving from the future towards the past.

While TSQM is a new conceptual point-of-view that has predicted novel, verified effects which {\bf seem} impossible according to standard QM, TSQM is in fact a {\em re-formulation} of QM.  Therefore, experiments cannot prove TSQM over QM (or vice-versa).  The motivation to pursue such re-formulations, then, depends on their usefulness.  The intention of this article is to answer this by discussing how TSQM fulfils several criterion which any re-formulation of QM should satisfy in order to be useful and interesting:  
\begin{itemize}
\item  TSQM is consistent with all the predictions made by standard QM (\S\ref{intro}),  
\item  TSQM has revealed new features and effects of QM that were missed before 
(\S\ref{WM}),
\item  TSQM has lead to new mathematics, simplifications in calculations,  and stimulated discoveries in other fields  (as occurred, e.g., with the Feynman re-formulation of QM)
\S\ref{simplify},
\item TSQM suggests generalizations of QM that could not be easily articulated in the old language (\S\ref{general})
\end{itemize}
A more conservative scientist may choose to utilize all the pragmatic, operational advantages listed above\footnote{While this paper focuses on theoretical issues, we emphasize that many of the novel predictions have been tested in quantum-optics laboratories utilizing Townes' laser technology~\cite{steinbergult}.  In addition, TSQM has suggested a number of innovative new technologies which could be implemented with lasers \S\ref{superoscsec}.}, but stick to the standard time-asymmetric QM formalism.  Our view is that these new effects form a logical, consistent, and intuitive pattern (in contrast to the traditional interpretation).  Therefore, we believe there are deeper reasons which underly  TSQMs success in predicting them.  
One generalization suggested by TSQM (\S\ref{eachmoment})  addresses the ``artificial" separation in theoretical physics between the kinematic and dynamical  descriptions~\cite{dgross}; another (\S\ref{destiny}) provides a novel solution to the measurement problem.
Consequently, 
we are able to change the meaning of uncertainty from ``capriciousness" to exactly what is needed in order that the future can be relevant for the present, without violating causality, thereby providing a new perspective to the
 question  ``Why
does God play dice?"  (\S\ref{dice})
In other words, TSQM suggests that two ``identical" particles are not really identical, but there is no way to find their differences based only on information coming from the past, one must also know the future.  
 We also show how the second generalization involving ``destiny" is consistent with free-will (\S\ref{timesymfw}).
Finally, we speculate on the novel perspectives that TSQM can offer for several other themes of this volume, such as emergence.
\section{Consistency of Time-symmetric Quantum Mechanics with Standard Quantum Mechanics}
\label{intro}
We first motivate TSQM with a paradox
concerning the relativistic covariance of the state-description in QM.
\subsection{\bf Motivation - A Relativistic Paradox}
\label{relparadox}
\begin{figure}[h]
\begin{picture}(200,100)(0,0)
\put(10,15){\vector(1,0){160}}
\put(25,0){\vector(0,1){100}}
\put(85,15){\vector(1,1){76}}
\put(85,15){\vector(-1,1){74}}
\put(5,0){\makebox(0,0){(a)}}
\put(85,0){\makebox(0,0){$\underbrace{\,\,\,\,\,\,\,\,\,\,\,\,\,\,\,\,\,\,\,\,\,\,\,\,\,\,\,\,\,\,\,\,\,\,\,\,\,\,}_{\mathrm{verify}\,\,\{ \mid\uparrow\rangle_{\mathrm{\textcolor{BlueViolet}{A}}}\mid\downarrow\rangle_{\mathrm{\textcolor{Red}{B}}}
-\mid\downarrow\rangle_{\mathrm{\textcolor{BlueViolet}{A}}}\mid\uparrow\rangle_{\mathrm{\textcolor{Red}{B}}} \}}$}}

\put(150,10){\makebox(0,0){$x=L$}}
\put(25,110){\makebox(0,0){$t$}}
\put(38,87){ \makebox(0,0){\bf $\textcolor{BlueViolet}{|\!\!\uparrow_z\rangle_{\mathrm{A}}}$ } }

\color{BlueViolet}
\put(20,70){\framebox(10,10)}

\color{Red}
\put(147,77){\framebox(10,10)}

\put(130,85){ \makebox(0,0){\bf $|\!\!\downarrow_z\rangle_{\mathrm{\textcolor{Red}{B}}}$ } }


\color{Black}
\put(55,15){\dashbox{1}(0,70)}

\put(85,15){\dashbox{1}(0,70)}
\put(115,15){\dashbox{1}(0,70)}
\put(145,15){\dashbox{1}(0,70)}
\color{OliveGreen}
\put(25,75){\line(1,0){130}}
\put(88,68){\makebox(0,0){\textcolor{OliveGreen}{\scriptsize lab plane of simultaneity}}}
\put(91,60){\makebox(0,0){\textcolor{OliveGreen}{$t_{\mathrm{lab}}=t_2$}}}

\color{Black}
\put(25,45){\dashbox{1}(140,0)}

\put(215,0){\vector(0,1){100}}

\put(200,15){\vector(1,0){160}}
\put(275,15){\vector(1,1){76}}
\put(275,15){\vector(-1,1){72}}
\put(205,0){\makebox(0,0){(b)}}
\color{BlueViolet}

\put(210,70){\framebox(10,10)}
\put(230,35){ \makebox(0,0){\bf $|\!\!\uparrow_z\rangle_{\mathrm{\textcolor{BlueViolet}{A}}}$ } }




\color{Red}
\put(335,54){ \makebox(0,0){\bf $|\!\!\downarrow_z\rangle_{\mathrm{\textcolor{Red}{B}}}$ } }

\color{Red}
\put(310,50){\framebox(10,10)}

\color{OliveGreen}
\put(215,75){\line(3,-1){140}}
\put(390,40){\makebox(0,0){\textcolor{OliveGreen}{\scriptsize rocket plane of simultaneity}}}
\put(395,30){\makebox(0,0){$t_{\mathrm{rocket}}=t_1<t_2$}}
\put(245,45){\line(1,0){60}}
\put(245,45){\circle{10}}
\put(305,45){\circle{10}}

\color{Black}
\put(245,15){\dashbox{1}(0,70)}
\put(215,15){\dashbox{1}(0,70)}

\put(275,15){\dashbox{1}(0,70)}
\put(305,15){\dashbox{1}(0,70)}
\put(335,15){\dashbox{1}(0,70)}

\put(275,0){\makebox(0,0){$\underbrace{\,\,\,\,\,\,\,\,\,\,\,\,\,\,\,\,\,\,\,\,\,\,\,\,\,\,\,\,\,\,\,\,\,\,\,\,\,\,}_{\mathrm{verify}\,\,\{ \mid\uparrow\rangle_{\mathrm{\textcolor{BlueViolet}{A}}}\mid\downarrow\rangle_{\mathrm{\textcolor{Red}{B}}}
-\mid\downarrow\rangle_{\mathrm{\textcolor{BlueViolet}{A}}}\mid\uparrow\rangle_{\mathrm{\textcolor{Red}{B}}} \}}$}}


\put(215,45){\dashbox{1}(30,0)}
\put(305,45){\dashbox{1}(60,0)}
\color{Black}
\put(215,75){\dashbox{1}(140,0)}

\end{picture}
\caption[Lack of Lorentz covariance for a single wavefunction]{\small Collapse of singlet state in 2-different frames of reference; a)  the $t_{lab}=0$ hypersurface intersects the \textcolor{Red}{$B$} worldline before \textcolor{Red}{$B$'s} measurement, b) the $t_{rocket}=0$ hypersurface intersects the \textcolor{Red}{$B$} worldline {\it after} \textcolor{Red}{$B$'s} measurement.}
\label{singcollapse}

\end{figure}
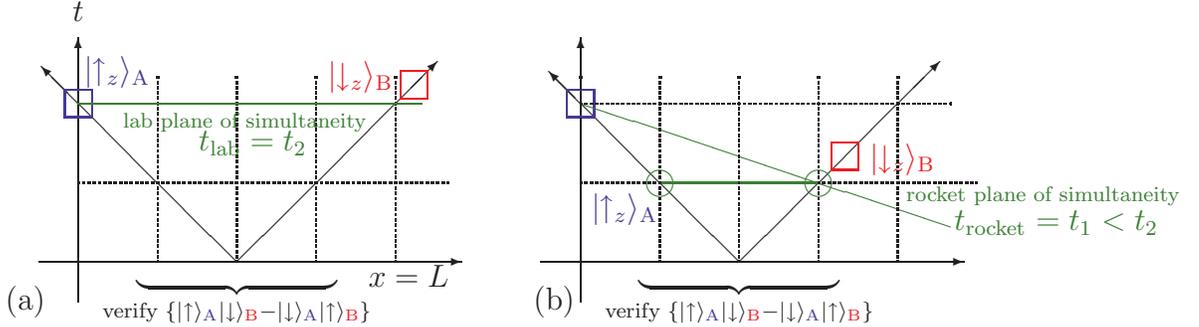

Consider two experimentalists \textcolor{BlueViolet}{A} and \textcolor{Red}{B} corresponding to two spin-1/2 particles prepared in a superposition with correlated spins ($\hat{\sigma}_{\mathrm{\textcolor{BlueViolet}{A}}} +\hat{\sigma}_{\mathrm{\textcolor{Red}{B}}}=0$), i.e. 
in an EPR state:
\begin{equation}
|\Psi_{EPR}(t=0)\rangle= \frac{1}{\sqrt{2}}\left\{\mid\uparrow\rangle_{\mathrm{\textcolor{BlueViolet}{A}}}\mid\downarrow\rangle_{\mathrm{\textcolor{Red}{B}}}
-\mid\downarrow\rangle_{\mathrm{\textcolor{BlueViolet}{A}}}\mid\uparrow\rangle_{\mathrm{\textcolor{Red}{B}}}\right\}
\label{14.1}
\end{equation}
Suppose at some later time $t_2$, the particles separate to a distance $L$ and \textcolor{BlueViolet}{A} measures his spin in the z-direction and obtains the outcome 
$\textcolor{BlueViolet}{\mid\uparrow_z\rangle_{\mathrm{A}}}$.
According to the usual interpretation, ideal-measurements  on 
either particle will instantly reduce the state from a superposition 
           $|\Psi_{EPR}(t_2-\varepsilon)\rangle=|\Psi_{EPR}(0)\rangle=\frac{1}{\sqrt{2}}\left\{\mid\uparrow\rangle_{\mathrm{\textcolor{BlueViolet}{A}}}\mid\downarrow\rangle_{\mathrm{\textcolor{Red}{B}}}
-\mid\downarrow\rangle_{\mathrm{\textcolor{BlueViolet}{A}}}\mid\uparrow\rangle_{\mathrm{\textcolor{Red}{B}}}\right\}$ into a direct product  
           $|\Psi(t_2+\varepsilon)=\textcolor{BlueViolet}{|\!\uparrow_z\rangle_{\mathrm{A}}}\textcolor{Red}{|\!\downarrow_z\rangle_{\mathrm{B}}}$.
{\bf I.e. after} \textcolor{BlueViolet}{A} performs his measurement at $t=t_2$, then the joint wavefunction collapses  so \textcolor{Red}{B}'s wavefunction also collapses to
   $\textcolor{Red}{|\downarrow_z\rangle_{\mathrm{B}}}$ which can be confirmed if \textcolor{Red}{B} actually performs a measurement.  When should \textcolor{Red}{B} perform this measurement?
 Consider a ``lab" frame-of-reference which is at rest relative to \textcolor{BlueViolet}{A} and \textcolor{Red}{B}, in which case the collapse is simultaneous (see figure \ref{singcollapse}.a) as indicated by the space-time coordinates:
   \begin{equation}\label{14.3}
      \begin{array}{cccc}
        \mathrm{\textcolor{BlueViolet}{A}}: & \left(%
                  \begin{array}{c}
                     ct_2 \\
                     0 \\
                  \end{array}%
               \right) &
        \mathrm{\textcolor{Red}{B}}: & \left(%
                \begin{array}{c}
                  ct_2 \\
                  L \\
                \end{array}%
              \right)
      \end{array}
   \end{equation}
  However, what is simultaneous in one frame-of-reference is not simultaneous in another: e.g. as we change to a rocket frame-of-reference which moves with velocity $\beta=\frac{v}{c}$ in the $x$ direction
   (with the same space-time origin), then the ``plane-of-simultaneity" changes (see figure \ref{singcollapse}.b) as can be seen with the new coordinates after a Lorentz-transformation:
   \begin{equation}\label{14.4}
      \begin{array}{c}
           \mathrm{\textcolor{BlueViolet}{A}}: \,\, \left(%
                  \begin{array}{cc}
                     \gamma & -\beta\gamma \\
                     -\beta\gamma & \gamma \\
                  \end{array}%
               \right)
               \left(%
                  \begin{array}{c}
                    ct_2 \\
                    0 \\
                  \end{array}%
               \right)
               =
               \left(%
                  \begin{array}{c}
                    \gamma ct_2 \\
                    -\beta\gamma ct_2 \\
                  \end{array}%
              \right) \\
            \mathrm{\textcolor{Red}{B}}: \,\, \left(%
                  \begin{array}{cc}
                     \gamma & -\beta\gamma \\
                     -\beta\gamma & \gamma \\
                  \end{array}%
               \right)
               \left(%
                  \begin{array}{c}
                    ct_2 \\
                    L \\
                  \end{array}%
               \right)
               =
               \left(%
                  \begin{array}{c}
                    \gamma ct_2-\beta\gamma L \\
                    -\beta\gamma ct_2 + \gamma L\\
                  \end{array}%
              \right) \\
        \end{array}
   \end{equation}
In the rocket frame-of-reference, the collapse of the wavefunction of \textcolor{Red}{B} happens at $t_1=\gamma t_2-\frac{\beta}{c}\gamma L<t_2$, i.e. the rocket frame-of-reference notices at $t_1<t_2$, that \textcolor{Red}{B} is in the state
   $\textcolor{Red}{|\downarrow_z\rangle_{\mathrm{B}}}$, implying that the joint EPR wavefunction had collapsed at $t_1$
   or before, so the state of $\textcolor{BlueViolet}{A}$ should be $\textcolor{BlueViolet}{|\uparrow_z\rangle_{\mathrm{A}}}$ no later than  $t_1$. 
If we transform back to our lab-frame-of-reference:
   \begin{equation}\label{14.6}
      \mathrm{\textcolor{BlueViolet}{A}}:\,\,\,\left(%
                  \begin{array}{cc}
                     \gamma & \beta\gamma \\
                     \beta\gamma & \gamma \\
                  \end{array}%
               \right)
               \left(%
                  \begin{array}{c}
                    ct_1 \\
                    -\beta ct_1 \\
                  \end{array}%
               \right)
               =
               \left(%
                  \begin{array}{c}
                    \gamma ct_2 -\beta L \\
                    0 \\
                  \end{array}%
              \right)
   \end{equation}
we see that the particle on \textcolor{BlueViolet}{A}'s side was in the $\textcolor{BlueViolet}{|\uparrow_z\rangle_{\mathrm{A}}}$ state even before \textcolor{BlueViolet}{A} made the measurement at $t_2$ (contradicting our notion that \textcolor{BlueViolet}{A}'s measurement supposedly caused the collapse in the first place.)
   
In summary, this paradox focuses on ``when did the collapse take place?". 
In the lab-frame, \textcolor{BlueViolet}{A}'s  measurement occurs first and then \textcolor{Red}{B}'s
measurement occurs (see figure \ref{singcollapse}.a).  However, in a rocket-frame, \textcolor{Red}{B's} measurement
occurs first and then \textcolor{BlueViolet}{A's} (see figure \ref{singcollapse}.b).  The lab-frame believes that \textcolor{BlueViolet}{A}'s measurement caused
the collapse whereas the rocket-frame disagrees and believes that \textcolor{Red}{B}'s measurement caused the collapse. 
While the two different versions give the same statistical results at the level of probabilities, they differ
completely on the state-description  during the intermediate times and there is
nothing in QM to suggest which version is the correct one.  A similar arrangement has been probed experimentally  producing results consistent with TSQMs hypothesis.~\cite{zb}
\footnote{This paradox can be sharpened in several ways~\cite{jt}.}
   This paradox has two possible resolutions:
   \begin{enumerate}
      \item Collapse cannot be described covariantly in a relativistic theory at the level of the state-description, only at the level of probabilities.  This thereby precludes progress on questions such as ``Why God plays dice."
      \item As first pointed out by Bell~\cite{bell,bub1,ar}, Lorentz-covariance in the state-description can be preserved in a theory like TSQM~\cite{jt}
(\S\ref{nonlocal}).  In addition, 
we believe it to be the most fruitful approach to probe deeper quantum realities beyond probabilities, thereby providing insight to questions like ``Why God plays dice."

   \end{enumerate}

\subsection{\bf The Main Idea}
\label{abl-main-idea}
\begin{figure}[h]
\vskip 5.9cm
\begin{picture}(600,90)(0,0)
\put(15,10){\vector(0,1){240}}
\put(10,70){\line(1,0){10}}
\put(0,70){\makebox(0,0){$t_{\mathrm{in}}$}}
\put(10,130){\line(1,0){10}}
\put(0,130){\makebox(0,0){$t$}}
\put(10,190){\line(1,0){10}}
\put(0,190){\makebox(0,0){$t_{\mathrm{fin}}$}}
\color{BlueViolet}
\put(50,100){\makebox(0,0){$U_{t_{\mathrm{in}}\rightarrow t}\!\mid\!\Psi_{\mathrm{in}}\rangle$}}

\put(30,80){\makebox(0,0){\Huge $\Uparrow$}}

\put(50,160){\makebox(0,0){$U_{t\rightarrow t_{\mathrm{fin}}}\!\mid\!a_j\rangle$}}

\put(30,140){\makebox(0,0){\Huge $\Uparrow$}}
\put(50,220){\makebox(0,0){$U_{t_{\mathrm{fin}}\rightarrow}\!\mid\!\Psi_{\mathrm{fin}}\rangle$}}

\put(30,200){\makebox(0,0){\Huge $\Uparrow$}}
\put(30,40){\makebox(0,0){\LARGE \bf ?}}
\color{OliveGreen}
\put(10,70){\dashbox{1}(430,0)}
\put(10,190){\dashbox{1}(430,0)}
\put(10,130){\dashbox{1}(430,0)}

\color{black}
\put(5,15){\makebox(0,0){(a)}}

\put(115,40){\makebox(0,0){\textcolor{OliveGreen}{\huge\bf+}}}

\put(165,10){\vector(0,1){240}}
\put(160,70){\line(1,0){10}}
\put(150,70){\makebox(0,0){$t_{\mathrm{in}}$}}
\put(160,130){\line(1,0){10}}
\put(150,130){\makebox(0,0){$t$}}
\put(160,190){\line(1,0){10}}
\put(150,190){\makebox(0,0){$t_{\mathrm{fin}}$}}
\color{RedViolet}
\put(200,40){\makebox(0,0){$\la U_{ t_{\mathrm{in}} \rightarrow}\Psi_{\mathrm{in}}\!\!\mid$}}

\put(180,60){\makebox(0,0){\Huge $\Downarrow$}}
\put(200,100){\makebox(0,0){$\la U_{t\rightarrow t_{\mathrm{in}} } a_j\!\!\mid$}}


\put(180,120){\makebox(0,0){\Huge $\Downarrow$}}
\put(200,160){\makebox(0,0){$\la U_{t_{\mathrm{fin}}\rightarrow t}\Psi_{\mathrm{fin}}\!\!\mid$}}


\put(180,180){\makebox(0,0){\Huge $\Downarrow$}}
\put(180,210){\makebox(0,0){\LARGE \bf ?}}
\color{black}
\put(155,15){\makebox(0,0){(b)}}

\put(270,40){\makebox(0,0){\textcolor{OliveGreen}{\huge\bf=}}}

\put(315,10){\vector(0,1){240}}
\put(310,70){\line(1,0){10}}
\put(300,70){\makebox(0,0){$t_{\mathrm{in}}$}}
\put(310,130){\line(1,0){10}}
\put(300,130){\makebox(0,0){$t$}}
\put(310,190){\line(1,0){10}}
\put(300,190){\makebox(0,0){$t_{\mathrm{fin}}$}}
\color{RedViolet}
\put(420,40){\makebox(0,0){$\la U_{ t_{\mathrm{in}}\rightarrow }\Psi_{\mathrm{in}}\!\!\mid$}}


\put(410,60){\makebox(0,0){\Huge $\Downarrow$}}
\put(420,100){\makebox(0,0){$\la U_{t\rightarrow t_{\mathrm{in}} }a_j\!\!\mid$}}


\put(410,120){\makebox(0,0){\Huge $\Downarrow$}}
\put(350,150){\makebox(0,0){$\la U_{t_{\mathrm{fin}}\rightarrow t}\Psi_{\mathrm{fin}}\!\!\mid$}}


\put(340,170){\makebox(0,0){\Huge $\Downarrow$}}
\color{Black}

\put(305,15){\makebox(0,0){(c)}}
\color{BlueViolet}

\put(350,110){\makebox(0,0){$U_{t_{\mathrm{in}}\rightarrow t}\mid\!\Psi_{\mathrm{in}}\rangle$}}

\put(340,90){\makebox(0,0){\Huge $\Uparrow$}}
\put(420,160){\makebox(0,0){$U_{t\rightarrow t_{\mathrm{fin}}}\!\mid\!a_j\rangle$}}



\put(410,140){\makebox(0,0){\Huge $\Uparrow$}}
\put(350,220){\makebox(0,0){$U_{t_{\mathrm{fin}}\rightarrow}\!\mid\!\Psi_{\mathrm{fin}}\rangle$}}
\put(340,200){\makebox(0,0){\Huge $\Uparrow$}}
\color{OliveGreen}
\put(318,75){\dashbox{1}(63,110)}

\color{black}
\end{picture}
\vskip -.75cm
\caption[Time-reversal symmetry in probability amplitudes.]
{\small Time-reversal symmetry in probability amplitudes.}
\label{timerev}
\end{figure}
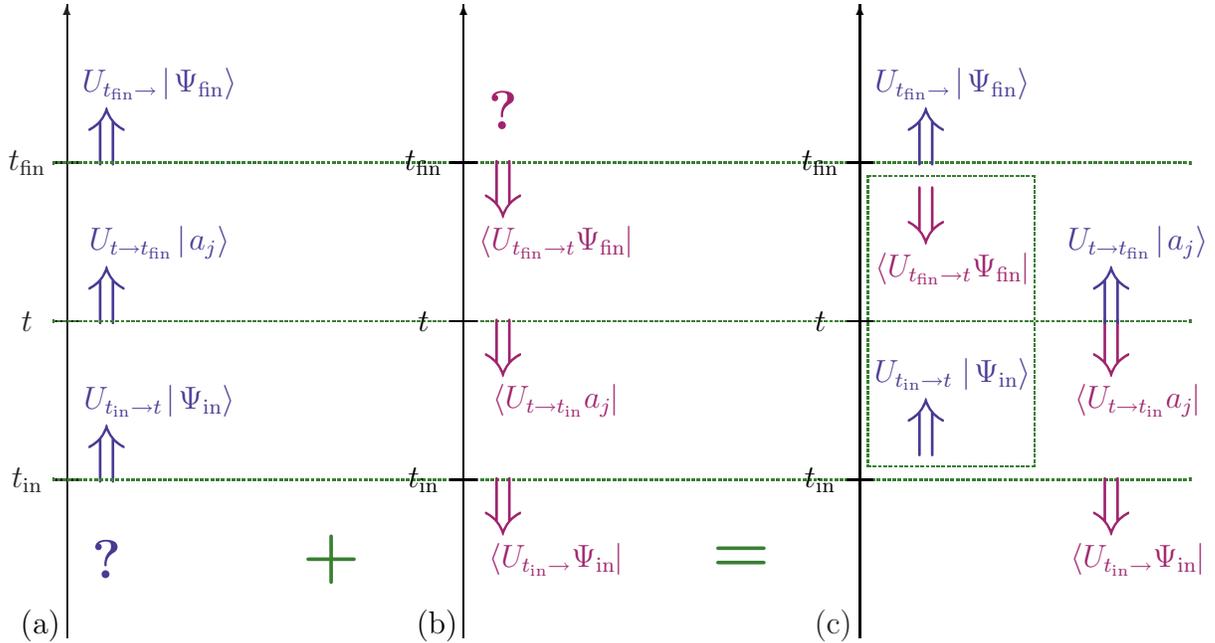

\begin{figure}[tbp] 
  \centering
  \includegraphics[width=5.14in,height=6.71in,keepaspectratio]{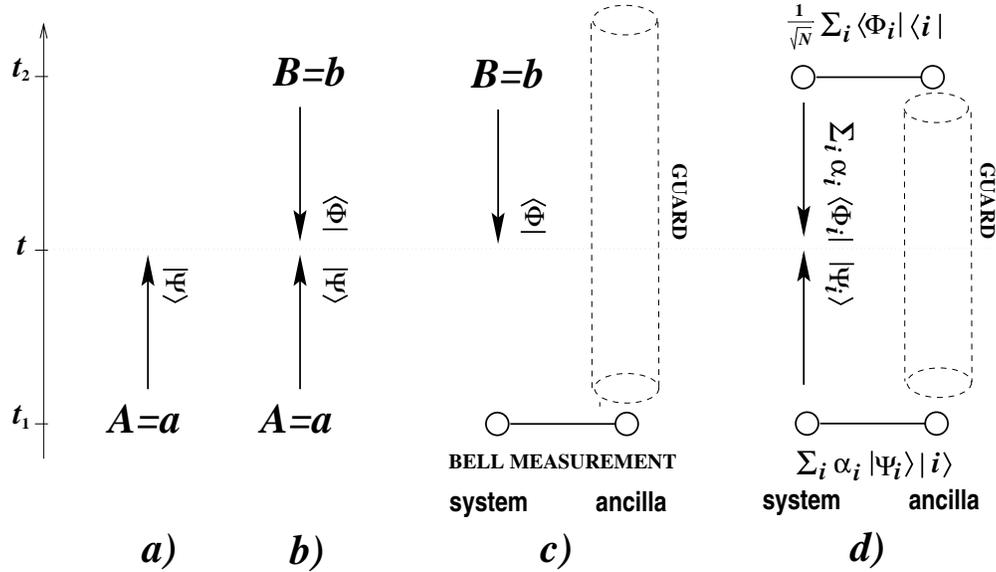}
\caption[]{\small {\bf Description of quantum systems:} (a) pre-selected, (b) pre- and
post-selected, (c) post-selected, and (d) generalized pre-and-post-selected. From \cite{av2v}}
\label{2vstates}
\end{figure}

TSQM contemplates measurements which occur at the present time $t$
 while the state is known both  at $t_{\mathrm{in}}<t$ (past) and at $t_{\mathrm{fin}}>t$ (future). 
More precisely, we start at $t=t_{\mathrm{in}}$ with a  measurement of  a nondegenerate operator $\hat{O}_{\mathrm{in}}$.   This yields as one potential outcome the state $\ket{\Psi_{\mathrm{in}}}$, i.e. we prepared the ``pre-selected" state
$\ket{\Psi_{\mathrm{in}}}$.  At the later time $t_{\mathrm{fin}}$, we perform another measurement of a nondegenerate
operator $\hat{O}_{\mathrm{fin}}$ which yields one possible outcome: the post-selected state $\ket{\Psi_{\mathrm{fin}}}$. 
 At an intermediate time $t\in [t_{\mathrm{in}},t_{\mathrm{fin}}]$, we measure a non-degenerate observable $\hat{A}$ (for simplicity), with eigenvectors $\{ \ket{a_j} \}$. We wish to
determine the conditional probability of $a_j$, given that we have both boundary conditions,
$\ket{\Psi_{\mathrm{in}}}$ and $\bra{\Psi_{\mathrm{fin}}}$.  \footnote{Such an arrangement has long been considered in actual experiments: consider a bubble chamber scattering experiment. The incoming particle, $\ket{\Psi_{\mathrm{in}}}$, interacts with a target and then evolves into various outgoing states,  $\ket{\Psi_{\mathrm{fin}}}_1$, $\ket{\Psi_{\mathrm{fin}}}_2$, etc.  Typically, photographs are not taken for every target-interaction, but only for certain ones that were triggered by subsequently interacting with detectors. In CM, there is (in principle) a one-to-one mapping between incoming states and outgoing states, whereas in QM, it is one-to-many.  By selecting a single outcome for the post-selection-measurement, we define the pre-and-post-selected-ensemble that has no classical analog.}
To answer this, we use the time displacement operator:
   $U_{t_{\mathrm{in}}\rightarrow t}=\exp\{-iH(t-t_{\mathrm{in}})\}$
   where $H$ is the Hamiltonian for the free system.  For simplicity, we assume $H$ is time
   independent and set $\hbar=1$. 
The standard 
theory of collapse states that the system collapses into an eigenstate $\ket{a_j}$
{\it after} the  measurement at $t$ with an amplitude $\textcolor{BlueViolet}{\bra{a_j} U_{t_{\mathrm{in}}\rightarrow t}\ket{\Psi_{\mathrm{in}}}}$.
The amplitude for our series of events is $\alpha_j\equiv\textcolor{BlueViolet}{\bra{\Psi_{\mathrm{fin}}}U_{t\rightarrow t_{\mathrm{fin}}}|a_j\rangle\langle a_j|U_{t_{\mathrm{in}}\rightarrow t}\ket{\Psi_{\mathrm{in}}}}$ which is illustrated in figure \ref{timerev}.a. 
This means that the conditional probability to measure $a_j$ given $\ket{\Psi_{\mathrm{in}}}$ is pre-selected
   and $\ket{\Psi_{\mathrm{fin}}}$ will be post-selected is given by the ABL formula~\cite{abl}
\footnote{ABL is intuitive:  $|\bra{a_j} U_{t_{\mathrm{in}}\rightarrow t}\ket{\Psi_{\mathrm{in}}}|^2$ is the probability to obtain $\ket{a_j}$ having started with $\ket{\Psi_{\mathrm{in}}}$.  If $\ket{a_j}$ was obtained, then the system collapsed to $\ket{a_j}$ and $|\bra{\Psi_{\mathrm{fin}}} U_{t\rightarrow t_{\mathrm{fin}}}\ket{a_j}|^2$ is then the probability to obtain $\ket{\Psi_{\mathrm{fin}}}$.  The probability to obtain $\ket{a_j}$ and $\ket{\Psi_{\mathrm{fin}}}$ then is $|\alpha_j|^2$.  This is not yet the conditional probability since the post-selection may yield outcomes other than $\bra{\Psi_{\mathrm{fin}}}$. The probability to obtain $\ket{\Psi_{\mathrm{fin}}}$ is
$\sum_j|\alpha_j|^2=|\bra{\Psi_{\mathrm{fin}}}\Psi_{\mathrm{in}}\ra|^2<1$.  
The question being investigated concerning probabilities of $a_j$ at $t$ assumes we are successful in obtaining
 the post-selection and therefore requires the denominator in eq.~\ref{ablnts}, $\sum_j|\alpha_j|^2$, which is a
 re-normalization to obtain a proper probability.}: 
\begin{equation}
Pr(a_j,t|\Psi_{\mathrm{in}},t_{\mathrm{in}}; \Psi_{\mathrm{fin}},t_{\mathrm{fin}})  =\textcolor{BlueViolet}{\frac{ |\bra{\Psi_{\mathrm{fin}}} U_{t\rightarrow t_{\mathrm{fin}}}\ket{a_j}\bra{a_j} U_{t_{\mathrm{in}}\rightarrow t}\ket{\Psi_{\mathrm{in}}}|^2 }{\sum_{n} |\bra{\Psi_{\mathrm{fin}}} U_{t\rightarrow t_{\mathrm{fin}}}\ket{a_n}\bra{a_n} U_{t_{\mathrm{in}}\rightarrow t}\ket{\Psi_{\mathrm{in}}}|^2}} 
\label{ablnts}
\end{equation}
As a first step toward understanding the underlying time-symmetry in the ABL formula, we consider the time-reverse of the numerator of eq.~\ref{ablnts} and figure \ref{timerev}.a.   First we 
apply $U_{t\rightarrow t_{\mathrm{fin}}}$ on $\bra{\Psi_{\mathrm{fin}}}$ instead of on $\bra{a_j}$.  We note that $\bra{\Psi_{\mathrm{fin}}}U_{t\rightarrow t_{\mathrm{fin}}}=\langle U_{t\rightarrow t_{\mathrm{fin}}}^\dag\Psi_{\mathrm{fin}}|$ by using the well-known QM symmetry $U_{t\rightarrow
      t_{\mathrm{fin}}}^\dag={\left\{e^{-iH(t_{\mathrm{fin}}-t)}\right\}}^\dag=e^{iH(t_{\mathrm{fin}}-t)}=e^{-iH(t-t_{\mathrm{fin}})}=U_{t_{\mathrm{fin}}\rightarrow t}$.   We also apply $U_{t_{\mathrm{in}}\rightarrow t}$ on $\bra{a_j}$ instead of on  $\ket{\Psi_{\mathrm{in}}}$ which yields the time-reverse re-formulation of the numerator of eq.~\ref{ablnts}, $\textcolor{RedViolet}{\langle U_{t_{\mathrm{fin}}\rightarrow t}\Psi_{\mathrm{fin}}|a_j\rangle\langle U_{t\rightarrow t_{\mathrm{in}}} a_j|\Psi_{\mathrm{in}}\rangle}$ as depicted in fig. \ref{timerev}.b.  
Further work is needed to formulate what we mean by the 2-vectors in TSQM.  E.g. if  
we are interested in the probability for possible outcomes of $a_j$ at time $t$, we must consider 
both $\textcolor{BlueViolet}{U_{t_{\mathrm{in}}\rightarrow t}\mid\!\Psi_{\mathrm{in}}\rangle}$ and $\textcolor{RedViolet}{\la U_{t_{\mathrm{fin}}\rightarrow t}\Psi_{\mathrm{fin}}\!\!\mid}$, 
since these expressions propagate the pre-and-post-selection to the present time $t$ (see the conjunction of both figures \ref{timerev}.a and \ref{timerev}.b giving \ref{timerev}.c which is re-drawn in figure \ref{2vstates}.b; these 2-vectors are not just the time-reverse of each other).  This represents the basic idea behind the  Time-Symmetric re-formulation of Quantum Mechanics (TSQM)\footnote{We note that because (full) collapses take place at the $t_{\mathrm{in}}$ and $t_{\mathrm{fin}}$ measurements, there is no meaning to information coming from $t>t_{\mathrm{fin}}$ or $t<t_{\mathrm{in}}$.  Therefore, at least in this context, there is no meaning to a ``multi-vector" formalism.}:
\begin{equation}
Pr(a_j,t|\Psi_{\mathrm{in}},t_{\mathrm{in}}; \Psi_{\mathrm{fin}},t_{\mathrm{fin}})  =\frac{ |\textcolor{RedViolet}{\langle U_{t_{\mathrm{fin}}\rightarrow t}\Psi_{\mathrm{fin}}}\ket{a_j}\bra{a_j} \textcolor{BlueViolet}{U_{t_{\mathrm{in}}\rightarrow t}\ket{\Psi_{\mathrm{in}}}}|^2 }{\sum_{n} |\textcolor{RedViolet}{\langle U_{t_{\mathrm{fin}}\rightarrow t}\Psi_{\mathrm{fin}}}\ket{a_n}\bra{a_n} \textcolor{BlueViolet}{U_{t_{\mathrm{in}}\rightarrow t}\ket{\Psi_{\mathrm{in}}}}|^2} 
\label{ablnts}
\end{equation}
While this mathematical manipulation clearly proves that TSQM is consistent with QM, it yields a very different interpretation.  For example, the action of $U_{t_{\mathrm{fin}}\rightarrow t}$ on $\bra{\Psi_{\mathrm{fin}}}$ (i.e. 
$\textcolor{RedViolet}{\langle U_{t_{\mathrm{fin}}\rightarrow t}\Psi_{\mathrm{fin}}}|$)
 can be interpreted to mean that the time displacement operator $U_{t_{\mathrm{fin}}\rightarrow t}$ sends $\bra{\Psi_{\mathrm{fin}}}$ back in time from the time $t_{\mathrm{fin}}$ to the present, $t$.  
A number of new categories of states (figure \ref{2vstates}) are suggested by the TSQM formalism and have proven useful in a wide variety of situations.

In summary, the ABL formulation clarified a number of issues in QM.  E.g.: in this formulation, both the probability and the amplitude are symmetric under the exchange of $\ket{\Psi_{\mathrm{in}}}$ and $\ket{\Psi_{\mathrm{fin}}}$. Therefore, the possibility of wavefunction collapse in QM does not necessarily imply irreversibility of an arrow of time at the QM  level.  
Nevertheless, the real litmus test of any re-formulation is whether conceptual shifts can teach us something fundamentally new or suggest generalizations of QM, etc.  
The re-formulation to TSQM suggested a number of new experimentally observable effects, one important example of which are {\em weak-measurements} (\S\ref{WM}), which we now begin to motivate by considering strange pre-and-post-selection effects.

\subsubsection{Pre-and-post-selection and Spin-1/2}
\label{spinhalf}
One of the simplest, surprising, example of pre-and-post-selection is to pre-select a spin-1/2 system with $\ket{\Psi_{\mathrm{in}}}=|\hat{\sigma}_x=+1\ra=\vert\!\!\uparrow_x\!\rangle $ at time $t_{\mathrm{in}}$.
After the pre-selection, spin measurements in the direction perpendicular  to $x$ yields complete uncertainty in the result \footnote{E.g. in the $z$-basis the state is $\frac{1}{\sqrt{2}}(|\uparrow_z\rangle+|\downarrow_z\rangle)$ which yields equal probability either spin-up or spin-down in the $z$-direction.}, so if we post-select at time $t_{\mathrm{fin}}$ in the $y$-direction, we obtain $\ket{\Psi_{\mathrm{fin}}}=|\hat{\sigma}_{\mathrm{y}}= +1\ra=\vert\uparrow_y\rangle$ half the time.  Since the particle is free, the spin is conserved in time and thus for any $t\in [t_{\mathrm{in}},t_{\mathrm{fin}}]$, an ideal-measurement of either $\hat{\sigma}_x$ or $\hat{\sigma}_y$, yields
   $+1$ for this pre-and-post-selection.  This by itself, two non-commuting observables known with certainty, is a most surprising property which no pre-selected-only-ensemble could possess.
\footnote{This is also evident from ABL: the probability to obtain $\hat{\sigma}_{\xi}= +1$ at the intermediate time if an ideal-measurement is performed is  $Pr(\hat{\sigma}_{\xi}=+1)=\frac{1+\cos (\xi) +\sin (\xi) + \cos (\xi) \sin(\xi)}{1+\cos(\xi)\sin(\xi)}$.
We see that if $\xi=0^\circ$ (i.e. $\hat{\sigma}_x$) then the intermediate ideal-measurement will yield $\hat{\sigma}_{\mathrm{x}}= +1$ with certainty  and when  $\xi=90^\circ$ (i.e. $\hat{\sigma}_y$), then the intermediate ideal-measurement will again yield $\hat{\sigma}_{\mathrm{y}}= +1$ with certainty.  E.g. $\hat{\sigma}_{\xi=45}=\pm 1$ is displayed in figure \ref{wmspin2}.a}

We now ask a slightly more complicated question about the spin in a
direction $\xi=45^{\circ}$ relative to the $x-y$ axis.  This yields:
\beq
\hat{\sigma}_{\xi}=\hat{\sigma}_x \cos 45^{\circ} +\hat{\sigma}_y\sin 45^{\circ}=\frac{\hat{\sigma}_x +\hat{\sigma}_y}{\sqrt{2}}
\label{spin45}
\eeq
From the results $Pr(\hat{\sigma}_{x}=+1)=1$ and $Pr(\hat{\sigma}_{y}=+1)=1$, one might wonder why we couldn't  insert both values, $\hat{\sigma}_{\mathrm{x}}= +1$ {\bf and} $\hat{\sigma}_{\mathrm{y}}= +1$ into eq.~\ref{spin45} and obtain $\hat{\sigma}_{\xi}=\frac{1+1}{\sqrt{2}}=\frac{2}{\sqrt{2}}=\sqrt{2}$.
Such a result is incorrect for an ideal-measurement because the eigenvalues of any spin operator, including $\hat{\sigma}_{\xi}$, must be $\pm 1$.
The inconsistency can also be seen by noting ${\left(\frac{\sigma_x+\sigma_y}{\sqrt{2}}\right)}^2=\frac{\sigma_x^2+\sigma_y^2+\sigma_x\sigma_y+\sigma_y\sigma_x}{2}=\frac{1+1+0}{2}=1$.
 By implementing the above argument, we would expect ${\left(\frac{\sigma_x+\sigma_y}{\sqrt{2}}\right)}^2={\left(\frac{1+1}{\sqrt{2}}\right)}^2=2\neq 1$.
Performing this step of replacing $\hat{\sigma}_{\mathrm{x}}= +1$ {\it and} $\hat{\sigma}_{\mathrm{y}}= +1$ in eq.~\ref{spin45} can only be done if $\hat{\sigma}_{\mathrm{x}}$ and $\hat{\sigma}_{\mathrm{y}}$ commute, which would allow both values  simultaneously to be definite.  Although it appears we have reached the end-of-the-line with this argument, 
nevertheless, it still seems that there should be some sense in which 
both  $Pr(\hat{\sigma}_{x}=+1)=1$ and $Pr(\hat{\sigma}_{y}=+1)=1$ manifest themselves simultaneously to produce $\hat{\sigma}_{\xi}=\sqrt{2}$. 

\subsubsection{Pre-and-post-selection and 3-Box-paradox}
\label{3boxesintro}
Another example of a surprising pre-and-post-selection effect is the 3-box-paradox~\cite{AAD}
which uses a {\it single} quantum particle
that is placed in a superposition of 3 closed, separated boxes. 
The particle is  pre-selected to be in the state $|\Psi_{\mathrm{in}} \rangle 
= 1/\sqrt 3 ~(|A\rangle + |B\rangle +|C\rangle)$, where $|A\rangle$, $|B\rangle$ and $|C\rangle$
denote the particle localized in boxes $A$, $B$, or $C$,  
respectively.  The particle 
is post-selected to be in the state $|\Psi_{\mathrm{fin}} \rangle = 1/\sqrt 3   ~(|A\rangle + 
|B\rangle -|C\rangle)$.
If an ideal-measurement is performed on box $A$ in the intermediate time (e.g. we open the box), then the particle is found in box $A$ with certainty.  This is confirmed by the ABL~\cite{abl} probability for projection in $A$:
$Pr(\mathrm{\bf \hat{P}}_{A})=
{\vert\langle\Psi_{\mathrm{fin}}\vert \mathrm{\bf \hat{P}}_{A}\vert\Psi_{\mathrm{in}}\rangle\vert^{2}
\over
\vert\langle\Psi_{\mathrm{fin}}\vert
  \mathrm{\bf \hat{P}}_{A}\vert\Psi_{\mathrm{in}}\rangle\vert^{2}+\vert\langle\Psi_{\mathrm{fin}}\vert
\mathrm{\bf \hat{P}}_{B}+\mathrm{\bf \hat{P}}_{C}\vert\Psi_{\mathrm{in}}\rangle\vert^{2}}
=1$.  
This can also be seen intuitively by contradiction: suppose we do not find the particle in box $|A\rangle$.  In that case, since we do not interact with box $|B\rangle$ or $|C\rangle$,  we would have to conclude that the state that remains after we didn't find it in $|A\rangle$ is proportional to $|B\rangle +|C\rangle$.
But this is orthogonal to the post-selection (which we know will definitely be obtained). 
Because this is a contradiction, we conclude that the particle
must be found in box $A$.  
Similarly, the probability to 
find the particle in box $B$ is  $1$, i.e. $Pr(\mathrm{\bf \hat{P}}_{\mathrm{B}}=1)=1$. 
The ``paradox" is, what ``sense" can these 2 definite statements be simultaneously true.  We cannot detect the distinction with ideal-measurements: e.g. $Pr(\mathrm{\bf \hat{P}}_{A}=1)=1$ if only box $A$ is opened, while $Pr(\mathrm{\bf \hat{P}}_{\mathrm{B}}=1)=1$ if only box $B$ is opened.  If ideal-measurements are performed on {\bf both} box $A$ and box $B$, then obviously the particle will not be found in both boxes, i.e. $\mathrm{\bf \hat{P}}_{\mathrm{A}}\mathrm{\bf \hat{P}}_{\mathrm{B}}=0$.  
\footnote{The mystery is increased by the fact that both $\mathrm{\bf \hat{P}}_{\mathrm{A}}$ and $\mathrm{\bf \hat{P}}_{\mathrm{B}}$ commute with each other, so one may ask ``how is it possible that measurement of one box can disturb the measurement of another?"}
\vskip -2cm
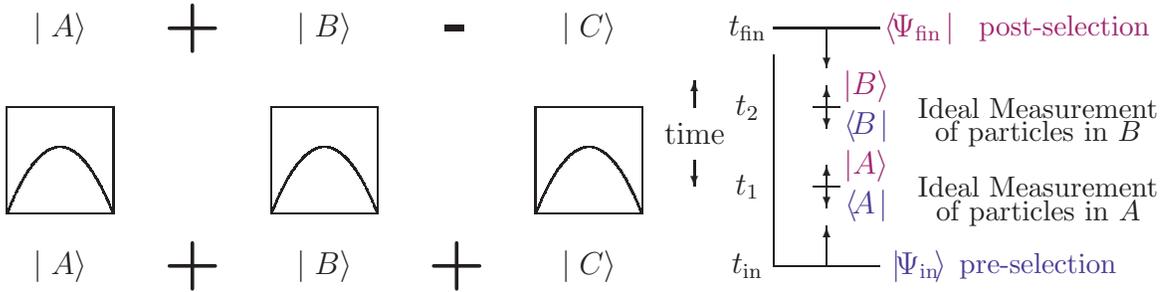
\begin{figure}[h]
\begin{picture}(200,150)(0,0)
\put(0,20){\framebox(40,40)}
\put(100,20){\framebox(40,40)}
\put(200,20){\framebox(40,40)}
\put(20,0){\makebox(0,0){$\mid A \rangle$}}
\put(70,0){\makebox(0,0){\huge\bf+}}
\put(120,0){\makebox(0,0){$\mid B \rangle$}}
\put(170,0){\makebox(0,0){\huge\bf+}}
\put(220,0){\makebox(0,0){$\mid C \rangle$}}

\put(20,90){\makebox(0,0){$\mid A \rangle$}}
\put(70,90){\makebox(0,0){\huge\bf+}}
\put(120,90){\makebox(0,0){$\mid B \rangle$}}
\put(170,90){\makebox(0,0){\huge\bf-}}
\put(220,90){\makebox(0,0){$\mid C \rangle$}}

\bezier{500}(0,20)(20,70)(40,20)
\bezier{500}(100,20)(120,70)(140,20)
\bezier{500}(200,20)(220,70)(240,20)

\put(290,0){\line(0,1){80}}
\put(290,0){\line(1,0){40}}
\put(305,60){\line(1,0){10}}
\put(310,60){\vector(0,1){8}}
\put(310,60){\vector(0,-1){8}}

\put(305,30){\line(1,0){10}}
\put(310,30){\vector(0,1){8}}
\put(310,30){\vector(0,-1){8}}

\put(290,90){\line(1,0){40}}
\put(310,0){\vector(0,1){15}}
\put(310,90){\vector(0,-1){15}}
\put(345,0){\makebox(0,0){\textcolor{BlueViolet}{$|\!\Psi_{\mathrm{in}}\!\rangle$}}}
\put(345,90){\makebox(0,0){\textcolor{RedViolet}{$\langle\!\Psi_{\mathrm{fin}}\!\mid$}}}
\put(325,68){\makebox(0,0){\textcolor{RedViolet}{$|B\rangle$}}}
\put(325,53){\makebox(0,0){\textcolor{BlueViolet}{$\langle\!B\!\mid$}}}

\put(325,38){\makebox(0,0){\textcolor{RedViolet}{$|A\rangle$}}}
\put(325,23){\makebox(0,0){\textcolor{BlueViolet}{$\langle\!A\!\mid$}}}

\put(260,50){\makebox(0,0){time}}
\put(280,0){\makebox(0,0){$t_{\mathrm{in}}$}}
\put(280,90){\makebox(0,0){$t_{\mathrm{fin}}$}}
\put(280,30){\makebox(0,0){$t_1$}}
\put(280,60){\makebox(0,0){$t_2$}}
\put(390,0){\makebox(0,0){\textcolor{BlueViolet}{\small pre-selection}}}
\put(390,60){\makebox(0,0){\small Ideal Measurement}}
\put(390,50){\makebox(0,0){\small of particles in $B$}}

\put(390,30){\makebox(0,0){\small Ideal Measurement}}
\put(390,20){\makebox(0,0){\small of particles in $A$}}

\put(400,90){\makebox(0,0){\textcolor{RedViolet}{\small post-selection}}}
\put(260,40){\vector(0,-1){10}}
\put(260,60){\vector(0,1){10}}

\end{picture}

\caption[One particle superposed in three boxes] 
{\small a) pre-selected vector $|\Psi_{\mathrm{in}} \rangle 
= 1/\sqrt 3 ~(|A\rangle + |B\rangle +|C\rangle)$ propagates
forwards in time from $t_{\mathrm{in}}$ to $t_1$, and post-selected vector $|\Psi_{\mathrm{fin}} \rangle = 1/\sqrt
3   ~(|A\rangle + 
|B\rangle -|C\rangle)$ propagating backwards in time from $t_{\mathrm{fin}}$
to $t_2$.  b)  ideal-measurement of $\mathrm{\bf \hat{P}}_{A}$ at $t_1$ and of $\mathrm{\bf \hat{P}}_{B}$ at $t_2$.}
\label{suppart}
\end{figure}

\subsubsection{Counterfactuals}
\label{counterfactual}
There is a 
widespread tendency  to  ``resolve" these paradoxes by pointing out that there is an element of counter-factual reasoning: the
 contradictions arise only because inferences are made that do not refer to actual experiments.  Had the experiment actually been performed, then standard measurement theory predicts that the system would have been disrupted so that no paradoxical implications arises.
Suppose we applied this to the 3-box-paradox: the resolution then is that there is no meaning to say that the particle is in both boxes without actually {\bf measuring} both boxes during the intermediate time.

We have proven~\cite{at2,jt} that one shouldn't be so quick in
 throwing away counter-factual reasoning; though indeed
 counter-factual statements have no observational meaning,
 such
 reasoning is actually a very good pointer towards interesting
 physical situations.  {\it Without invoking
 counter-factual reasoning}, we have shown that the apparently paradoxical
 reality implied counter-factually has new, {\it
 experimentally accessible} consequences. These observable
 consequences become evident in terms of {\em weak
 measurements},
 which allow us to test - to some extent -  assertions that have
 been otherwise regarded as counter-factual.

 The main argument against counter-factual statements is that
 if we
 actually perform ideal-measurements to test them, we disturb the
 system significantly, and such disturbed conditions hide the counter-factual situation, so no
 paradox
 arises. 
TSQM also provides some novel insights for this ``disturbance-based-argument".
E.g., for the spin-1/2 case (\S\ref{spinhalf}),  
if we verify $\hat{\sigma}_{\mathrm{x}}$ at $t=t_1$ and $\hat{\sigma}_{\mathrm{y}}$ at $t=t_2$, 
$t_{\mathrm{in}}<t_1<t_2<t_{\mathrm{fin}}$, then   $Pr(\hat{\sigma}_{x}=+1)=1$ and $Pr(\hat{\sigma}_{y}=+1)=1$ are
 simultaneously true.  But if we switch the order and perform $\hat{\sigma}_y$ before $\hat{\sigma}_x$, then 
$Pr(\hat{\sigma}_{x}=+1)=1$ and $Pr(\hat{\sigma}_{y}=+1)=1$ are not simultaneously true, since measuring 
$\hat{\sigma}_y$ at time $t=t_1$ would not allow the information from the earlier ($t_{\mathrm{in}}<t$) pre-selection of  
$\hat{\sigma}_{\mathrm{x}}=+1$ to propagate to the later time ($t_2>t_1>t_{\mathrm{in}}$) of the 
$\hat{\sigma}_{\mathrm{x}}$ measurement.  As a consequence, the $\hat{\sigma}_{\mathrm{x}}$ measurement at time $t_2$ would yield both outcomes $\hat{\sigma}_x=\pm 1$\footnote{The same argument applies in the reverse direction of time.  The 4-outcomes are consistent with $\ket{\Psi_{\mathrm{in}}}=|\hat{\sigma}_x=+1\ra$ and $\ket{\Psi_{\mathrm{fin}}}=|\hat{\sigma}_{\mathrm{y}}= +1\ra$.
Physically, the ideal-measurement of $\hat{\sigma}_{\xi}$ exposes the particle to
   a magnetic field with a {\it strong} gradient in the $\xi=45^\circ$ direction, which causes the spin to revolve
   around this axis in an uncertain fashion.}.
So, in general, the finding that $\hat{\sigma}_{\mathrm{x}}= +1$ with certainty or $\hat{\sigma}_{\mathrm{y}}= +1$ with certainty in the pre-and-post-selected ensemble only held when {\it one} of these two measurements was performed in the intermediate time, not both.  Therefore, we should not expect both $\hat{\sigma}_{\mathrm{y}}= +1$ and $\hat{\sigma}_{\mathrm{x}}= +1$ when measured simultaneously through $\hat{\sigma}_{\xi=45^\circ}$.

For the spin-1/2 case, the ABL-assignment relied on only the pre-{\bf or}-post-selection, while in the 3-box-paradox, the ABL assignment relies on both the pre-{\bf and}-post-selection.  However, ABL still only gives an answer for one actual ideal-measurement.  What happens if we tried to obtain two answers for the 3-box-paradox? In order to deduce $\mathrm{\bf \hat{P}}_{\mathrm{A}}=1$, we used information from both
 pre-and-post-selected vectors. 
When we actually measured  $\mathrm{\bf \hat{P}}_{\mathrm{A}}$, then this ideal-measurement will  limit the ``propagation" of the 2-vectors that were relied on to make this determination (see fig. \ref{suppart}.b).
If we subsequently were to measure  $\mathrm{\bf \hat{P}}_{\mathrm{B}}$, then the
necessary information from both the pre-and-post-selected vectors is no longer available (i.e. information from $t_{\mathrm{in}}$ cannot propagate beyond the ideal-measurement of $\mathrm{\bf \hat{P}}_{A}$ at time $t_1$ due to the disturbance caused by the ideal-measurement of $\mathrm{\bf \hat{P}}_{\mathrm{A}}$).  {\it Thus, even though $\mathrm{\bf \hat{P}}_{\mathrm{A}}$ and $\mathrm{\bf \hat{P}}_{\mathrm{B}}$ commute, ideal-measurements of one can disturb ideal-measurement of the other.}\footnote{This is related to a violation of the product rule. In general, if $|\Psi_1\rangle$ is an eigenvector of $\hat{A}$  with eigenvalue $a$ and $|\Psi_2\rangle$ is an eigenvector of $\hat{B}$  with eigenvalue $b$ and $[\hat{A},\hat{B}]=0$, then if $\hat{A}$ and $\hat{B}$ are known only by either pre-selection {\bf or} post-selection, then the product rule is valid, i.e. $\hat{A}\hat{B}=ab$.  However if $\hat{A}$ and $\hat{B}$ are known by both pre-selection {\bf and} post-selection, then the product rule is not valid, i.e. $\hat{A}\hat{B}\neq ab$, i.e. they can still disturb each other, even though they commute.~\cite{vaidman1993b})}

Since we have understood the reason why both statements are not simultaneously true as a result of disturbance, we can now see the ``sense" in which the definite ABL assignments can be simultaneously relevant. Our main argument is 
 that if one doesn't perform
 absolutely precise (ideal) measurements but is willing to accept
 some
 finite accuracy, then one can bound the disturbance on the system. For example, according to Heisenberg's
 uncertainty
 relations, a precise measurement of position
 reduces the
 uncertainty in position to zero $\Delta x=0$ but produces an
 infinite uncertainty in momentum $\Delta p=\infty$. On the
 other
 hand, if we measure the position only up to some finite
 precision
 $\Delta x=\Delta$ we can limit the disturbance of momentum
 to a
 finite amount $\Delta p\geq \hbar/\Delta$. 
By replacing precise measurements with a bounded-measurement paradigm, counter-factual thought experiments become experimentally accessible.
What we often find is
 that
 the paradox remains - measurements produce surprising and often
 strange, but nevertheless consistent structures.
With limited-disturbance measurements, there {\bf is} a sense in which both $Pr(\hat{\sigma}_{x}=+1)=1$ {\it and} $Pr(\hat{\sigma}_{y}=+1)=1$
are simultaneously relevant because measurement of one  does {\it not} disturb the other. Since measurement of $\hat{\sigma}_{\xi}$ also can be understood as a simultaneous measurement of $\hat{\sigma}_x$ {\it and} $\hat{\sigma}_y$, we will see that with limited-disturbance measurements, we can simultaneously use both $\hat{\sigma}_{x}=+1$ and $\hat{\sigma}_{y}=+1$ to obtain $(\hat{\sigma}_{\xi=45^\circ})_{\mathrm{w}} =\frac{\langle \uparrow_y\vert \frac{\hat{\sigma}_y+\hat{\sigma}_x}{\sqrt{2}} 
\vert\!\uparrow_x
\rangle}{\langle{\uparrow_y}\vert{\uparrow_x}\rangle}=
\frac{{\left\{\langle \uparrow_y\vert \hat{\sigma}_y\right\}+\left\{\hat{\sigma}_x\vert\!\uparrow_x
\rangle\right\}}}  {\sqrt{2}\langle{\uparrow_y}\vert{\uparrow_x}\rangle}=\frac{\langle \uparrow_y\vert 1+1 
\vert\!\uparrow_x
\rangle}{\sqrt{2}\langle{\uparrow_y}\vert{\uparrow_x}\rangle}= \sqrt{2}$.

\section{\bf TSQM has revealed new features and effects: Weak-Measurements}
\label{sec.13.3}
\label{WM}

ABL  considered the situation of measurements {\em between} two successive ideal-measurements where one transitions from a pre-selected state $\ket{\Psi_{\mathrm{in}}}$ to a post-selected state $\ket{\Psi_{\mathrm{fin}}}$.  
The state of the system at a time $t\in [t_{\mathrm{in}},t_{\mathrm{fin}}]$, i.e. after $t_{\mathrm{in}}$ when the state is $\ket{\Psi_{\mathrm{in}}}$ and before $t_{\mathrm{fin}}$ when the state is $\ket{\Psi_{\mathrm{fin}}}$ is  generally disturbed by an intermediate ideal-measurement.  
A subsequent theoretical development arising out of the ABL work was the introduction of the weak-value of an observable which can be probed by a  new type of measurement called the weak-measurement  \cite{av}. 
The motivation behind these measurements is to explore the relationship between $\ket{\Psi_{\mathrm{in}}}$ and $\ket{\Psi_{\mathrm{fin}}}$  by reducing the disturbance on the system at the intermediate time.  
This is useful in many ways, e.g. if a weak-measurement of $\hat{A}$ is performed at the intermediate time $t\in [t_{\mathrm{in}},t_{\mathrm{fin}}]$ then, in contrast to the ABL situation,  the basic object in the entire interval $t_{\mathrm{in}}\rightarrow t_{\mathrm{fin}}$ for the purpose of calculating {\it other} weak-values for other measurements 
is the pair of states $\ket{\Psi_{\mathrm{in}}}$ and $\ket{\Psi_{\mathrm{fin}}}$.   

\subsection{\bf Quantum Measurements}
\label{IM}
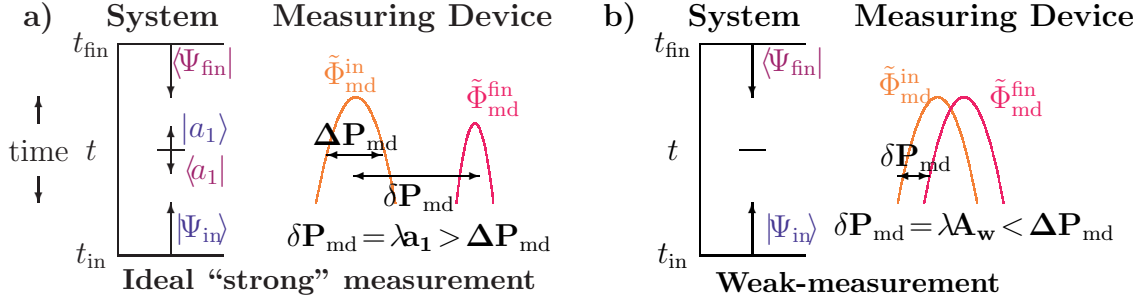
\begin{figure}[h]
\begin{picture}(500,110)(0,0)
\put(40,10){\line(0,1){80}}
\put(40,10){\line(1,0){40}}
\put(55,50){\line(1,0){10}}
\put(60,50){\vector(0,1){9}}
\put(60,50){\vector(0,-1){9}}
\put(40,90){\line(1,0){40}}
\put(60,10){\vector(0,1){20}}
\put(60,90){\vector(0,-1){20}}
\put(72,20){\makebox(0,0){\textcolor{BlueViolet}{$|\!\Psi_{\mathrm{in}}\!\rangle$}}}
\put(72,83){\makebox(0,0){\textcolor{RedViolet}{$\langle\!\Psi_{\mathrm{fin}}\!\!\mid$}}}
\put(73,42){\makebox(0,0){\textcolor{RedViolet}{$\langle\!a_1\!\!\mid$}}}
\put(73,58){\makebox(0,0){\textcolor{BlueViolet}{$|a_1\rangle$}}}
\put(10,50){\makebox(0,0){time}}
\put(30,10){\makebox(0,0){$t_{\mathrm{in}}$}}
\put(30,90){\makebox(0,0){$t_{\mathrm{fin}}$}}
\put(30,50){\makebox(0,0){$t$}}
\put(10,40){\vector(0,-1){10}}
\put(10,60){\vector(0,1){10}}
\put(10,100){\makebox(0,0){\bf a)}}
\put(120,100){\makebox(0,0){\bf System \,\,\,\,\,\,\,Measuring Device}}
\put(120,0){\makebox(0,0){\bf \small Ideal ``strong" measurement}}

\put(128,79){\makebox(0,0){{\textcolor{Orange}{\bf $\tilde{\Phi}^{\mathrm{in}}_{\mathrm{md}}$}}}}

\color{Orange} 

\bezier{500}(115,30)(130,110)(145,30)

\color{WildStrawberry}
\put(182,70){\makebox(0,0){\bf $\tilde{\Phi}^{\mathrm{fin}}_{\mathrm{md}}$}}

\bezier{500}(168,30)(175,90)(182,30)

\color{black}
\put(154,32){\makebox(0,0){$\bf \delta P_{\mathrm{md}}$}}
\put(130,48){\vector(1,0){10}}
\put(130,48){\vector(-1,0){11}}
\put(130,55){\makebox(0,0){$\bf \Delta P_{\mathrm{md}}$}}

\put(154,40){\vector(1,0){23}}
\put(154,40){\vector(-1,0){25}}
\put(154,17){\makebox(0,0){$\bf \delta P_{\mathrm{md}}\!=\!\lambda \!a_1\!>\!\Delta P_{\mathrm{md}}$}}


\put(338,75){\makebox(0,0){{\textcolor{Orange}{\bf $\tilde{\Phi}^{\mathrm{in}}_{\mathrm{md}}$}}}}

\color{Orange} 

\bezier{500}(335,30)(350,110)(365,30)

\color{WildStrawberry}
\put(380,70){\makebox(0,0){\bf $\tilde{\Phi}^{\mathrm{fin}}_{\mathrm{md}}$}}

\bezier{500}(345,30)(360,110)(375,30)


\color{black}
\put(342,47){\makebox(0,0){$\bf \delta P_{\mathrm{md}}$}}
\put(341,40){\vector(1,0){6}}
\put(341,40){\vector(-1,0){6}}
\put(364,21){\makebox(0,0){$\bf \delta P_{\mathrm{md}}\!=\!\lambda \!A_w\!<\!\Delta P_{\mathrm{md}}$}}

\put(320,0){\makebox(0,0){\bf \small Weak-measurement}}
\put(230,100){\makebox(0,0){\bf b)}}
\put(340,100){\makebox(0,0){\bf System \,\,\,\,\,\,\,Measuring Device}}

\put(260,10){\line(0,1){80}}
\put(260,10){\line(1,0){40}}
\put(275,50){\line(1,0){10}}
\put(260,90){\line(1,0){40}}
\put(280,10){\vector(0,1){20}}
\put(280,90){\vector(0,-1){20}}
\put(295,20){\makebox(0,0){\textcolor{BlueViolet}{$|\!\Psi_{\mathrm{in}}\!\rangle$}}}
\put(295,83){\makebox(0,0){\textcolor{RedViolet}{$\langle\!\Psi_{\mathrm{fin}}\!\!\mid$}}}
\put(250,10){\makebox(0,0){$t_{\mathrm{in}}$}}
\put(250,90){\makebox(0,0){$t_{\mathrm{fin}}$}}
\put(250,50){\makebox(0,0){$t$}}

\end{picture}

\caption[Two vector formulation with strong and weak-measurements]{\small  a) 
with an ideal or ``strong" measurement at $t$ (characterized e.g. by $\delta P_{\mathrm{md}}\!=\!\lambda \!a_1\!\gg\!\Delta P_{\mathrm{md}}$), then ABL gives the probability to obtain a collapse onto eigenstate $a_1$ by propagating \textcolor{RedViolet}{$\langle\!\Psi_{\mathrm{fin}}\!\!\mid$} backwards in time from $t_{\mathrm{fin}}$
to $t$ and \textcolor{BlueViolet}{$|\!\Psi_{\mathrm{in}}\!\rangle$} forwards in time from $t_{\mathrm{in}}$ to $t$; in addition, the collapse caused by ideal-measurement at $t$ creates a new boundary condition 
\textcolor{BlueViolet}{$|a_1\rangle$}
$\!\!$\textcolor{RedViolet}{$\langle\!a_1\!\!\mid$} at time $t\in [t_{\mathrm{in}},t_{\mathrm{fin}}]$; b) 
if a weak-measurement is performed at $t$ (characterized e.g. by $\delta P_{\mathrm{md}}\!=\!\lambda \!A_w\!\ll\!\Delta P_{\mathrm{md}}$), then the outcome of the weak-measurement, the weak-value, can be calculated by propagating the
state \textcolor{RedViolet}{$\langle\!\Psi_{\mathrm{fin}}\!\!\mid$} backwards in time from $t_{\mathrm{fin}}$
to $t$ and the state \textcolor{BlueViolet}{$|\!\Psi_{\mathrm{in}}\!\rangle$} 
forwards in time from 
$t_{\mathrm{in}}$ to $t$; the weak-measurement does not cause a collapse and thus no new boundary condition is created at time $t$.}
\label{seqm1}
\end{figure}

Weak-measurements~\cite{av} originally grew out of the quantum measurement theory developed by von Neumann~\cite{vn}\footnote{Weak-measurements and their outcome, weak-values, can be derived in all approaches to quantum measurement theory.  E.g. the usual projective measurement typically utilized in quantum experiments is a special case of these weak-measurements~\cite{brun}.}.
First we consider ideal-measurements of observable $\hat{A}$ by using an interaction Hamiltonian $H_{\mathrm{int}}$ of the form 
$H_{\mathrm{int}}=-\lambda(t)\hat{Q}_{\mathrm{md}}\hat{A}$
where $\hat{Q}_{\mathrm{md}}$ is an observable of the measuring-device (e.g. the position of the pointer) and $\lambda(t) $ is a coupling constant which determines the duration and strength  of the measurement.
For an impulsive  measurement we need the coupling to be strong and short and thus take $\lambda (t)\neq0$ only for
   $t\in(t_0-\varepsilon,t_0+\varepsilon)$ and set
   $\lambda=\int_{t_0-\varepsilon}^{t_0+\varepsilon}\lambda(t)dt$. We may then neglect the time evolution
   given by $H_{\mathrm{s}}$ and $H_{\mathrm{md}}$ in the complete Hamiltonian $H=H_{\mathrm{s}}+_{\mathrm{md}}+H_{\mathrm{int}}$.
Using the Heisenberg equations-of-motion for the momentum $\hat{P}_{\mathrm{md}}$ of the measuring-device (conjugate to the position $\hat{Q}_{\mathrm{md}}$), we see that $\hat{P}_{\mathrm{md}}$ evolves according to $\frac{d\hat{P}_{\mathrm{md}}}{d t}=\lambda (t) \hat{A}$.
Integrating this, we see that $P_{\mathrm{md}}(T)-P_{\mathrm{md}}(0)=\lambda \hat{A}$, where $P_{\mathrm{md}}(0)$ characterizes the initial state of the measuring-device and $P_{\mathrm{md}}(T)$ characterizes the final.  
To make a more precise determination of $\hat{A}$ requires that the shift in $P_{\mathrm{md}}$, i.e. $\delta P_{\mathrm{md}}=P_{\mathrm{md}}(T)-P_{\mathrm{md}}(0)$, be distinguishable from it's uncertainty, $\Delta P_{\mathrm{md}}$.  This occurs, e.g., if $P_{\mathrm{md}}(0)$ and $P_{\mathrm{md}}(T)$ are more precisely defined and/or if $\lambda$ is sufficiently large (see figure \ref{seqm1}.a).  
However, under these conditions (e.g. if the measuring-device approaches a delta function in $P_{\mathrm{md}}$),
 then the disturbance or back-reaction on the system is increased due to a larger  $H_{\mathrm{int}}$, the result of the larger $\Delta Q_{\mathrm{md}}$ ($\Delta Q_{\mathrm{md}}\geq\frac{1}{\Delta P_{\mathrm{md}}}$).
When $\hat{A}$ is 
measured in this way, then any operator $\hat{O}$ ($[\hat{A},\hat{O}]\neq 0$) is disturbed because it evolved according to $\frac{d}{dt}{\hat{O}}=i\lambda(t)[\hat{A},\hat{O}]\hat{Q}_{\mathrm{md}}$, and since $\lambda\Delta Q_{\mathrm{md}}$ is not zero, $\hat{O}$ changes in an uncertain way proportional to $\lambda\Delta Q_{\mathrm{md}}$.\footnote{E.g. in the spin-1/2 example, the conditions for an ideal-measurement $\delta P_{\mathrm{md}}^{\xi}=\lambda \hat{\sigma}_{\xi}\gg\Delta P_{\mathrm{md}}^{\xi}$ 
will also necessitate $\Delta Q_{\mathrm{md}}^{\xi} \gg \frac{1}{\lambda\hat{\sigma}_{\xi}}$ which will thereby create a back-reaction causing a precession in the spin such that $\Delta \Theta \gg 1$ (i.e. more than one revolution), thereby destroying (i.e. making completely uncertain) the information that in the past we had $\hat{\sigma}_x=+1$,
   and in the future we will have $\hat{\sigma}_y=+1$.}  

In the Schroedinger picture, the time evolution operator for the
   complete system from $t=t_0-\varepsilon$ to $t=t_0+\varepsilon$ is $\exp\{-i\int_{t_0-\varepsilon}^{t_0+\varepsilon}H(t)dt\}=\exp\{-i\lambda\hat{Q}_{\mathrm{md}}\hat{A}\}$.  This shifts $P_{\mathrm{md}}$ (see figure \ref{seqm1}.a).  If before the measurement the system was in a superposition of eigenstates of $\hat{A}$, then the measuring-device will also be superposed proportional
   to the system. This leads to the ``quantum measurement problems," discussed in ~\cite{leggett}.  A conventional solution to this problem is to argue that because the measuring-device is macroscopic, it cannot be in a
   superposition, and so it will ``collapse" into one of these states and the system will collapse
   with it.

\subsubsection{Weakening the interaction between system and measuring device}
\label{weakennopost}
Following our intuition we now perform measurements which do not disturb either the pre-or-post-selections. The interaction $H_{\mathrm{int}}\!=\!-\lambda(t)\hat{Q}_{\mathrm{md}}\hat{A}$ is weakened 
 by minimizing $\lambda \Delta Q_{\mathrm{md}}$.  For simplicity, we consider $\lambda\ll 1$ (assuming without lack of generality that the state of the measuring-device is a Gaussian with spreads $\Delta P_{\mathrm{md}}\!=\!\Delta Q_{\mathrm{md}}\!=\!1$).   We may then set $e^{ -i \lambda  \hat{Q}_{\mathrm{md}} \hat{A}
 }\!\approx\! 1-i\lambda \hat{Q}_{\mathrm{md}} \hat{A}$ and use a theorem~\cite{avg}\footnote{where $\la\hat{A}\ra = \ave{\Psi}{\hat{A}}$, $|\Psi\rangle$ is any vector in Hilbert space, $\Delta A^2 = \ave{\Psi}{(\hat{A} - \la\hat{A}\ra)^2}$, and $\ket{\Psi_\perp}$ is a state such that $\amp {\Psi}{\Psi_\perp} = 0$.}:
\beq\label{identity}
\hat{A} |\Psi \rangle = \la\hat{A}\ra \ket{\Psi}  + \Delta A \ket{\Psi_\perp}\, ,
\label{thm1}
\eeq
 to show that before the post-selection, the system state is:
   \begin{equation}
      e^{ -i  \lambda\hat{Q}_{\mathrm{md}} \hat{A}}|\Psi_{\mathrm{in}}\ra\!=\!
(1\!-\!i\lambda \hat{Q}_{\mathrm{md}}\hat{A})|\Psi_{\mathrm{in}}\rangle\!=\!(1\!-i\!\lambda \hat{Q}_{\mathrm{md}}\langle\hat{A}\rangle)|\Psi_{\mathrm{in}}\rangle\!-i\!\lambda \hat{Q}_{\mathrm{md}}\Delta\hat{A}|\Psi_{\mathrm{in}\perp}\rangle
   \end{equation}

\noindent Using the norm of this state ${\parallel (1-i\lambda \hat{Q}_{\mathrm{md}}\hat{A})|\Psi_{\mathrm{in}}\rangle
      \parallel}^2=1+{\lambda ^2\hat{Q}_{\mathrm{md}}^2}\langle \hat{A}^2\rangle$,
 the probability to leave $|\Psi_{\mathrm{in}}\rangle$ un-changed after the measurement is:
   \begin{equation}
      \frac{1+{\lambda ^2\hat{Q}_{\mathrm{md}}^2}{\langle\hat{A}\rangle}^2}
   {1+{\lambda ^2\hat{Q}_{\mathrm{md}}^2}\langle
   \hat{A}^2\rangle}\longrightarrow 1\,\,\,\,\,\,(\lambda \rightarrow 0)
   \end{equation}
while the probability to disturb the state (i.e. to obtain $|\Psi_{in\perp}\rangle$) is:
   \begin{equation}\label{14.11}
      \frac{{\lambda ^2\hat{Q}_{\mathrm{md}}^2}{\Delta\hat{A}}^2}
   {1+{\lambda ^2\hat{Q}_{\mathrm{md}}^2}\langle
   \hat{A}^2\rangle}\longrightarrow 0\,\,\,\,\,\,(\lambda \rightarrow 0)
\label{collprob}
   \end{equation}
The final state of the measuring-device is now a superposition of many substantially overlapping Gaussians with probability distribution given by $Pr(P_{\mathrm{md}})=\sum_i |\la a_i|\Psi_{\mathrm{in}}\ra|^2 \exp\left\{{-\frac{(P_{\mathrm{md}}-\lambda a_i)^{2}} {2\Delta P_{\mathrm{md}}^{2}}}\right\} $.  
This sum is a Gaussian mixture, so it can be approximated by a single Gaussian 
 $\tilde{\Phi}^{\mathrm{fin}}_{\mathrm{md}}(P_{\mathrm{md}})\approx\langle P_{\mathrm{md}}|e^{-i\lambda \hat{Q}_{\mathrm{md}}\la\hat{A}\ra}|\Phi^{\mathrm{in}}_{\mathrm{md}}\rangle\approx\exp\left\{-{{(P_{\mathrm{md}}-\lambda\la\hat
 A\ra)^2}\over{\Delta P_{\mathrm{md}}^2}}\right\}$ centered on $\lambda\la\hat{A}\ra$.

\subsubsection{Information gain without disturbance: safety in numbers}
\label{infogain}
It follows from 
eq.~\ref{collprob} that the probability for a collapse  decreases as $O(\lambda ^2)$, but the measuring-device's shift grows linearly $O(\lambda)$, so $\delta P_{\mathrm{md}}=\lambda a_i$.
For a sufficiently weak interaction (e.g. $\lambda\ll 1$), the probability for a collapse can be made arbitrarily small, while the measurement still yields information but becomes less precise because 
the shift in the measuring-device is much smaller than its uncertainty $\delta P_{\mathrm{md}}\ll\Delta P_{\mathrm{md}}$ (figure \ref{seqm1}.b).
If we perform this measurement on a single particle, then 
two non-orthogonal states will be indistinguishable.
If this were possible, it would violate unitarity because these states could time evolve into orthogonal states $|\Psi_1\ra|\Phi^{\mathrm{in}}_{\mathrm{md}}\ra \longrightarrow |\Psi_1\ra|\Phi^{\mathrm{in}}_{\mathrm{md}}(1)\ra$ and 
$|\Psi_2\ra|\Phi^{\mathrm{in}}_{\mathrm{md}}\ra \longrightarrow |\Psi_2\ra|\Phi^{\mathrm{in}}_{\mathrm{md}}(2)\ra$, with $|\Psi_1\ra|\Phi^{\mathrm{in}}_{\mathrm{md}}(1)\ra$ orthogonal to $|\Psi_2\ra|\Phi^{\mathrm{in}}_{\mathrm{md}}(2)\ra$.  With weakened measurement interactions, this does not happen because the measurement of these two non-orthogonal states causes a smaller shift in the measuring-device than it's uncertainty.  We conclude that the shift $\delta P_{\mathrm{md}}$ of the measuring-device is  a measurement error because $ \tilde{\Phi}_{\mathrm{fin}}^{\mathrm{MD}}(P_{\mathrm{md}})=\langle
      P_{\mathrm{md}}-\lambda \la\hat{A}\ra|\Phi^{\mathrm{in}}_{\mathrm{md}}\rangle\approx\langle P_{\mathrm{md}}|\Phi^{\mathrm{in}}_{\mathrm{md}}\rangle$ for $\lambda \ll 1$.
Nevertheless, if a large ($N\geq\frac{N'}{\lambda }$) ensemble of  particles is used, then the shift of all the measuring-devices ($\delta P^{tot}_{\mathrm{md}}\approx\lambda \la\hat{A}\ra\frac{N'}{\lambda}=N'\la\hat{A}\ra$) becomes distinguishable because of repeated integrations, while  the collapse probability   still goes to zero.
That is, for a large ensemble of particles which are all either $|\Psi_2\ra$ or
   $|\Psi_1\ra$, this measurement can distinguish between them even if $|\Psi_2\ra$ and
   $|\Psi_1\ra$ are not orthogonal\footnote{because the scalar product $\langle\Psi_1\upn|\Psi_2\upn\rangle=\cos^n\theta\longrightarrow 0$}. 

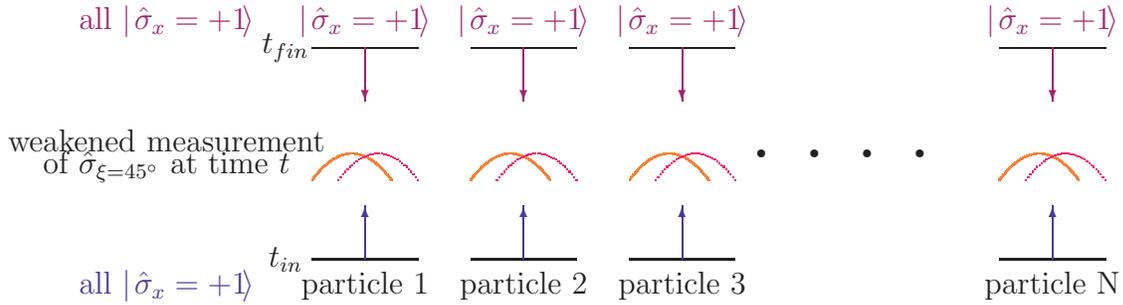
\begin{figure}[h] 
\vskip 1cm
\begin{picture}(400,90)(0,0)
\put(90,10){\line(1,0){40}}
\put(90,90){\line(1,0){40}}
\color{BlueViolet}
\put(35,0){\makebox(0,0){all $\mid\!\hat{\sigma}_x =+1\!\rangle$}}
\put(110,10){\vector(0,1){20}}
\color{RedViolet}
\put(110,90){\vector(0,-1){20}}
\put(35,100){\makebox(0,0){all $\mid\!\hat{\sigma}_x =+1\!\rangle$}}

\put(110,100){\makebox(0,0){$\mid\!\hat{\sigma}_x =+1\!\rangle$}}

\put(170,100){\makebox(0,0){$\mid\!\hat{\sigma}_x =+1\!\rangle$}}

\put(230,100){\makebox(0,0){$\mid\!\hat{\sigma}_x =+1\!\rangle$}}

\put(370,100){\makebox(0,0){$\mid\!\hat{\sigma}_x =+1\!\rangle$}}

\color{OliveGreen}


\color{Black}

\put(110,0){\makebox(0,0){particle 1}}
\put(150,10){\line(1,0){40}}
\put(150,90){\line(1,0){40}}
\color{BlueViolet}

\put(170,10){\vector(0,1){20}}
\color{RedViolet}

\put(170,90){\vector(0,-1){20}}

\color{Black}
\put(170,0){\makebox(0,0){particle 2}}
\put(210,10){\line(1,0){40}}
\put(210,90){\line(1,0){40}}
\color{BlueViolet}

\put(230,10){\vector(0,1){20}}
\color{RedViolet}

\put(230,90){\vector(0,-1){20}}

\color{Black}
\put(230,0){\makebox(0,0){particle 3}}

\put(260,50){\circle*{3}}
\put(280,50){\circle*{3}}
\put(300,50){\circle*{3}}
\put(320,50){\circle*{3}}
\put(350,10){\line(1,0){40}}
\put(350,90){\line(1,0){40}}
\color{BlueViolet}

\put(370,10){\vector(0,1){20}}
\color{RedViolet}

\put(370,90){\vector(0,-1){20}}
\color{Black}
\put(370,0){\makebox(0,0){particle N}}
\put(80,10){\makebox(0,0){$t_{in}$}}
\put(80,90){\makebox(0,0){$t_{fin}$}}
\put(35,55){\makebox(0,0){weakened measurement}}
\put(35,45){\makebox(0,0){of $\hat{\sigma}_{\xi=45^\circ}$ at time $t$}}

\color{Orange}
\bezier{500}(90,40)(105,60)(120,40)

\color{WildStrawberry}
\bezier{35}(100,40)(115,60)(130,40)

\color{Orange}
\bezier{500}(150,40)(165,60)(180,40)
\color{WildStrawberry}
\bezier{35}(160,40)(175,60)(190,40)

\color{Orange}
\bezier{500}(210,40)(225,60)(240,40)
\color{WildStrawberry}
\bezier{35}(220,40)(235,60)(250,40)

\color{Orange}
\bezier{500}(350,40)(365,60)(380,40)
\color{WildStrawberry}
\bezier{35}(360,40)(375,60)(390,40)

\end{picture}

\caption[]{Obtaining the average for an ensemble.}
{\small }
\label{fig1}
\end{figure}

\begin{figure}[tbp] 
  \centering
  \includegraphics[bb=144 146 466 639,width=3.14in,height=4.71in,keepaspectratio]{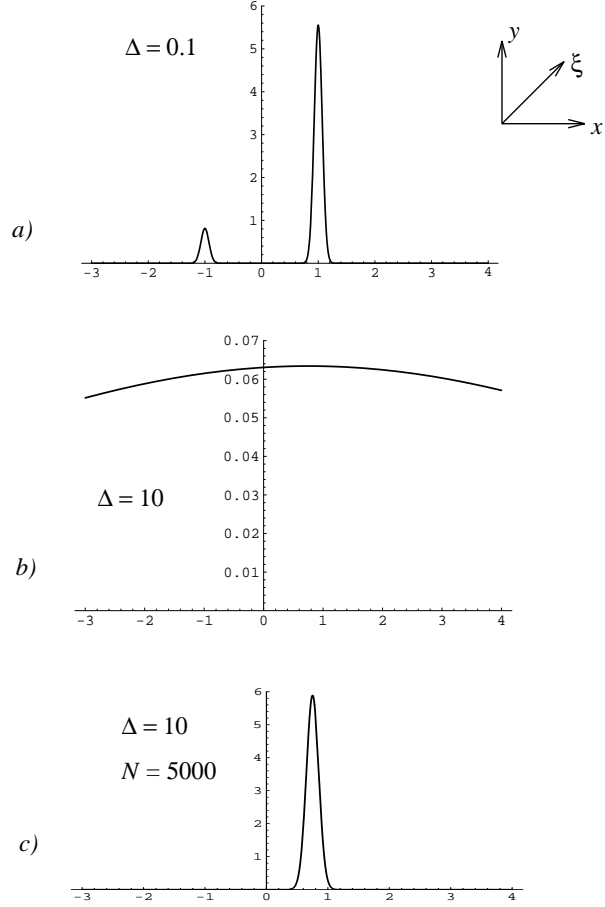}
\caption[Spin component measurement without post-selection]{\small {\bf
~ ``Spin component measurement without post-selection.}~ Probability distribution of the pointer variable for measurement of
$\sigma_\xi$ when the particle is pre-selected in the state $\vert
{\uparrow_x} \rangle$.  ($a$) Strong measurement, $\Delta = 0.1.$
($b$) Weak  measurement, $\Delta = 10$.  ($c$) Weak-measurement on the ensemble
of 5000 particles. The original width of the peak, 10, is reduced to 
$10/\sqrt 5000 \simeq 0.14$. In the strong measurement ($a$) the pointer is localized
around the eigenvalues $\pm1$, while in the weak-measurements ($b$) and ($c$)
the peak is located  in the expectation value $\langle {\uparrow_x}|
\sigma_\xi|{\uparrow_x} \rangle = 1/\sqrt2$." From \cite{av2v}}
\label{wmspin}
\end{figure}

Using these observations, we now emphasize that 
the average of any operator $\hat{A}$, i.e. $\la\hat{A}\ra\equiv\la\Psi|\hat{A}|\Psi\ra$, can be obtained in three distinct cases~\cite{at3,spie-nswm}:  
\begin{enumerate}
\item {\bf Statistical method with disturbance: } 
 the traditional approach is to perform ideal-measurements of $\hat{A}$ on each particle, obtaining a variety of different eigenvalues, and then manually calculate   
 the usual statistical average to obtain $\la\hat{A}\ra$.
\item {\bf Statistical method without disturbance } as demonstrated by using $\hat{A}|\Psi\rangle=\la\hat{A}\ra\ket{\Psi}  + \Delta A \ket{\Psi_\perp}$.
We can also verify that there was no disturbance: consider the spin-1/2 example (\S\ref{spinhalf}), pre-selecting an ensemble, $|{\uparrow_x} \rangle$, 
then performing  
a weakened-measurement of $\hat{\sigma}_\xi$ and finally a post-selection again in the $x$-direction (figure \ref{fig1}).  For every post-selection, we will again find $|{\uparrow_x} \rangle$ with greater and greater certainty (in the weakness limit), verifying our claim of no disturbance.  Each measuring device is centered on 
$\langle \uparrow_x\vert \sigma_{\xi} \vert\uparrow_x\rangle=\frac{1}{\sqrt{2}}$
and the whole ensemble can be used to reduce the spread (figure \ref{wmspin}.c).  The weakened interaction for $\hat{\sigma}_\xi$ means that the inhomogeneity in the magnetic field induces a shift in momentum which is less than the uncertainty 
$\delta P_{\mathrm{md}}^\xi < \Delta P_{\mathrm{md}}^\xi$ and thus a wave packet corresponding to $\frac{\hat{\sigma}_x+\hat{\sigma}_y}{\sqrt{2}}=1$ will be broadly overlapping with the wave packet corresponding to $\frac{\hat{\sigma}_x+\hat{\sigma}_y}{\sqrt{2}}=-1$.
A particular example is depicted in fig. \ref{wmspin}.a. (following~\cite{av2v}) with 
$\Phi^{\mathrm{md}}_{\mathrm{in}} (P_{\mathrm{md}}) =(\Delta ^2 \pi )^{-1/4} \exp\{{ -{{P_{\mathrm{md}}^2} /{2\Delta ^2}}}\}$ and $\Delta\equiv\Delta P_{\mathrm{md}}$ now parametrizes the ``weakness" of the interaction instead of $\lambda$. 
In the ideal-measurement regime of $\Delta <<1$, the probability distribution of the measuring-device is a sum of 2 distributions centered on  eigenvalues $\pm 1$, figure \ref{wmspin}.a.
\begin{equation}
Pr(P_{\mathrm{md}})=\cos ^2(\pi/8) e^{ -(P_{\mathrm{md}}-1)^2 /{\Delta}^2}
+ \sin ^2(\pi/8) e^{ -(P_{\mathrm{md}}+1)^2 /{\Delta} ^2}
\end{equation}
The weak regime occurs when $\Delta$ is larger than the separation between the eigenvalues of $\pm 1$ (i.e. $\Delta >> 1$); e.g. \ref{wmspin}.b.

\item {\bf Non-statistical method without disturbance } is the case where $\la\Psi|\hat{A}|\Psi\ra$ is the ``eigenvalue" of a single 
``collective operator," $\hat{A}\upn\equiv \frac{1}{N} \sum_{\mathrm{i=1}}^{N} \hat{A}_i$ (with  $\hat{A}_i $ the same operator $\hat{A}$ acting on the $i$-th particle).
Using this, we are able to obtain information about $\la\Psi|\hat{A}|\Psi\ra$  without causing disturbance (or a collapse) and without using a statistical approach 
because any product state
  $|\Psi \upn\rangle$ becomes an eigenstate of the operator $\hat{A}\upn$. 
To see this, we apply the theorem $\hat{A} |\Psi \rangle = \la\hat{A}\ra \ket{\Psi}  + \Delta A \ket{\Psi_\perp}$~\cite{avg} to ${\hat{A}}\upn\ket{\Psi\upn}$, i.e.:  
\beq
\hat{A}\upn\ket{\Psi\upn}  = \frac{1}{N}\left[ N \la\hat{A}\ra\ket{\Psi\upn} + \Delta A \sum_i
|\Psi\upn_\perp(i) \rangle \right]
\label{avgop}
\eeq
where $\la\hat{A}\ra$ is the average for any one particle and the states $|\Psi\upn_\perp(i) \rangle$ are mutually orthogonal and are given by
$|\Psi\upn_\perp(i) \rangle = \ket{\Psi}_1\ket{\Psi}_2...\ket{\Psi_\perp}_i...\ket{\Psi}_N$.
That is, the $i$th state has particle $i$ changed to an orthogonal state and all the other particles
remain in the same state.  If we further define a normalized state
 $|\Psi\upn_{\perp} \rangle = \sum_{i}\frac{1}{\sqrt{N}}|\Psi\upn_\perp(i) \rangle$ 
then the last term of eq.~\ref{avgop} is $\frac{\Delta A}{\sqrt{N}}|\Psi\upn_{\perp} \rangle$ and it's size is $|\frac{\Delta A}{\sqrt{N}}
|\Psi\upn_{\perp} \rangle|^2\propto\frac{1}{N}\rightarrow 0$.
Therefore, $|\Psi\upn\rangle$ becomes an eigenstate of $\hat{A}\upn$, with the value
$\la\hat{A}\ra$ and not even a single particle has been disturbed (as $\hat{N} \rightarrow \infty$). 
\end{enumerate}
In the last case, the average for a single particle becomes a robust property over the entire ensemble, so a single experiment is sufficient to
determine the average with great precision. There is no longer
any need to average over results obtained in multiple experiments.

Tradition has dictated that when measurement interactions are limited so there is no disturbance on the system, then no information can be gained. 
However, we have now shown that when considered as a limiting process, the disturbance goes to zero more quickly than the shift in the measuring-device, which means for a large enough ensemble, information (e.g. the expectation value) can be obtained even though not even a single particle is disturbed.  
This viewpoint thereby shifts the standard perspective on two fundamental postulates of QM.\footnote{This is also helpful to understand the quantum to classical transition because typical classical interactions involve these collective observables which do not disturb each other.}

\subsubsection{Adding a post-selection to the weakened interaction: Weak-Values and Weak-Measurements}
Having established a new measurement paradigm -information gain without disturbance- it is fruitful to inquire whether this type of measurement reveals new values or properties.  
With weak-measurements (which involve adding a post-selection to this ordinary -but weakened- von Neumann measurement),  the measuring-device registers a new value, the weak-value.
As an indication of this, we insert a complete set of states $\{ \ket{\Psi_{\mathrm{fin}}}_j \}$ into the outcome of the weak interaction of \S\ref{weakennopost} (i.e. the expectation value $\la\hat{A}\ra$):
\beq
 \la\hat{A}\ra =  \bra{\Psi_{\mathrm{in}}} { \left[\sum_j  \ket{\Psi_{\mathrm{fin}}}_j\bra{\Psi_{\mathrm{fin}}}_j\right]\hat{A}} \ket{\Psi_{\mathrm{in}}}
= \sum_j |\langle \Psi _{\mathrm{fin}} \!\mid_j \!\Psi _{\mathrm{in}}
\rangle|^2\ 
{ {\langle \Psi _{\mathrm{fin}}\! \mid_j \hat{A} \mid \!\Psi _{\mathrm{in}}
\rangle} \over {\langle \Psi _{\mathrm{fin}} \!\mid_j \!\Psi _{\mathrm{in}}
\rangle}}
\label{expweak}
\eeq
If we interpret the states $ \ket{\Psi_{\mathrm{fin}}}_j $ as the outcomes of a final ideal-measurement on the system (i.e. a post-selection)
then performing  a weak-measurement (e.g. with $\lambda\Delta Q_{\mathrm{md}}\rightarrow 0$) during the intermediate time $t\in [t_{\mathrm{in}},t_{\mathrm{fin}}]$, provides the coefficients for $|\langle \Psi_{\mathrm{fin}}\! |_j \Psi_{\mathrm{in}}
\rangle |^2$ which gives  the probabilities $Pr(j)$ 
for obtaining a pre-selection of $\bra{\Psi_{\mathrm{in}}}$ and a post-selection of  $ \ket{\Psi_{\mathrm{fin}}}_j $.  The intermediate weak-measurement does not disturb these states  
and the  quantity
$A_{\mathrm{w}}(j) \equiv { {\langle \Psi _{\mathrm{fin}} \!\mid_j \hat{A} \mid \!\Psi _{\mathrm{in}}
\rangle} \over {\langle \Psi _{\mathrm{fin}} \!\mid_j \!\Psi _{\mathrm{in}}
\rangle}}$
 is the weak-value of $\hat{A}$ given a particular final post-selection $\langle \Psi _{\mathrm{fin}} \!\mid_j$. 
Thus, from the definition  
$\la\hat{A}\ra = \sum_j Pr(j)\,  A_{\mathrm{w}}(j)$,
 one can think of  $\la\hat{A}\ra$ for the whole ensemble as being constructed out of sub-ensembles of pre-and-post-selected-states in which the weak-value is multiplied by a probability for a post-selected-state.

The weak-value arises naturally from a weakened measurement with post-selection:  taking  
 $\lambda<<1$, the final state of measuring-device in the momentum representation becomes: 
\begin{eqnarray}
\bra{P_{\mathrm{md}}} \bra{\Psi_{\mathrm{fin}}}e^{ -i \lambda  \hat{Q}_{\mathrm{md}} \hat{A}
 }\ket{\Psi_{\mathrm{in}}}\ket{\Phi_{\mathrm{in}}^{\mathrm{MD}}}& \approx & 
\bra{P_{\mathrm{md}}} \bra{\Psi_{\mathrm{fin}}}1+i\lambda \hat{Q}_{\mathrm{md}} \hat{A}\ket{\Psi_{\mathrm{in}}}\ket{\Phi_{\mathrm{in}}^{\mathrm{MD}}}\nonumber\\
&\approx& \bra{P_{\mathrm{md}}}\langle\Psi_{\mathrm{fin}}\!\mid
\Psi_{\mathrm{in}} \rangle \lbrace 1+i\lambda \hat{Q} \weakv {\Psi_\mathrm{fin}}{\hat{A}}{\Psi_\mathrm{in} }\rbrace\ket{\Phi_{\mathrm{in}}^{\mathrm{MD}}}\nonumber\\
&\approx & \langle\Psi_{\mathrm{fin}}\ket{\Psi_{\mathrm{in}}}    
\bra{P_{\mathrm{md}}} 
e^{ -i \lambda  \hat{Q}A_{\mathrm{w}}
 }\ket{\Phi_{\mathrm{in}}^{\mathrm{MD}}}\nonumber\\
&\rightarrow & \langle\Psi_{\mathrm{fin}}\ket{\Psi_{\mathrm{in}}}\exp\left\{{-{{(P_{\mathrm{md}}-\lambda \,
 A_{\mathrm{w}})^2}
}}\right\}\\
where \,\,\,A_{\mathrm{w}}&=&\weakv {\Psi_\mathrm{fin}}{\hat{A}}{\Psi_\mathrm{in} }\nonumber
\label{wv1}
 \label{post_selected}
\end{eqnarray}
The final
 state
 of the measuring-device is almost un-entangled with the
 system; it is shifted by a very unusual quantity, the weak-value, $A_{\mathrm{w}}$, 
which is not in general an eigenvalue of $\hat{A}$\footnote{Thereby challenging another fundamental postulate of QM.}.  We have used such limited
 disturbance measurements to explore many 
 paradoxes (see, e.g. ~\cite{at2,jt}).  A number of experiments have been performed to test the predictions made by weak-measurements and results have proven to be in very good agreement with
 theoretical predictions \cite{RSH,Ahnert,Pryde,Wiseman,Parks}.  Since eigenvalues or expectation values can be {\bf derived} from weak-values~\cite{ab}, we believe that the weak-value is indeed of fundamental importance in QM.  In addition, the weak-value is the relevant quantity for all generalized weak interactions with an environment, not just measurement interactions.  The only requirement being that the 2-vectors, i.e. the pre-and-post-selection, are not significantly disturbed by the environment.

\subsection{\bf Fundamentally new features of weak-values}

\subsubsection{ Weak-values and 3-box-paradox}\label{sec.14.5}

\noindent Returning to the 3-box-paradox (\S\ref{3boxesintro}), we can calculate the weak-values of the number of particles in each box, e.g.:
   \begin{displaymath}
      \begin{array}{c}
        (\ket{A}\bra{A})_w=\frac{\bra{\Psi_{\mathrm{fin}}}A\rangle\langle A\ket{\Psi_{\mathrm{in}}}}{\bra{\Psi_{\mathrm{fin}}} {\Psi_{\mathrm{fin}} }\rangle}
        =\frac{\frac{1}{\sqrt{3}} \left\{\langle A|+\langle B|-\langle C|\right\} |A\rangle \langle A|\frac{1}{\sqrt{3}}
        \left\{|A\rangle+|B\rangle+|C\rangle\right\} }{ \frac{1}{\sqrt{3}} \left\{\langle A|+\langle B|-\langle C|\right\} \frac{1}{\sqrt{3}}\left\{|A\rangle+|B\rangle+|C\rangle\right\}}= 
        \frac{\frac{1}{3}1\cdot1}{\frac{1}{3}(1+1-1)}=1\\
      \end{array}
   \end{displaymath}
\noindent However, we can more easily ascertain the weak-values without calculation due to the following theorems:
\bigskip

\noindent {\bf Theorem 1}: The sum of the weak-values is equal to the weak-value of the sum~\cite{proofthm1}:
\beq
if \,\,\, (\mathrm{\bf \hat{P}}_{A})_{\mathrm{w}}=(\mathrm{\bf \hat{P}}_{B} +\mathrm{\bf \hat{P}}_{C} )_{\mathrm{w}} \,\,\, then \,\,\, (\mathrm{\bf \hat{P}}_{A})_{\mathrm{w}}=(\mathrm{\bf \hat{P}}_{B})_{\mathrm{w}} +(\mathrm{\bf \hat{P}}_{C})_{\mathrm{w}}
\eeq

\bigskip

\noindent {\bf Theorem 2}~\cite{proofthm2}: If a single ideal-measurement of an observable $\mathrm{\bf \hat{P}}_{A}$ is performed between the pre-and-post-selection, then if the outcome is definite (e.g. $Prob(\mathrm{\bf \hat{P}}_{A}=1$)=1) then the weak-value is equal to
this eigenvalue (e.g. ($\mathrm{\bf \hat{P}}_{A})_{\mathrm{w}}=1$)~\cite{jmav}. 

\bigskip

This also provides a direct link to the counterfactual statements  (\S\ref{counterfactual}) because all
counterfactual statements which claim that something occurs with
certainty, and which can actually be experimentally verified by
{\bf separate} ideal-measurements, continue to remain true when tested by weak-measurements. However,
given that weak-measurements do not disturb each other, all these
statements can be measured {\bf simultaneously}.

\bigskip

Applying Theorem 2 to the 3-box-paradox, we know the following weak-values with certainty:
\begin{equation}
  \label{psi1}
(\mathrm{\bf \hat{P}}_{\mathrm{A}})_{\mathrm{w}} =1,~~ (\mathrm{\bf \hat{P}}_{\mathrm{B}})_{\mathrm{w}} =1,~~ \mathrm{\bf \hat{P}}_{total}=(\mathrm{\bf \hat{P}}_{\mathrm{A}}+ \mathrm{\bf \hat{P}}_{\mathrm{B}} +\mathrm{\bf \hat{P}}_{\mathrm{C}})_{\mathrm{w}} =1 . 
\end{equation}
Using theorem 1, we obtain:
\begin{eqnarray}
  \label{psi2}
(\mathrm{\bf \hat{P}}_{\mathrm{C}})_{\mathrm{w}}  &=& {\langle\Psi_{\mathrm{fin}}\vert \mathrm{\bf \hat{P}}_{total}-\mathrm{\bf \hat{P}}_{\mathrm{A}}-\mathrm{\bf \hat{P}}_{\mathrm{B}}\vert\Psi_{\mathrm{in}}\rangle
  \over \langle\Psi_{\mathrm{fin}}\vert\Psi_{\mathrm{in}}\rangle}\nonumber\\
&=& (\mathrm{\bf \hat{P}}_{\mathrm{A}}+ \mathrm{\bf \hat{P}}_{\mathrm{B}} +\mathrm{\bf \hat{P}}_{\mathrm{C}})_{\mathrm{w}}- (\mathrm{\bf \hat{P}}_{\mathrm{A}})_{\mathrm{w}} -(\mathrm{\bf \hat{P}}_{\mathrm{B}})_{\mathrm{w}}
= -1 . 
\end{eqnarray}
This surprising theoretical prediction of TSQM has been verified experimentally using photons~\cite{stein1}.
What interpretation should be given to $(\mathrm{\bf \hat{P}}_{\mathrm{C}})_{\mathrm{w}}  =-1$?  
Any weak-measurement which is sensitive to the projection operator $\mathrm{\bf \hat{P}}_{\mathrm{C}}$  will register the opposite effect from those cases in which the projection operator is positive, e.g. a weak-measurement of the amount of charge in  box $C$ in
the intermediate 
time will yield a negative charge (assuming it is a positively charged particle).  For numerous reasons, we believe the most natural interpretation is: there are $-1$ particles in box $C$.

\subsubsection{How the weak-value of a spin-1/2 can be 100} 
\label{spin100}
\begin{figure}[h] 
\vskip 1cm
\begin{picture}(400,90)(0,0)
\put(90,10){\line(1,0){40}}
\put(90,90){\line(1,0){40}}
\color{BlueViolet}
\put(35,0){\makebox(0,0){all $\mid\!\hat{\sigma}_x =+1\!\rangle$}}
\put(110,10){\vector(0,1){20}}
\color{RedViolet}
\put(110,90){\vector(0,-1){20}}
\put(35,113){\makebox(0,0){either}}
\put(35,100){\makebox(0,0){$\mid\!\hat{\sigma}_y =+1\!\rangle$}}
\put(35,87){\makebox(0,0){or $\mid\!\hat{\sigma}_y =-1\!\rangle$}}

\put(110,100){\makebox(0,0){$\mid\!\hat{\sigma}_y =-1\!\rangle$}}

\put(170,100){\makebox(0,0){$\mid\!\hat{\sigma}_y =+1\!\rangle$}}

\put(230,100){\makebox(0,0){$\mid\!\hat{\sigma}_y =-1\!\rangle$}}

\put(370,100){\makebox(0,0){$\mid\!\hat{\sigma}_y =+1\!\rangle$}}

\color{OliveGreen}

\put(170,50){\oval(70,120)}
\put(370,50){\oval(70,120)}

\color{Black}

\put(110,0){\makebox(0,0){particle 1}}
\put(150,10){\line(1,0){40}}
\put(150,90){\line(1,0){40}}
\color{BlueViolet}

\put(170,10){\vector(0,1){20}}
\color{RedViolet}

\put(170,90){\vector(0,-1){20}}

\color{Black}
\put(170,0){\makebox(0,0){particle 2}}
\put(210,10){\line(1,0){40}}
\put(210,90){\line(1,0){40}}
\color{BlueViolet}

\put(230,10){\vector(0,1){20}}
\color{RedViolet}

\put(230,90){\vector(0,-1){20}}

\color{Black}
\put(230,0){\makebox(0,0){particle 3}}

\put(260,50){\circle*{3}}
\put(280,50){\circle*{3}}
\put(300,50){\circle*{3}}
\put(320,50){\circle*{3}}
\put(350,10){\line(1,0){40}}
\put(350,90){\line(1,0){40}}
\color{BlueViolet}

\put(370,10){\vector(0,1){20}}
\color{RedViolet}

\put(370,90){\vector(0,-1){20}}
\color{Black}
\put(370,0){\makebox(0,0){particle N}}
\put(80,10){\makebox(0,0){$t_{in}$}}
\put(80,90){\makebox(0,0){$t_{fin}$}}
\put(35,55){\makebox(0,0){weak-measurement}}
\put(35,45){\makebox(0,0){of $\hat{\sigma}_{\xi=45^\circ}$ at time $t$}}

\color{Orange}
\bezier{500}(90,40)(105,60)(120,40)

\color{WildStrawberry}
\bezier{35}(100,40)(115,60)(130,40)

\color{Orange}
\bezier{500}(150,40)(165,60)(180,40)
\color{WildStrawberry}
\bezier{35}(160,40)(175,60)(190,40)

\color{Orange}
\bezier{500}(210,40)(225,60)(240,40)
\color{WildStrawberry}
\bezier{35}(220,40)(235,60)(250,40)

\color{Orange}
\bezier{500}(350,40)(365,60)(380,40)
\color{WildStrawberry}
\bezier{35}(360,40)(375,60)(390,40)

\end{picture}

\caption[Complete correlation between N particles]{Statistical weak-measurement ensemble.}
{\small }
\label{multparticlesprepost}
\end{figure}
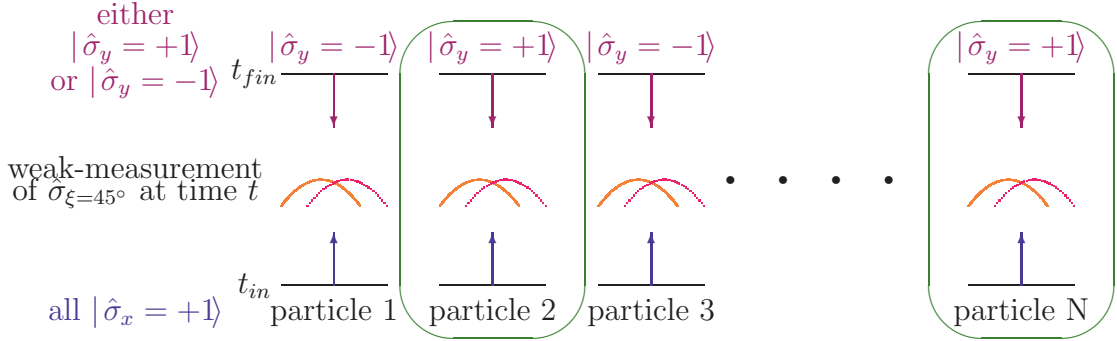

The weak-value for the spin-1/2 considered in \S\ref{spinhalf}  (which was confirmed experimentally for an analogous observable, the polarization~\cite{RSH}) is:
\begin{equation}
(\hat{\sigma}_{\xi=45^\circ})_{\mathrm{w}} =\frac{\langle \uparrow_y\vert \frac{\hat{\sigma}_y+\hat{\sigma}_x}{\sqrt{2}} 
\vert\!\uparrow_x
\rangle}{\langle{\uparrow_y}\vert{\uparrow_x}\rangle}=
\frac{{\left\{\langle \uparrow_y\vert \hat{\sigma}_y\right\}+\left\{\hat{\sigma}_x\vert\!\uparrow_x
\rangle\right\}}}  {\sqrt{2}\langle{\uparrow_y}\vert{\uparrow_x}\rangle}=\frac{\langle \uparrow_y\vert 1+1 
\vert\!\uparrow_x
\rangle}{\sqrt{2}\langle{\uparrow_y}\vert{\uparrow_x}\rangle}= \sqrt{2}
\end{equation}
Normally, the component of spin
$\hat{\sigma}_{\hat{\xi}}$ is an eigenvalue, $\pm 1$, but the weak-value
$(\hat{\sigma}_{\hat{\xi}})_w=\sqrt{2}$ is $\sqrt{2}$ times
bigger,
 (i.e. lies outside the range of 
eigenvalues of ${\bf \hat{\sigma} \cdot n}$)
\footnote{Weak-values even further outside the eigenvalue spectrum can be obtained by post-selecting states which are more anti-parallel to the pre-selection: e.g. if we post-select the $+1$ eigenstate of $(\cos\alpha)\sigma_x + (\sin\alpha)\sigma_z$, then $(\hat{\sigma}_z)_{\mathrm{w}}=\lambda\tan \frac{\alpha}{2}$, yielding arbitrarily large values such as spin-100.}.  
How do we obtain this?  Instead of post-selecting $\hat{\sigma}_x=1$ (figure \ref{fig1}), we post-select $\hat{\sigma}_y=1$ which will be satisfied in one-half the trials (figure \ref{multparticlesprepost}).\footnote{If a post-selection does not satisfy $\hat{\sigma}_y=+1$, then that member of the sub-ensemble must be discarded.  This highlights a fundamental difference between pre-and-post-selection due to the macrosopic arrow-of-time: in contrast to post-selection, if the pre-selection does not satisfy the criteria, then a subsequent unitary transformation can transform to the proper criteria.}

\begin{figure}[tbp] 
  \centering
  \includegraphics[bb=144 146 466 639,width=3.14in,height=4.71in,keepaspectratio]{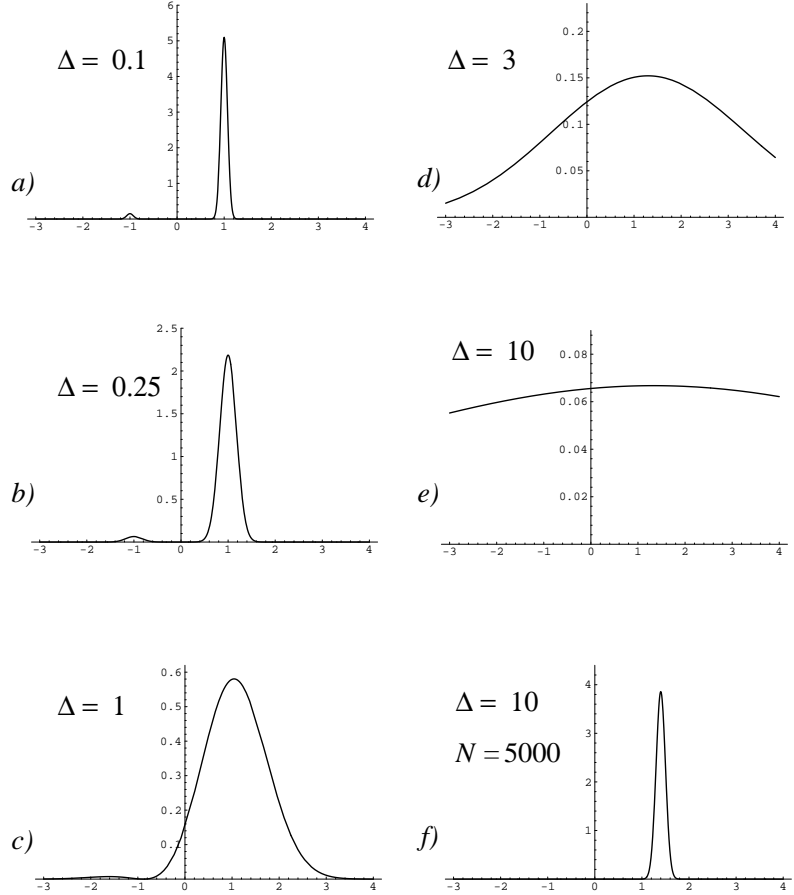}
\caption[Measurement on pre-and-post-selected ensemble of single spins]{\small {\bf
~ ``Measurement on pre-and-post-selected ensemble.}~
 Probability distribution of the pointer variable for measurement of
$\sigma_\xi$ when the particle is pre-selected in the state $\vert
{\uparrow_x} \rangle$ and post-selected in  the state $\vert
{\uparrow_y} \rangle$. The  strength of the
measurement  is 
parameterized by the width of the distribution $\Delta$.
 ($a$) $\Delta = 0.1$; ($b$) $\Delta = 0.25$; ($c$) $\Delta =
1$; ($d$) $\Delta = 3$; ($e$) $\Delta = 10$.
  ($f$) Weak-measurement on the ensemble
of 5000 particles; the original width of the peak, $\Delta = 10$, is reduced to
$10/\sqrt 5000 \simeq 0.14$. In the strong measurements ($a$)-($b$)
the pointer is localized
around the eigenvalues $\pm1$, while in the weak-measurements ($d$)-($f$)
the peak of the distribution is located in the weak-value
$(\sigma_\xi)_w = \langle {\uparrow_y}|
\sigma_\xi |{\uparrow_x} \rangle/\langle {\uparrow_y}|{\uparrow_x} \rangle
= \sqrt2$. The outcomes of the weak-measurement on the ensemble
of 5000 pre-and-post-selected particles, ($f$), are clearly outside the
range of the eigenvalues, (-1,1)." From \cite{av2v}}
\label{wmspin2}
\end{figure}

To show this in an actual calculation, we
use eq.~\ref{post_selected} and the post-selected state of the quantum system in the  $\sigma_{\xi}$ basis ($ \vert\uparrow_y\rangle\equiv \cos (\pi/8) \vert\uparrow_{\xi}\rangle - \sin (\pi/8) 
\vert\downarrow_{\xi}\rangle$), the measuring-device probability 
distribution is:
\begin{equation}
Pr(P_{\mathrm{md}})=N^2 [\cos ^2(\pi/8) e^{ -(P_{\mathrm{md}}-1)^2 /{\Delta} ^2}
- \sin ^2(\pi/8) e^{ -(P_{\mathrm{md}}+1)^2 /{\Delta} ^2}]^2
\end{equation}
With a strong or ideal-measurement, $\Delta\ll1$, the distribution is localized again around the eigenvalues $\pm1$, as illustrated in figures \ref{wmspin2}.a and 
\ref{wmspin2}.b, similar to what occured in figure \ref{wmspin}.a.   
What is different, however, is that when the measurement is weakened, i.e. $\Delta$ 
is made 
larger, then the distribution changes to one single distribution centered around 
$\sqrt{2}$, the weak-value, as illustrated in figures \ref{wmspin2}.c-f, (the width again is reduced with an ensemble \ref{wmspin2}.f).  
Using  eq.~\ref{expweak}, we can see that the weak-value is just the pre-and-post-selected sub-ensemble arising from within the pre-selected-only ensembles.  That is, \ref{wmspin2}.f is a sub-ensemble from the full ensemble represented by  the expectation value, figure \ref{wmspin}.c.
\noindent

\begin{figure}[tbp] 
  \centering
  \includegraphics[bb=144 146 466 639,width=3.14in,height=4.71in,keepaspectratio]{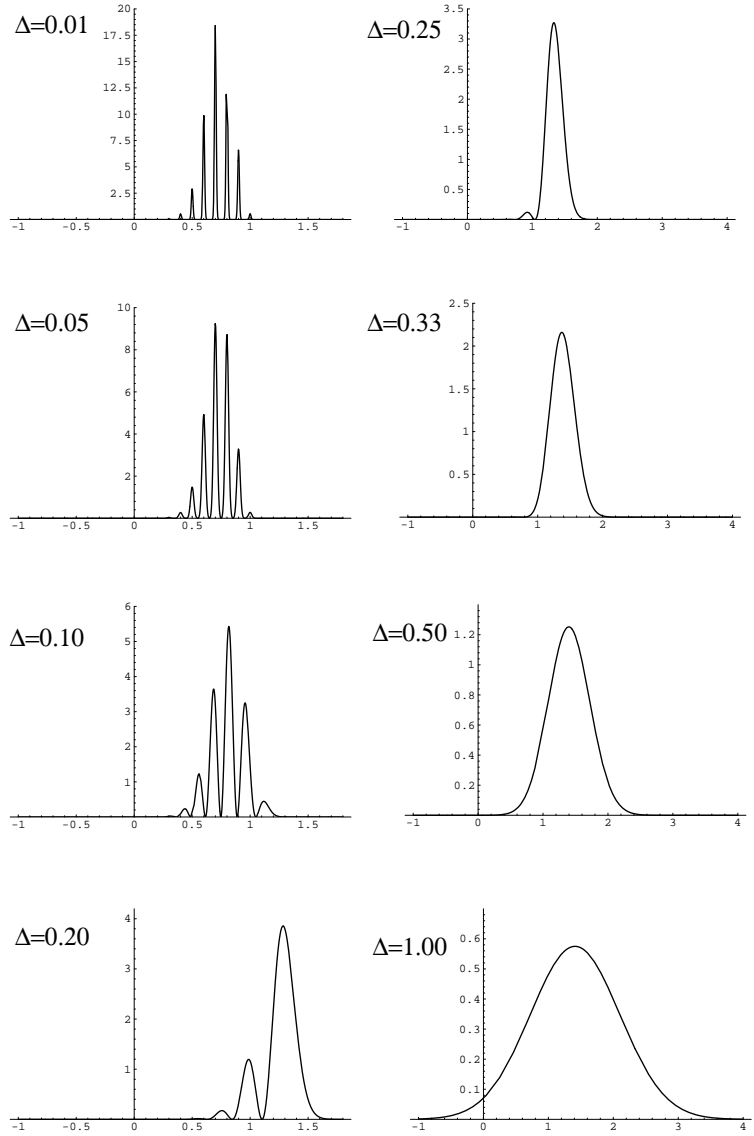}
\vskip 1cm
\caption[Measurement on pre-and-post-selected ensemble of a single large spin]{\small{\bf ~ ``Measurement on a single system.}~ Probability
  distribution of the pointer variable for the measurement of $A
  =(\sum_{i=1}^{20} (\sigma_i)_\xi)/20$ when the system of 20 spin-$1\over 2$
  particles is pre-selected in the state $|\Psi_1\rangle =
  \prod_{i=1}^{20} |{\uparrow_x} \rangle _i$ and post-selected in the
  state ${|\Psi_2\rangle = \prod_{i=1}^{20} |{\uparrow_y} \rangle _i}$.
  While in the very strong measurements, $\Delta = 0.01-0.05$, the
  peaks of the distribution located at the eigenvalues, starting from
  $\Delta = 0.25$ there is essentially a single peak at the location
  of the weak-value, $A_w = \sqrt 2$." From \cite{av2v}}
\label{wmspin3}
\end{figure}

The non-statistical aspect mentioned in case-3 (\S\ref{infogain}) can also be explored by changing the problem slightly.  Instead of considering an ensemble of spin-1/2 particles, we now consider ``particles" which are composed of many ($N$) spin-1/2 particles,
and perform  a weak-measurement of the collective observable $\hat{\sigma}_\xi\upn \equiv \frac{1}{N}
\sum_{\mathrm{i=1}}^{N} \hat{\sigma}_\xi^i$ in the $45^{\circ}$-angle to the $x-y$ plane. Using 
$H_{\mathrm{int}} = -{{\lambda  \delta(t)}\over N}  \hat{Q}_{\mathrm{md}} \sum_{\mathrm{i=1}}^N  \hat{\sigma}^i_\xi$, a particular pre-selection of $|{\uparrow_x} \rangle$ (i.e. $|\Psi_{\mathrm{in}}\upn\rangle = \prod_{\mathrm{j=1}}^N |{\uparrow_x} \rangle_j$) and post-selection
 $|{\uparrow_y}\rangle$
(i.e. 
$\langle\Psi_{\mathrm{fin}}\upn|  = \prod_{\mathrm{k=1}}^N \langle{\uparrow_y}|_k=\prod_{\mathrm{n=1}}^N \left\{\langle{\uparrow_z} |_n+i\langle{\downarrow_z}
|_n\right\}$), 
the final state of the measuring-device is:
\beq
|\Phi_{\mathrm{fin}}^{\mathrm{MD}}\ra=\prod_{j=1}^N \langle{\uparrow_y}|_j \exp\left\{{{\lambda}\over N}  \hat{Q}_{\mathrm{md}} \sum_{\mathrm{k=1}}^N  \hat{\sigma}^k_\xi\right\}  \prod_{i=1}^N|{\uparrow_x} \rangle_i |\Phi_{\mathrm{in}}^{\mathrm{MD}}\ra
\label{bigspinb}
\eeq
Since the spins do not interact with each other, we can calculate one of the products and take the result to the $Nth$ power:
\beq
|\Phi_{\mathrm{fin}}^{\mathrm{MD}}\ra=\prod_{j=1}^N \langle{\uparrow_y}|_j \exp\left\{{{\lambda}\over N}  \hat{Q}_{\mathrm{md}}  \hat{\sigma}^j_\xi\right\} |{\uparrow_x} \rangle_j |\Phi_{\mathrm{in}}^{\mathrm{MD}}\ra=\left\{\langle{\uparrow_y}| \exp\left\{{{\lambda}\over N}  \hat{Q}_{\mathrm{md}}  \hat{\sigma}_\xi\right\} |{\uparrow_x} \rangle\right\}^N\!\!\!\! |\Phi_{\mathrm{in}}^{\mathrm{MD}}\ra
\label{bigspinb}
\eeq
Using the following identity 
   $\exp\left\{{i\alpha\hat{\sigma}_{\vec{n}}}\right\}=\cos\alpha+i\hat{\sigma}_{\vec{n}}\sin\alpha$~\cite{proof}, 
this becomes:
\begin{eqnarray}
\ket{\Phi_{\mathrm{fin}}^{\mathrm{MD}}}&=&\left\{\langle{\uparrow_y}| \left[\cos \frac{{\lambda} \hat{Q}_{\mathrm{md}}}{N}-i\hat{\sigma}_\xi\sin \frac{{\lambda} \hat{Q}_{\mathrm{md}}}{N}\right]  |{\uparrow_x} \rangle\right\}^N |\Phi_{\mathrm{in}}^{\mathrm{MD}}\ra\nonumber\\
&=&
{\left[\langle{\uparrow_y}|{\uparrow_x} \rangle\right]^N }
\left\{\cos \frac{{\lambda} \hat{Q}_{\mathrm{md}}}{N}-i\alpha_w\sin \frac{{\lambda} \hat{Q}_{\mathrm{md}}}{N}\right\}^N  |\Phi_{\mathrm{in}}^{\mathrm{MD}}\ra
\label{bigspina}
\end{eqnarray}
where we have substituted $\alpha_w\equiv(\hat{\sigma}_\xi)_w=\weakv {\uparrow_y}{\hat{\sigma}_\xi}{\uparrow_x }$.    We consider only the second part (the first bracket, a number, can be neglected since it does not depend on $\hat{Q}$ and thus can only affect the normalization):
\beq
\ket{\Phi_{\mathrm{fin}}^{\mathrm{MD}}}=
\left\{1-\frac{{\lambda}^2 (\hat{Q}_{\mathrm{md}})^2}{N^2} - \frac{i{\lambda} \alpha_w\hat{Q}_{\mathrm{md}}}{N} \right\}^N|\Phi_{\mathrm{in}}^{\mathrm{MD}}\ra\approx e^{i\lambda\alpha_w  \hat{Q}_{\it md}^{(N)} } |\Phi_{\mathrm{in}}^{\mathrm{MD}}\ra\nonumber\\
\label{bigspinc}
\eeq
When\footnote{The last approximation was obtained as $N\rightarrow\infty$, using 
$(1+\frac{a}{N})^N=(1+\frac{a}{N})^{\frac{N}{a}a} \approx e^a$.} projected onto $P_{\mathrm{md}}$, i.e. the pointer, we see that the pointer is robustly shifted by the
the same weak-value obtained with the previous statistical method, i.e. $\sqrt{2}$:
\beq
  \label{wvC}
  (\hat{\sigma}_\xi)_{\mathrm{w}} = {{\prod_{k=1}^N \langle{\uparrow_y}|_k ~ \sum_{\mathrm{i=1}}^N
\left\{\hat{\sigma}^i_x + \hat{\sigma}^i_y\right\} ~ \prod_{\mathrm{j=1}}^N |{\uparrow_x} \rangle_j}
\over { \sqrt 2 ~ N(\langle{\uparrow_y} |{\uparrow_x} \rangle)^N}}=
\sqrt2 \pm O(\frac{1}{\sqrt N}).
\label{wvlargespin}
\eeq
A single experiment is now sufficient to
determine the weak-value with great precision and there is no longer
any need to average over results obtained in multiple experiments as we did in the previous section. 
Therefore, if we repeat the experiment with different measuring-devices, then each measuring-device will
show the very same weak-values, up to an insignificant spread of $\frac{1}{\sqrt
{ N}}$ 
and the information from {\it both} boundary conditions, i.e. 
$|\Psi_{\mathrm{in}}\rangle = \prod_{\mathrm{i=1}}^N |{\uparrow_x} \rangle_i$
 and 
$\langle\Psi_{\mathrm{fin}}|  = \prod_{\mathrm{i=1}}^N \langle{\uparrow_y}|_i$, describes the entire interval of time between pre-and-post-selection.
Following~\cite{av2v}, we consider an example with $N=20$.  The probability distribution of the measuring-device after the post-selection is:
\begin{equation}
  \label{probfin}
 prob (Q_{\it md}^{(N)}) ={\cal N}^2 \Bigl (\sum_{i=1}^N (-1)^i \bigl(\cos  ^2
(\pi/8)\bigr)^{N-i} \bigl(\sin  ^2 (\pi/8)\bigr)^i
e^{-(Q_{\it md}^{(N)}-{{(2N-i)}\over N})^2/{2\Delta^2}}\Bigr)^2 .  
\end{equation}
and is drawn for different values of $\Delta$ in figure \ref{wmspin3}. While this result is rare, we have recently shown~\cite{at3} how any ensemble can yield robust weak-values like this in a way that is not rare and for a much stronger regime of interaction. We have thereby shown that weak-values are a general property of every pre-and-post-selected ensemble.\footnote{We have also proposed this as another innovative new laser-technology, e.g. in the amplification of small non-random signals by minimizing uncertainties in determining the weak value and by minimizing sample size.\cite{at3}
}
\noindent

\subsubsection{Hardy's Paradox}
\begin{figure}[h!] 
\begin{flushleft}
  \includegraphics[width=2.24in,height=3.71in,keepaspectratio]{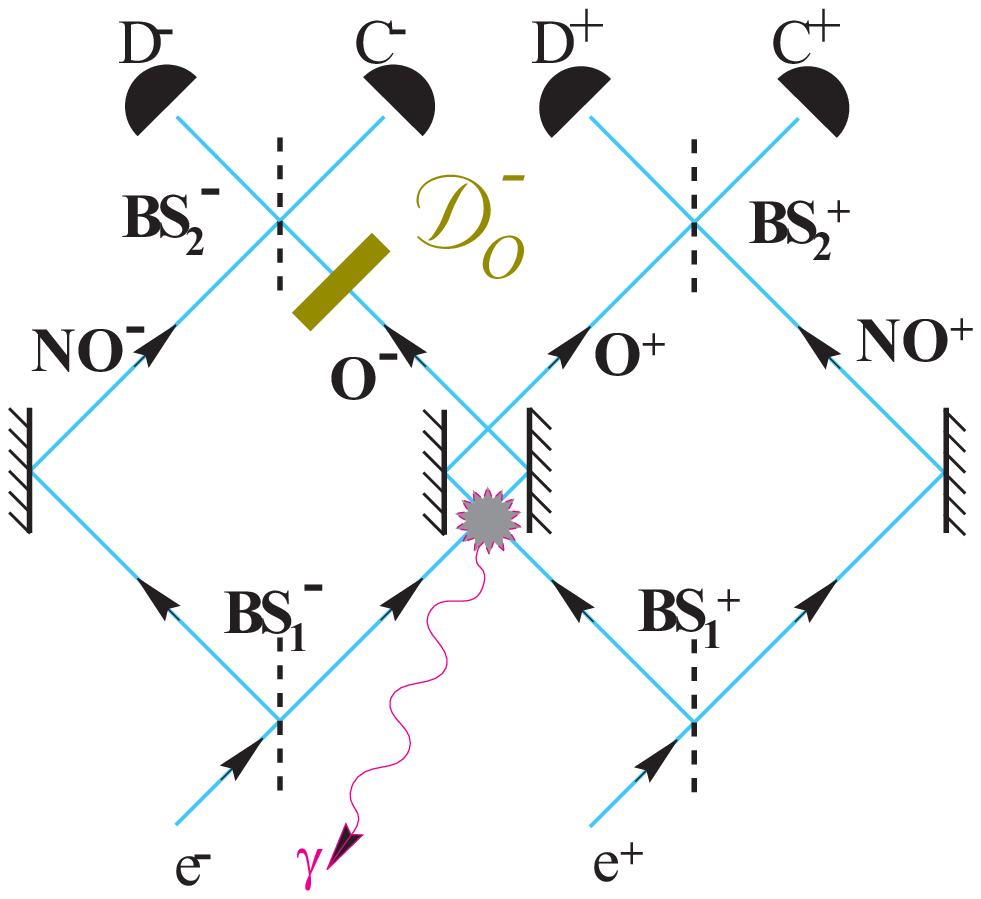}
\end{flushleft}
\vskip -1cm
\caption{\small a) counterfactual resolution: ${\cal D}_O^-$
 disturbs the electron and the electron could end up in the
 $D^-$
 detector even if no positron were present in the overlapping arm, b) electron must be on the overlapping path $\hat N^-_{O}=1$, c) positron also must be on overlapping path $\hat N^+_{O}=1$}
\label{hardy1}
\end{figure}

Another surprising pre-and-post-selection effect is 
Hardy's
 gedanken-experiment which 
 is a variation of interaction-free
  measurements (IFM)~\cite{ElVaid},
  consisting of two ``superposed" Mach-Zehnder
 interferometers
  (MZI)(figure \ref{hardy1}), one with a positron and one
 with an
  electron. Consider first a single interferometer, for instance
 that of
  the positron (labeled  by +). By  adjusting the arm lengths, it
  is possible to arrange specific relative phases in  the
 propagation
  amplitudes  for  paths between the beam-splitters $BS1^+$
 and $BS2^+$
  so that the positron 
  can only emerge towards the
 detector $C^+$. However, the phase difference can be
 altered by the
  presence of an object, for instance in the lower arm,  in
 which case
    detector $D^+$ may be triggered. In the usual IFM setup,
 this is
  illustrated by the dramatic example of a sensitive bomb that
 absorbs
  the  particle with unit probability and subsequently explodes.
  In this way, if $D^+$ is triggered, it is then possible  to infer
   the presence of the bomb without ``touching" it, i.e., to know
 both
  that there was a bomb and that the  particle went through the
 path where
 there was no bomb.

\begin{figure}[h!] 
\vskip -7.05cm
\centering
  \includegraphics[width=2.24in,height=3.71in,keepaspectratio]{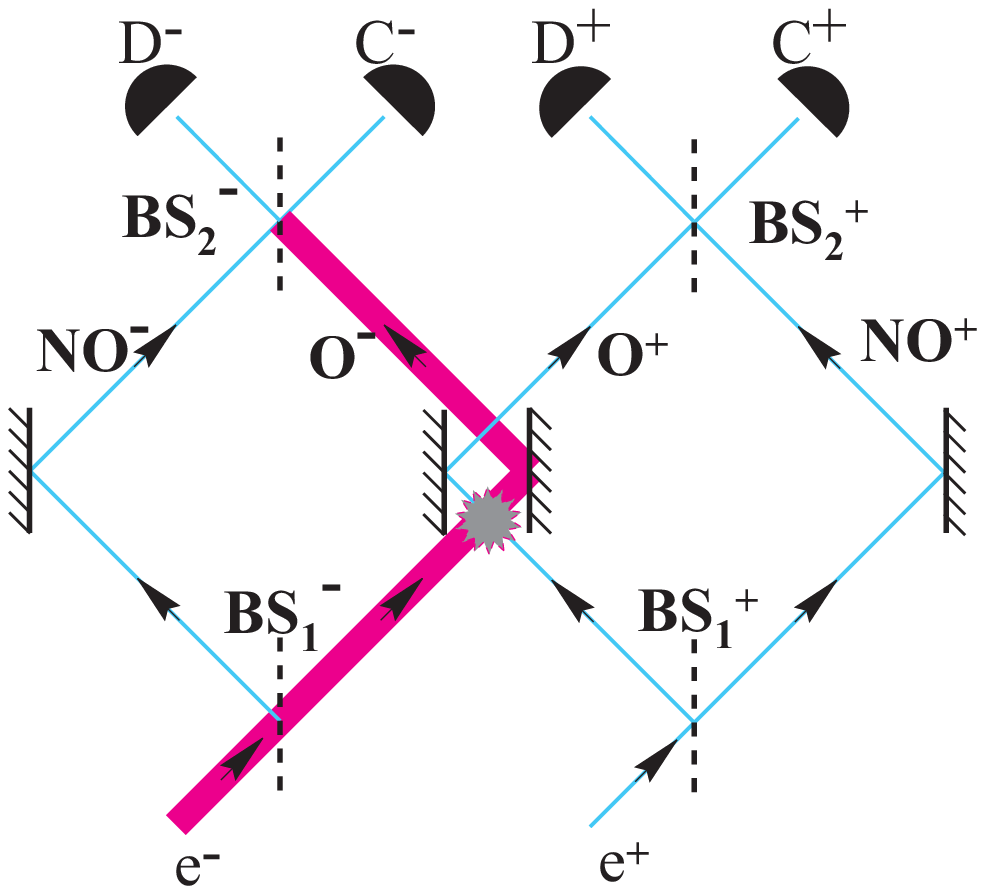}
\label{hardy}
\end{figure}
\begin{figure}[h!] 
\vskip -5.55cm
\begin{flushright}
  \includegraphics[width=2.24in,height=3.71in,keepaspectratio]{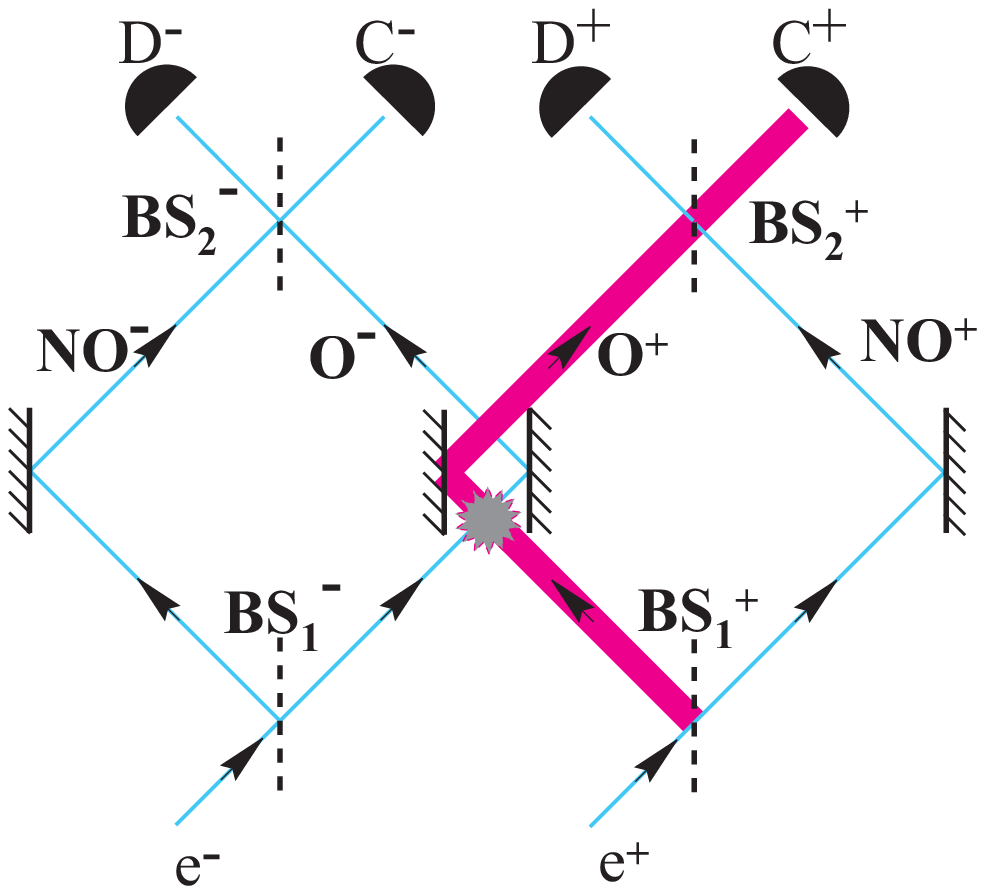}
\end{flushright}
\label{hardy}
\vskip .55cm
\end{figure}

 In the double-MZI, things are arranged so that if each
 MZI
 is considered separately,  the electron can only be detected
 at
 $C^-$ and the positron only at $C^+$. However, because
 there is
 now a region where the two particles overlap, there is also the
 possibility that they will annihilate each other. We assume
 that
 this occurs with unit probability if both particles happen to be
 in this region.

   According to QM, the
 presence of this interference-destroying alternative allows for
 a situation similar to IFM in which  detectors $D^-$ and
 $D^+$ may click in coincidence (in which case, obviously,
 there is no annihilation).

Suppose $D^-$ and $D^+$ do click. Trying to
 ``intuitively" understand this situation leads to paradox. For example, we should infer from the clicking
 of
 $D^-$ that the positron must have gone through the
 overlapping
 arm; otherwise nothing would have disturbed the electron,
 and the
 electron couldn't have ended in $D^-$. Conversely, the same
 logic
 can be applied starting from the clicking of $D^+$, in which
 case
 we deduce that the electron must have also gone through the
 overlapping arm. But then they should have annihilated, and
 couldn't have reached the detectors. Hence the paradox.

These statements, however, are counter-factual, i.e. we haven't actually
 measured
 the positions. Suppose we actually measured the
 position of the electron by inserting a detector ${\cal D}_O^-$ in
 the overlapping arm of the electron-MZI. Indeed, the electron is always
 in the overlapping arm.  
But, we can no longer infer from a click at $D^-$ that a positron should have traveled through the
 overlapping arm of the positron MZI in order to disturb the electron (figure \ref{hardy1}.a). The paradox
 disappears.

\begin{figure}[h!] 
\begin{flushleft}
  \includegraphics[width=2.24in,height=3.71in,keepaspectratio]{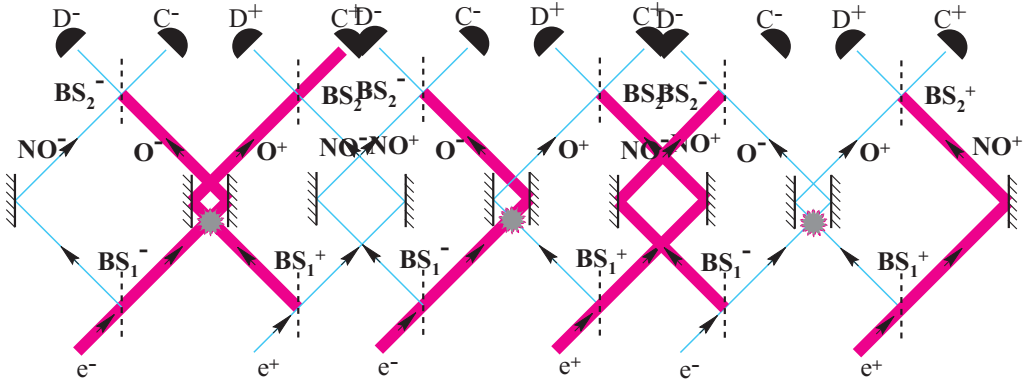}
\end{flushleft}
\vskip -.6cm
\caption{\small a)  $\hat N^{+,-}_{O, O}=0$, b) $N^{+,-}_{O,NO w}=1, N^{+,-}_{NO,O w}=1$, c) $N^{+,-}_{\rm NO\, ,\rm NO w} = -1$}
\label{hardy2}
\end{figure}

As we mentioned (\S\ref{counterfactual}), weak-measurements 
produce only limited disturbance and therefore can be performed simultaneously,
allowing us to {\bf experimentally} test such counter-factual 
statements. Therefore we would like to test~\cite{at2,jt} questions such as ``Which way does
the electron go?", ``Which way does the positron go?", ``Which way
does the positron go when the electron goes through the
overlapping arm?" etc. In other words, we would like to measure
the single-particle ``occupation" operators
 \begin{eqnarray}
 \hat N^+_{\rm NO} = \ket{\rm NO}_p\bra{\rm NO}_p \ \ & \ \
  \hat N^+_{\rm O} = \ket{\rm O}_p\bra{\rm O}_p
 \nonumber\\
 \hat N^-_{\rm NO} = \ket{\rm NO}_e\bra{\rm NO}_e \ \  &\ \
 \
 \, \hat N^-_{\rm O} = \ket{\rm O}_e\bra{\rm O}_e
 \end{eqnarray}
which tell us separately about the electron and the positron.  We note a most important fact, which is
essential in what follows: the weak-value of a product of
observables is {\it not} equal to the product of their weak-values. Hence, we have to measure the single-particle occupation-numbers independently from the pair occupation-operators:
 \begin{eqnarray}
 \hat{N}^{+,-}_{\rm NO\, ,\rm O} = \hat N^+_{\rm NO}
 \hat N^-_{\rm O} \ \ & \ \ \hat{N}^{+,-}_{\rm O\, ,\rm NO} =
 \hat N^+_{\rm O} \hat N^-_{\rm NO} \nonumber\\
 \hat{N}^{+,-}_{\rm O\, ,\rm O} = \hat N^+_{\rm O}
 \hat N^-_{\rm O} \ \ & \ \ \hat{N}^{+,-}_{\rm NO\, ,\rm NO} =
 \hat N^+_{\rm NO} \hat N^-_{\rm NO}
 \end{eqnarray} 
These tell us about the simultaneous locations of
the electron and positron. 
 The
results of all our weak-measurements on the above quantitites, echo, to some extent, the
counter-factual statements, but go far beyond that. They are now
true observational statements (and experiments have successfully verified these results~\cite{stein2}).  In addition,  weak-values obey an intuitive logic
of their own which allows us to deduce them directly.
While this full-intuition is left to published articles~\cite{at2,jt}, we discuss the essence of the paradox which is defined by three counterfactual statements:
\begin{itemize}
\item The electron is always in the overlapping arm.
\item The positron is always in the overlapping arm.
\item The electron and the positron are never both in the
overlapping arms.
\end{itemize}
To these counterfactual statements correspond the following {\it
observational} facts.  In the cases when the electron and positron end up at
$D^-$ and $D^+$ respectively and if we perform a single ideal-measurement of:
\begin{itemize}
\item  $\hat N^-_{O}$, we always find $\hat N^-_{O}=1$ (figure \ref{hardy1}.b).
\item $\hat N^+_{O}$, we always find $\hat N^+_{O}=1$ (figure \ref{hardy1}.c).
\item $\hat N^{+,-}_{O, O}$, we always find $\hat N^{+,-}_{O, O}=0$ (figure \ref{hardy2}.a).
\end{itemize}
The above statements seem paradoxical but, of course, they are
valid only if we perform the measurements separately; they do not
hold if the measurements are made simultaneously. However, Theorem 2 says that when measured weakly all these results
remain true simultaneously:
 \beq N^-_{O w}=1,~~~~~~~ N^+_{O w}=1\label{singleoverlap}\eeq
Using theorems 1 and 2, all other weak-values can be trivially deduced:

 \beq N^-_{NO w}=0,~~~~~~~ N^+_{NO w}=0\label{singlenonoverlap}\eeq

 \beq N^{+,-}_{O,O w}=0\eeq

 \beq N^{+,-}_{O,NO w}=1,~~~~~~~ N^{+,-}_{NO,O w}=1\eeq

 \beq N^{+,-}_{NO, NO w}=-1.\label{pairnonoverlap}\eeq

What do all these results tell us?

 First of all, the single-particle occupation numbers
 (\ref{singleoverlap})
 are consistent with the intuitive statements that
 ``the positron must have been in the overlapping arm
 otherwise the
 electron couldn't have ended at $D^-$" (figure \ref{hardy1}.c) and also that ``the
 electron must have been in the overlapping arm otherwise the
 positron couldn't have ended at $D^+$" (figure \ref{hardy1}.b). But then what
 happened to
 the fact that they could not be both in the overlapping arms
 since
 this will lead to annihilation? QM is
 consistent
 with this too - the pair occupation number  $N^{+,-}_{\rm O\, ,\rm O w}
 =
 0$ shows that there are zero electron-positron pairs in the
 overlapping arms (figure \ref{hardy2}.a)!

 We also feel intuitively that ``the positron must have been in
 the
 overlapping arm otherwise the electron couldn't have ended
 at
 $D^-$, and furthermore, the electron must have gone through
 the
 non-overlapping arm since there was no annihilation" (figure \ref{hardy2}.b). This is
 confirmed by $N^{+,-}_{\rm O\, ,\rm NO} =1$. But we also have the
 statement ``the electron must have been in the overlapping
 arm
 otherwise the positron couldn't have ended at $D^-$ and
 furthermore the positron must have gone through the
 non-overlapping arm since there was no annihilation". This is
 confirmed too, $N^{+,-}_{\rm NO\, ,\rm O w} =1$. But these two
 statements
 together are at odds with the fact that there is in fact just one
 electron-positron pair in the interferometer. QM 
 solves the paradox in a remarkable way - it tells us that  $N^{+,-}_{\rm NO\, ,\rm NO w} = -1$, i.e. that there
 is
 also {\it minus} one electron-positron pair in the
 non-overlapping arms which
 brings the total down to a single pair (figure \ref{hardy2}.c)!

\begin{figure}[h!] 
\vskip -7.1cm
\centering
  \includegraphics[width=2.24in,height=3.71in,keepaspectratio]{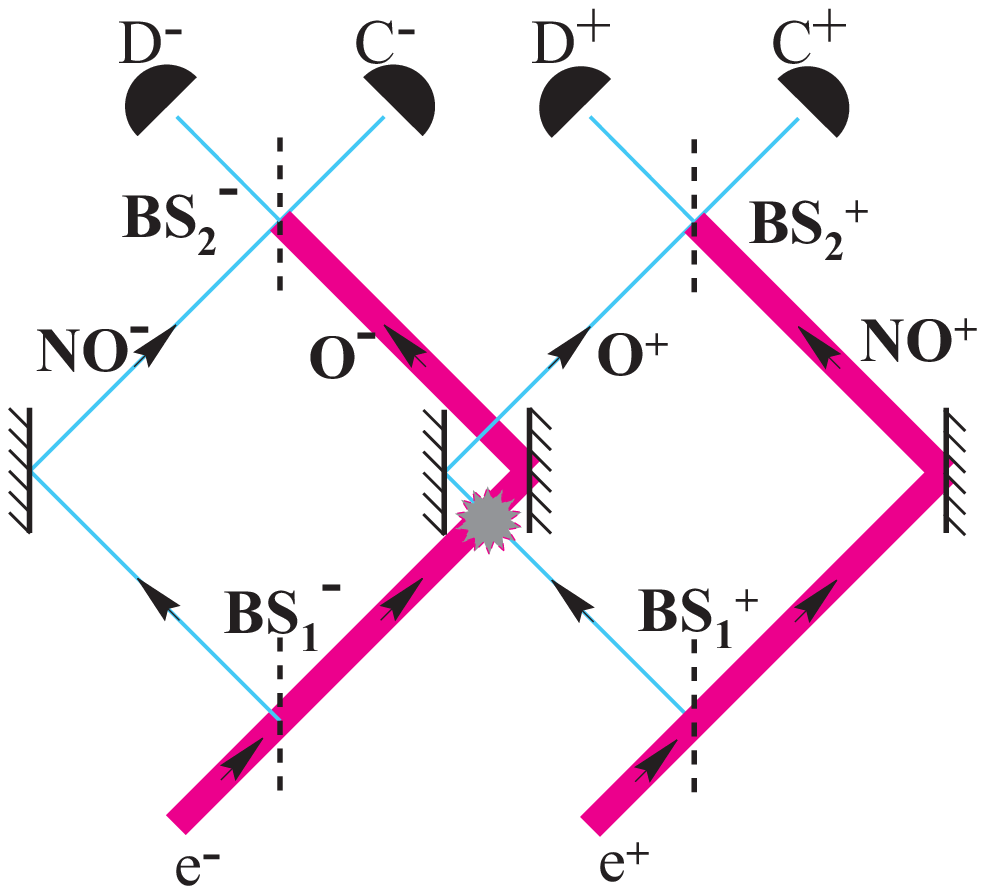}

\label{hardy}
\end{figure}
\begin{figure}[h!] 
\vskip -6.15cm
\begin{flushright}
  \includegraphics[width=2.24in,height=3.71in,keepaspectratio]{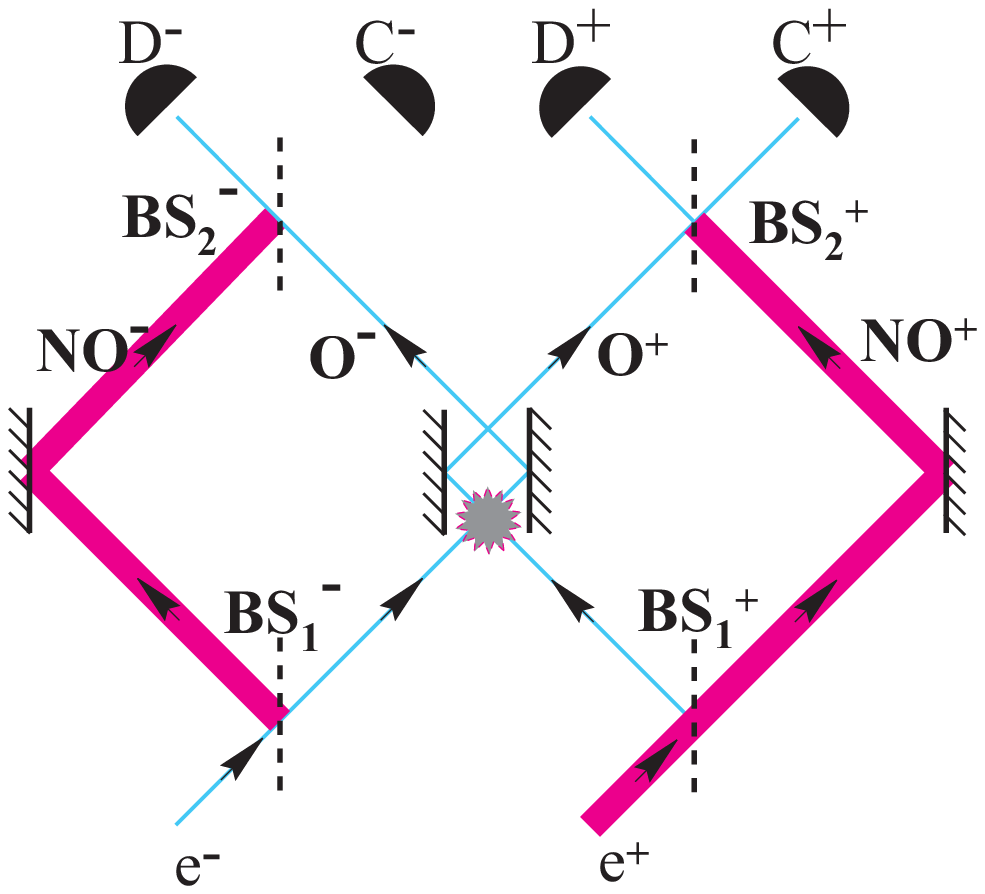}
\end{flushright}
\label{hardy}
\end{figure}

\subsection{\bf Contextuality}
\label{contextuality}
TSQM and WMs have proven very useful in exploring many un-settled aspects of QM.  For example, using TSQM, we have shown that it is possible to assign  definite values to observables in a new way in situations involving ``contextuality."  
Traditionally, contextuality was thought to be a requirement for certain hypothetical modifications of QM. 
However, using pre-and-post-selection and weak-measurements, we have shown that QM implies contextuality directly~\cite{jtcontextuality,jtcontextuality2,jt}. 

What is contextuality?
Bell-Kochen-Specker (BKS) proved that one cannot assign unique answers (i.e. a Hidden-Variable-Theory, HVT) to yes-no questions in such a way that one can think that  measurement  simply reveals the answer as a pre-existing property  that was intrinsic solely to the quantum system itself.
BKS assumed that the specification of the HVT, i.e. $V_{\vec{\psi}}(\hat{A})$, should satisfy: $V_{\vec{\psi}}(F\{\hat{A}\})=F\{V_{\vec{\psi}}(\hat{A})\}
$, i.e. 
any functional relation of an operator that is a member of a commuting subset of observables must also be satisfied if one substitutes the values for the observables into the functional relations.  A consequence of this is satisfaction of the sum and product rules and therefore 
BKS showed that with any system (of dimension greater than 2) the $2^n$ possible ``yes-no" assignments (to the n projection operators representing the yes-no questions) cannot be compatible with the sum and product rules  for all orthogonal resolutions of the identity. 
Thus, a HVT-the hypothetical modification of QM-must be contextual.

In \cite{jt} it was first pointed out and extensively discussed and later proven~\cite{sl2}, that whenever there is a logical pre-and-post-selection-paradox (as in the 3-box-paradox\S\ref{3boxesintro}), 
then there is a related proof of contextuality. However, the elements in the proof are all counter-factual.
TSQM has taught us that by applying theorems 1 and 2, we can for the first time obtain an {\em experimental} meaning to the proof. 

By way of example, 
we consider Mermin's  version of BKS with a set of 9 observables.  It is intuitive~\cite{merminKS} to represent all the 
``functional relationships between mutually commuting subsets of the observables," i.e.  $V_{\vec{\psi}}(F\{\hat{A}\})=F\{V_{\vec{\psi}}(\hat{A})\}$,
by drawing them in fig. 13 and arranging them so that all the observables
in each row (and column) commute with all the other observables in the same row (or column). 
\vskip -1cm
\singlespacing
\singlespacing
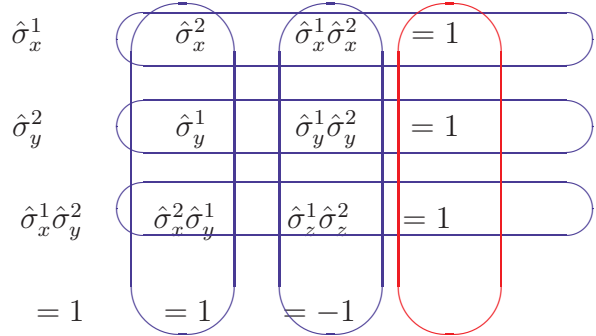
\begin{figure}[h] 
\begin{center}
\begin{eqnarray*}
& {\hat{\sigma}_x}^1  \;\;\;  \;\;\;\;\;\;\;\;\;\;\;\; {\hat{\sigma}_x}^2 
 \;\;\; \;\;\;\;\;\;\; {\hat{\sigma}_x}^1{\hat{\sigma}_x}^2 \;\; \;\;\;=1\nonumber \\
\nonumber \\
& {\hat{\sigma}_y}^2 \;\;\; \;\;\;\;\;\;\;\;\;\;\;\; {\hat{\sigma}_y}^1 
 \;\;\; \;\;\;\;\;\;\; {\hat{\sigma}_y}^1{\hat{\sigma}_y}^2 \;\; \;\;\;=1\nonumber \\ 
\nonumber \\
& {\hat{\sigma}_x}^1{\hat{\sigma}_y}^2\;\;\;  \;\;\;\;\; {\hat{\sigma}_x}^2 {\hat{\sigma}_y}^1
\;\;\; \;\; \;\;\; {\hat{\sigma}_z}^1{\hat{\sigma}_z}^2\;\; \;\;\;=1 \nonumber \\ 
\nonumber \\
& \;\;=1 \;\; \;\;\;\;\;\; =1 \;\; \;\;\;\;\; =-1 \;\;\;\;\;\; \;\;\;\;\;\nonumber \\ 
\end{eqnarray*}
\end{center}
\begin{picture}(400,90)(0,0)
\color{BlueViolet}

\put(240,164){\oval(180,20)}
\put(240,195){\oval(180,20)}
\put(240,228){\oval(180,20)}

\put(175,179){\oval(39,125)}
\put(231,179){\oval(39,125)}

\color{Red}
\put(276,179){\oval(39,125)}

\end{picture}
\vskip -10pc

\label{merm4dov}
\caption[]{4-D BKS example}

\end{figure}

$V_{\vec{\psi}}(F\{\hat{A}\})=F\{V_{\vec{\psi}}(\hat{A})\}$ requires that 
the value assigned to the product of all three observables in any row or column must obey the same identities that the observables themselves satisfy, i.e. the product of the values assigned to the observables in each oval yields a result of $+1$ except in the last column which gives $-1$.\footnote{The value assignments are given by $V_{\vec{\psi}}({\hat{\sigma}_x}^1 )=\langle \hat{\sigma}^1_x\bigotimes I^2\rangle$, $V_{\vec{\psi}}({\hat{\sigma}_x}^2 )=\langle I^1\bigotimes \hat{\sigma}^2_x\rangle$... $V_{\vec{\psi}}({\hat{\sigma}_x}^1 )=\langle \hat{\sigma}^1_z\bigotimes \hat{\sigma}^2_z\rangle$.}. E.g. computing column  3 of fig. 13: 
\begin{eqnarray}
\{\hat{\sigma}^1_x\hat{\sigma}^2_x\}\{\hat{\sigma}^1_y\hat{\sigma}^2_y\}\{\hat{\sigma}^1_z\hat{\sigma}^2_z\}&=&\hat{\sigma}^1_x\underbrace{\hat{\sigma}^2_x\hat{\sigma}^1_y}_{commute\,so\,\hookrightarrow}\hat{\sigma}^2_y\hat{\sigma}^1_z\hat{\sigma}^2_z=
\underbrace{\hat{\sigma}^1_x\hat{\sigma}^1_y}_{=i\hat{\sigma}^1_z}\underbrace{\hat{\sigma}^2_x\hat{\sigma}^2_y}_{=i\hat{\sigma}^2_z}\hat{\sigma}^1_z\hat{\sigma}^2_z\nonumber\\
&=&
i\hat{\sigma}^1_z\underbrace{i\hat{\sigma}^2_z\hat{\sigma}^1_z}_{commute\,so\,\hookrightarrow}\hat{\sigma}^2_z= i\hat{\sigma}^1_zi\hat{\sigma}^1_z\hat{\sigma}^2_z\hat{\sigma}^2_z= -1
\end{eqnarray}
Computing the product of the observables in the third row, i.e.:
\begin{eqnarray}
\{\hat{\sigma}^1_x\hat{\sigma}^2_y\}\{\hat{\sigma}^2_x\hat{\sigma}^1_y\}\{\hat{\sigma}^1_z\hat{\sigma}^2_z\}&=&\hat{\sigma}^1_x\underbrace{\hat{\sigma}^2_y\hat{\sigma}^2_x}_{=-i\hat{\sigma}^2_z}\hat{\sigma}^1_y\{\hat{\sigma}^1_z\hat{\sigma}^2_z\}=\underbrace{\hat{\sigma}^1_x\hat{\sigma}^1_y}_{=i\hat{\sigma}^1_z}\underbrace{\{-i\hat{\sigma}^2_z\}\{\hat{\sigma}
^1_z}_{commute\,so\,\hookrightarrow}\hat{\sigma}^2_z\}\nonumber\\
&=&\underbrace{i\hat{\sigma}^1_z \hat{\sigma}^1_z}_{=i}
\underbrace{\{-i\hat{\sigma}^2_z\}\{\hat{\sigma}^2_z\}}_{=-i}=+1,
\end{eqnarray}
If the product rule  is applied to the value assignments made in the rows, then:
\begin{equation}
\underbrace{V_{\vec{\psi}}({\hat{\sigma}_x}^1 )V_{\vec{\psi}}({\hat{\sigma}_x}^2 )V_{\vec{\psi}}({\hat{\sigma}_x}^1{\hat{\sigma}_x}^2)}_{row\,\, 1}=
\underbrace{V_{\vec{\psi}}({\hat{\sigma}_y}^2)V_{\vec{\psi}}({\hat{\sigma}_y}^1 )V_{\vec{\psi}}( {\hat{\sigma}_y}^1{\hat{\sigma}_y}^2)}_{row\,\, 2}=
\underbrace{V_{\vec{\psi}}({\hat{\sigma}_x}^1{\hat{\sigma}_y}^2)V_{\vec{\psi}}({\hat{\sigma}_x}^2 {\hat{\sigma}_y}^1)V_{\vec{\psi}}({\hat{\sigma}_z}^1{\hat{\sigma}_z}^2)}_{row\,\, 3}=+1
\label{rows}
\end{equation}
while the column identities require:
\begin{eqnarray}
\underbrace{V_{\vec{\psi}}({\hat{\sigma}_x}^1 )V_{\vec{\psi}}({\hat{\sigma}_y}^2)V_{\vec{\psi}}({\hat{\sigma}_x}^1{\hat{\sigma}_y}^2)}_{column\,\, 1}=
\underbrace{V_{\vec{\psi}}({\hat{\sigma}_x}^2 )V_{\vec{\psi}}({\hat{\sigma}_y}^1 )V_{\vec{\psi}}({\hat{\sigma}_x}^2 {\hat{\sigma}_y}^1)}_{column\,\, 2}&=&+1\nonumber\\
\underbrace{V_{\vec{\psi}}({\hat{\sigma}_x}^1{\hat{\sigma}_x}^2)V_{\vec{\psi}}( {\hat{\sigma}_y}^1{\hat{\sigma}_y}^2)V_{\vec{\psi}}({\hat{\sigma}_z}^1{\hat{\sigma}_z}^2)}_{column\,\, 3}&=&-1
\label{columns}
\end{eqnarray}
However, it is easy to see that the 9 numbers $V_{\vec{\psi}}$ cannot satisfy all 6 constraints  because  multiplying all $9$ observables together gives 2 different results, a $\bf +1$ when it is done row by row and a $\bf -1$  when it is done column by column: 
\begin{equation}
\underbrace{V_{\vec{\psi}}({\hat{\sigma}_x}^1 )V_{\vec{\psi}}({\hat{\sigma}_x}^2 )V_{\vec{\psi}}({\hat{\sigma}_x}^1{\hat{\sigma}_x}^2)}_{row\,\, 1} \underbrace{V_{\vec{\psi}}({\hat{\sigma}_y}^2)V_{\vec{\psi}}({\hat{\sigma}_y}^1 )V_{\vec{\psi}}( {\hat{\sigma}_y}^1{\hat{\sigma}_y}^2)}_{row\,\, 2}  \underbrace{V_{\vec{\psi}}({\hat{\sigma}_x}^1{\hat{\sigma}_y}^2)V_{\vec{\psi}}({\hat{\sigma}_x}^2 {\hat{\sigma}_y}^1)V_{\vec{\psi}}({\hat{\sigma}_z}^1{\hat{\sigma}_z}^2)}_{row\,\, 3}={\bf +1}
\label{1sthvt}
\end{equation}
\begin{equation}
\underbrace{V_{\vec{\psi}}({\hat{\sigma}_x}^1 )V_{\vec{\psi}}({\hat{\sigma}_y}^2)V_{\vec{\psi}}({\hat{\sigma}_x}^1{\hat{\sigma}_y}^2)}_{column\,\, 1}\underbrace{V_{\vec{\psi}}({\hat{\sigma}_x}^2 )V_{\vec{\psi}}({\hat{\sigma}_y}^1 )V_{\vec{\psi}}({\hat{\sigma}_x}^2 {\hat{\sigma}_y}^1)}_{column\,\, 2}
\underbrace{V_{\vec{\psi}}({\hat{\sigma}_x}^1{\hat{\sigma}_x}^2)V_{\vec{\psi}}( {\hat{\sigma}_y}^1{\hat{\sigma}_y}^2)V_{\vec{\psi}}({\hat{\sigma}_z}^1{\hat{\sigma}_z}^2)}_{column\,\, 3}={\bf -1}
\label{2ndhvt}
\end{equation}
There obviously is no consistent solution to eqs. \ref{2ndhvt} and \ref{1sthvt} since they contain the same set of numbers, simply ordered differently.
Therefore the values assigned to the observables cannot obey the same identities that the observables themselves obey, $V_{\vec{\psi}}(F\{\hat{A}\})\neq F\{V_{\vec{\psi}}(\hat{A})\}$, 
 and an HVT would have to assign  values to observables in a
way that depended on the choice of which of 2 mutually commuting sets of observables that were
also chosen to measure, i.e. the values assigned are contextual. For example, the assignment ${\hat{\sigma}_z}^1{\hat{\sigma}_z}^2=\pm 1$ depends on whether we associate ${\hat{\sigma}_z}^1{\hat{\sigma}_z}^2$ with row-3 or with column-3.

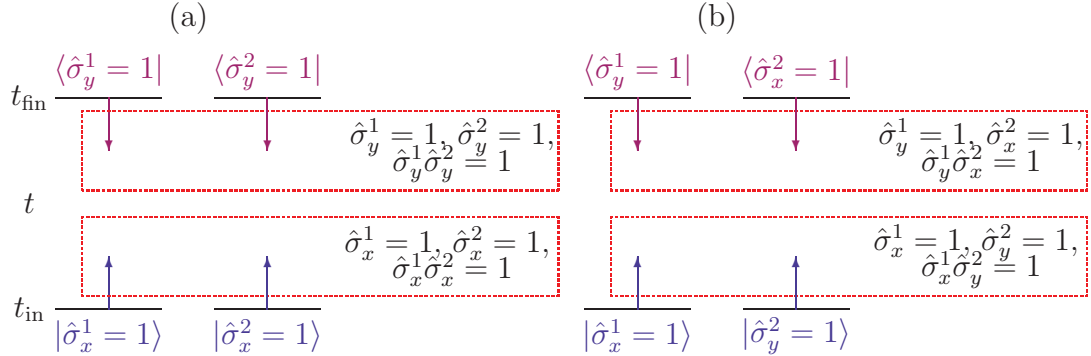
\begin{figure}[h]
 \epsfxsize=4.5truein
\begin{picture}(400,90)(0,0)
\put(40,10){\line(1,0){40}}
\put(40,90){\line(1,0){40}}
\color{BlueViolet}
\put(60,10){\vector(0,1){20}}
\color{RedViolet}
\put(60,90){\vector(0,-1){20}}
\color{Red}
\put(50,55){\dashbox{1}(180,30)}
\put(50,15){\dashbox{1}(180,30)}
\put(250,55){\dashbox{1}(180,30)}
\put(250,15){\dashbox{1}(180,30)}
\color{OliveGreen}

\color{Black}
\put(100,10){\line(1,0){40}}
\put(100,90){\line(1,0){40}}
\color{BlueViolet}
\put(120,10){\vector(0,1){20}}
\color{RedViolet}
\put(120,90){\vector(0,-1){20}}
\color{Black}
\put(240,10){\line(1,0){40}}
\put(240,90){\line(1,0){40}}
\color{BlueViolet}

\put(260,10){\vector(0,1){20}}
\color{RedViolet}

\put(260,90){\vector(0,-1){20}}
\color{OliveGreen}

\color{BlueViolet}
\put(260,0){\makebox(0,0){$|\hat{\sigma}_x^1=1\ra$}}
\put(120,0){\makebox(0,0){$|\hat{\sigma}_x^2=1\ra$}}
\put(60,0){\makebox(0,0){$|\hat{\sigma}_x^1=1\ra$}}
\put(320,0){\makebox(0,0){$|\hat{\sigma}_y^2=1\ra$}}

\color{RedViolet}
\put(260,100){\makebox(0,0){$\la\hat{\sigma}_y^1=1|$}}
\put(120,100){\makebox(0,0){$\la\hat{\sigma}_y^2=1|$}}
\put(60,100){\makebox(0,0){$\la\hat{\sigma}_y^1=1|$}}
\put(320,100){\makebox(0,0){$\la\hat{\sigma}_x^2=1|$}}
\color{Black}
\put(300,10){\line(1,0){40}}
\put(300,90){\line(1,0){40}}
\color{BlueViolet}

\put(320,10){\vector(0,1){20}}
\color{RedViolet}

\put(320,90){\vector(0,-1){20}}
\color{Black}
\put(30,10){\makebox(0,0){$t_{\mathrm{in}}$}}
\put(30,90){\makebox(0,0){$t_{\mathrm{fin}}$}}
\put(30,50){\makebox(0,0){$t$}}
\put(90,120){\makebox(0,0){(a)}}
\put(290,120){\makebox(0,0){(b)}}
\put(190,35){\makebox(0,0){$\hat{\sigma}^1_x=1$, $\hat{\sigma}^2_x=1$, }}
\put(190,25){\makebox(0,0){$\hat{\sigma}^1_x\hat{\sigma}^2_x=1$}}
\put(190,75){\makebox(0,0){$\hat{\sigma}^1_y=1$, $\hat{\sigma}^2_y=1$,}}
\put(190,65){\makebox(0,0){$\hat{\sigma}^1_y\hat{\sigma}^2_y=1$}}

\put(390,35){\makebox(0,0){$\hat{\sigma}^1_x=1$, $\hat{\sigma}^2_y=1$, }}
\put(390,25){\makebox(0,0){$\hat{\sigma}^1_x\hat{\sigma}^2_y=1$}}
\put(390,75){\makebox(0,0){$\hat{\sigma}^1_y=1$, $\hat{\sigma}^2_x=1$,}}
\put(390,65){\makebox(0,0){$\hat{\sigma}^1_y\hat{\sigma}^2_x=1$}}

\color{OliveGreen}

\end{picture}

\caption{\small pre-and-post-selection states for Mermin example.}
\label{2vmermin}
\end{figure}

\begin{figure}[h] \epsfxsize=3.5truein
\begin{picture}(400,130)(0,0)
\put(40,10){\line(1,0){40}}
\put(40,90){\line(1,0){40}}
\color{BlueViolet}
\put(60,10){\vector(0,1){20}}
\color{RedViolet}
\put(60,90){\vector(0,-1){20}}
\color{OliveGreen}

\put(60,50){\oval(25,80)}

\put(55,37){\makebox(0,0){\bf \small correlated}}
\put(95,50){\makebox(0,0){\bf \huge +}}

\color{Black}
\put(100,10){\line(1,0){40}}
\put(100,90){\line(1,0){40}}
\color{BlueViolet}

\put(120,10){\vector(0,1){20}}
\color{RedViolet}

\put(120,90){\vector(0,-1){20}}
\color{OliveGreen}

\put(320,50){\oval(25,80)}

\put(120,37){\makebox(0,0){\bf \small correlated}}
\put(155,50){\makebox(0,0){\bf \huge +}}
\put(300,50){\makebox(0,0){\bf \huge +}}

\color{Black}
\put(160,10){\line(1,0){40}}
\put(160,90){\line(1,0){40}}
\color{BlueViolet}

\put(180,10){\vector(0,1){20}}
\color{RedViolet}

\put(180,90){\vector(0,-1){20}}
\color{OliveGreen}
\put(120,50){\oval(25,80)}
\put(180,50){\oval(25,80)}

\color{BlueViolet}
\put(180,0){\makebox(0,0){$|\Phi_3\ra$}}
\put(120,0){\makebox(0,0){$|\Phi_2\ra$}}
\put(60,0){\makebox(0,0){$|\Phi_1\ra$}}
\put(320,0){\makebox(0,0){$|\Phi_n\ra$}}

\color{RedViolet}
\put(180,100){\makebox(0,0){$\la\Psi_3|$}}
\put(120,100){\makebox(0,0){$\la\Psi_2|$}}
\put(60,100){\makebox(0,0){$\la\Psi_1|$}}
\put(320,100){\makebox(0,0){$\la\Psi_n|$}}
\color{Black}
\put(210,50){\circle*{3}}
\put(230,50){\circle*{3}}
\put(250,50){\circle*{3}}
\put(270,50){\circle*{3}}
\put(300,10){\line(1,0){40}}
\put(300,90){\line(1,0){40}}
\color{BlueViolet}

\put(320,10){\vector(0,1){20}}
\color{RedViolet}

\put(320,90){\vector(0,-1){20}}
\color{Black}
\put(30,10){\makebox(0,0){$t_{\mathrm{in}}$}}
\put(30,90){\makebox(0,0){$t_{\mathrm{fin}}$}}
\put(30,50){\makebox(0,0){$t$}}
\end{picture}

\caption[Generalized State: superpositions of 2-vectors]
{Generalized State: superpositions of 2-vectors.}
\label{mult2particles}
\end{figure}
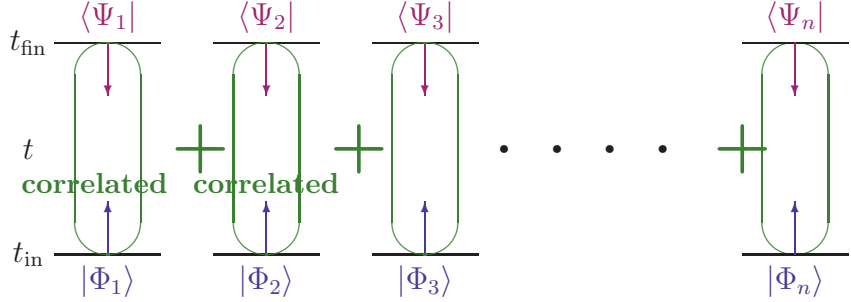

We briefly summarize application of TSQM to the Mermin example\cite{jtcontextuality,jtcontextuality2,jt}.  
A single pre-and-post-selection (figure \ref{2vmermin}) allows us to assign a definite value to any single observable in figure 13.  That by itself is new and surprising. 
\vskip 1cm
Moreover, for ``contextuality," 
we must determine how many of the {\it products} of the 9 observables 
in figure 13 can be ascertained together with certainty.
In order to ascertain the products of any 2 pairs, the generalized state is required, an outcome that Mermin describes as ``intriguing"~\cite{mermin2v}.  The generalized state is defined
by\cite{jmav}:
$\Psi = \sum _{i} \alpha _{i} \langle \Psi _{i} \mid \mid \Phi _{i}\rangle$.
\footnote{This correlated state can be created by preparing
at $t_\mathrm{in}$ a correlated state $\sum_i \alpha_i|\Psi_i\rangle|i\rangle$ with $|i\rangle$ an orthonormal
set of states of an ancilla.  
Then the ancilla is ``guarded" so there are no interactions with the ancilla during the time
$(t_{in}, t_{fin})$.  At $t_{fin}$ we post select  on the particle and ancilla the state
$\frac{1}{\sqrt{N}}\sum_i|\Phi_i\rangle|i\rangle$.  If we
are successful in obtaining this state for the post-selection, then the state of 2 particles is
described in the intermediate
time  by the entangled state (see figure \ref{mult2particles}).  This is yet another example of a useful generalization of QM.}
The outcome for the product of the first two observables in  column 3 of figure 13
with the pre-and-post-selection of fig. \ref{2vmermin}.a is $\sigma^1_x\sigma^2_x\sigma^1_y\sigma^2_y =+1$.
However, if we measure the operators corresponding to the first 2 observables of row 3 in figure 13,
i.e.
$\hat{\sigma}^1_x\hat{\sigma}^2_y\hat{\sigma}^2_x\hat{\sigma}^1_y$,
given this particular pre-and-post-selection shown in fig. \ref{2vmermin}.a, then the sequence of measurements interfere with each other (as represented by the slanted ovals in
figure \ref{2vmermindiag}.a).
To see this, consider that $\hat{\sigma}^1_x\hat{\sigma}^2_y\hat{\sigma}^2_x\hat{\sigma}^1_y$ corresponds to the sequence of measurements
represented in figure \ref{2vmermin2}.a.  While the pre-selection of
particle 2 is $\hat{\sigma}^2_x=1$ at $t_{\mathrm{in}}$, the next measurement after the pre-selection at $t_2$ is for $\hat{\sigma}^2_y$ and only
{\it after} that a measurement of $\hat{\sigma}^2_x$ is performed at $t_3$.  Thus, there is no guarantee  that the $\hat{\sigma}^2_x$  measurement at $t_3$ will give the same  value as the pre-selected state of
$\hat{\sigma}^2_x=1$ or that the $\hat{\sigma}^2_y$ measurement will give the same value as the post-selected state of
$\hat{\sigma}^2_y=1$.
In TSQM, this is  due to the disturbance of the 2-vector boundary conditions which is created by the ideal-measurement (\S\ref{IM}): the initial pre-selected vector $\hat{\sigma}^2_x=1$ from $t_{\mathrm{in}}$ is ``destroyed" when the $\hat{\sigma}^2_y$  measurement at time $t_2$ is performed and therefore cannot inform the later $\hat{\sigma}^2_x$ measurement at time $t_3$. In other words, with the particular pre-and-post-selection  given in fig. \ref{2vmermin}.a and \ref{2vmermin2}.a, the operator, 
$\hat{\sigma}^1_x\hat{\sigma}^2_y\hat{\sigma}^2_x\hat{\sigma}^1_y$ depends on information from both the
pre-selected vector $\hat{\sigma}^1_x=1$, $\hat{\sigma}^2_x=1$ and the post-selected vector
$\hat{\sigma}^1_y=1$, $\hat{\sigma}^2_y=1$ in a ``diagonal-pre-and-post-selection" sense.   We call this diagonal-pre-and-post-selection because a line connecting $\hat{\sigma}^1_x(t_1)$ with $\hat{\sigma}^2_x(t_3)$ will be diagonal or will cross the line connecting $\hat{\sigma}^2_y(t_2)$ with $\hat{\sigma}^1_y(t_4)$, where $t_{\mathrm{in}}<t_1<t_2 ...<t_{\mathrm{fin}}$, see fig. \ref{2vmermindiag}.a).   
\begin{figure}[h] \epsfxsize=4.5truein
\vskip 1cm
\begin{picture}(400,90)(0,0)
\put(40,10){\line(1,0){40}}
\put(40,90){\line(1,0){40}}
\color{BlueViolet}
\put(60,10){\vector(0,1){10}}
\color{RedViolet}
\put(60,90){\vector(0,-1){10}}

\color{Black}
\put(100,10){\line(1,0){40}}
\put(100,90){\line(1,0){40}}
\color{BlueViolet}

\put(120,10){\vector(0,1){25}}
\color{RedViolet}

\put(120,90){\vector(0,-1){25}}
\color{OliveGreen}

\color{Black}
\put(240,10){\line(1,0){40}}
\put(240,90){\line(1,0){40}}
\color{BlueViolet}

\put(260,10){\vector(0,1){10}}
\color{RedViolet}

\put(260,90){\vector(0,-1){25}}
\color{OliveGreen}

\color{BlueViolet}
\put(260,0){\makebox(0,0){$|\hat{\sigma}_x^1=1\ra$}}
\put(120,0){\makebox(0,0){$|\hat{\sigma}_x^2=1\ra$}}
\put(60,0){\makebox(0,0){$|\hat{\sigma}_x^1=1\ra$}}
\put(320,0){\makebox(0,0){$|\hat{\sigma}_y^2=1\ra$}}

\color{RedViolet}
\put(260,100){\makebox(0,0){$\la\hat{\sigma}_y^1=1|$}}
\put(120,100){\makebox(0,0){$\la\hat{\sigma}_y^2=1|$}}
\put(60,100){\makebox(0,0){$\la\hat{\sigma}_y^1=1|$}}
\put(320,100){\makebox(0,0){$\la\hat{\sigma}_x^2=1|$}}
\color{Black}
\put(300,10){\line(1,0){40}}
\put(300,90){\line(1,0){40}}
\color{BlueViolet}

\put(320,10){\vector(0,1){25}}
\color{RedViolet}

\put(320,90){\vector(0,-1){10}}
\color{Black}
\put(30,10){\makebox(0,0){$t_{\mathrm{in}}$}}
\put(30,90){\makebox(0,0){$t_{\mathrm{fin}}$}}
\put(30,50){\makebox(0,0){$t$}}
\put(90,120){\makebox(0,0){(a)}}
\put(290,120){\makebox(0,0){(b)}}

\put(80,20){\makebox(0,0){$\hat{\sigma}^1_x=1$}}
\put(110,40){\makebox(0,0){$\hat{\sigma}^2_y=?$}}
\put(110,60){\makebox(0,0){$\hat{\sigma}^2_x=?$}} 
\put(80,80){\makebox(0,0){$\hat{\sigma}^1_y=1$}}

\put(150,20){\makebox(0,0){$t_1$}}
\put(150,40){\makebox(0,0){$t_2$}}
\put(150,60){\makebox(0,0){$t_3$}} 
\put(150,80){\makebox(0,0){$t_4$}}

\put(280,20){\makebox(0,0){$\hat{\sigma}^1_x=1$}}
\put(340,40){\makebox(0,0){$\hat{\sigma}^2_x=?$}}
\put(280,60){\makebox(0,0){$\hat{\sigma}^1_y=1$}} 
\put(340,80){\makebox(0,0){$\hat{\sigma}^2_y=?$}}

\put(370,20){\makebox(0,0){$t_1$}}
\put(370,40){\makebox(0,0){$t_2$}}
\put(370,60){\makebox(0,0){$t_3$}} 
\put(370,80){\makebox(0,0){$t_4$}}

\end{picture}

\caption{\small Time sequence of pre-and-post-selection measurements for Mermin example.}
\label{2vmermin2}
\end{figure}
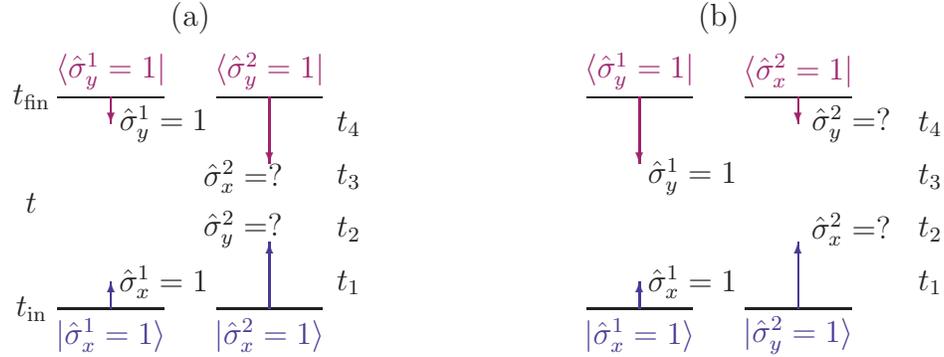
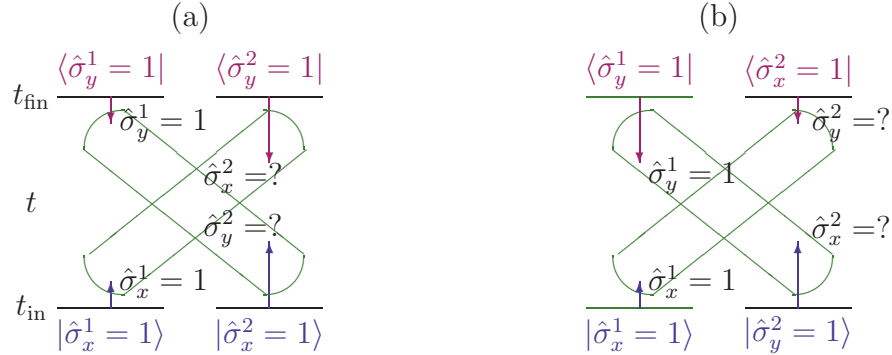
\begin{figure}[h] \epsfxsize=4.5truein
\vskip 1.2cm
\begin{picture}(400,90)(0,0)
\put(40,10){\line(1,0){40}}
\put(40,90){\line(1,0){40}}
\color{BlueViolet}
\put(60,10){\vector(0,1){10}}
\color{RedViolet}
\put(60,90){\vector(0,-1){10}}

\color{Black}
\put(100,10){\line(1,0){40}}
\put(100,90){\line(1,0){40}}
\color{BlueViolet}

\put(120,10){\vector(0,1){25}}
\color{RedViolet}

\put(120,90){\vector(0,-1){25}}
\color{OliveGreen}

\color{Black}
\color{OliveGreen}

\put(65,70){\oval(30,30)[tl]}
\put(50,70){\line(5,-4){69}}
\put(118,30){\oval(30,30)[br]}
\put(65,85){\line(5,-4){68}}

\put(118,70){\oval(30,30)[tr]}
\put(134,70){\line(-5,-4){69}}
\put(65,30){\oval(30,30)[bl]}
\put(120,85){\line(-5,-4){68}}

\put(265,70){\oval(30,30)[tl]}
\put(250,70){\line(5,-4){69}}
\put(318,30){\oval(30,30)[br]}
\put(265,85){\line(5,-4){68}}

\put(318,70){\oval(30,30)[tr]}
\put(334,70){\line(-5,-4){69}}
\put(265,30){\oval(30,30)[bl]}
\put(320,85){\line(-5,-4){68}}

\put(240,10){\line(1,0){40}}
\put(240,90){\line(1,0){40}}
\color{BlueViolet}

\put(260,10){\vector(0,1){10}}
\color{RedViolet}

\put(260,90){\vector(0,-1){25}}
\color{OliveGreen}

\color{BlueViolet}
\put(260,0){\makebox(0,0){$|\hat{\sigma}_x^1=1\ra$}}
\put(120,0){\makebox(0,0){$|\hat{\sigma}_x^2=1\ra$}}
\put(60,0){\makebox(0,0){$|\hat{\sigma}_x^1=1\ra$}}
\put(320,0){\makebox(0,0){$|\hat{\sigma}_y^2=1\ra$}}

\color{RedViolet}
\put(260,100){\makebox(0,0){$\la\hat{\sigma}_y^1=1|$}}
\put(120,100){\makebox(0,0){$\la\hat{\sigma}_y^2=1|$}}
\put(60,100){\makebox(0,0){$\la\hat{\sigma}_y^1=1|$}}
\put(320,100){\makebox(0,0){$\la\hat{\sigma}_x^2=1|$}}
\color{Black}
\put(300,10){\line(1,0){40}}
\put(300,90){\line(1,0){40}}
\color{BlueViolet}

\put(320,10){\vector(0,1){25}}
\color{RedViolet}

\put(320,90){\vector(0,-1){10}}
\color{Black}
\put(30,10){\makebox(0,0){$t_{\mathrm{in}}$}}
\put(30,90){\makebox(0,0){$t_{\mathrm{fin}}$}}
\put(30,50){\makebox(0,0){$t$}}
\put(90,120){\makebox(0,0){(a)}}
\put(290,120){\makebox(0,0){(b)}}

\put(80,20){\makebox(0,0){$\hat{\sigma}^1_x=1$}}
\put(110,40){\makebox(0,0){$\hat{\sigma}^2_y=?$}}
\put(110,60){\makebox(0,0){$\hat{\sigma}^2_x=?$}} 
\put(80,80){\makebox(0,0){$\hat{\sigma}^1_y=1$}}

\put(280,20){\makebox(0,0){$\hat{\sigma}^1_x=1$}}
\put(340,40){\makebox(0,0){$\hat{\sigma}^2_x=?$}}
\put(280,60){\makebox(0,0){$\hat{\sigma}^1_y=1$}} 
\put(340,80){\makebox(0,0){$\hat{\sigma}^2_y=?$}}

\end{picture}

\caption{\small a) Measurement of $\hat{\sigma}^1_x\hat{\sigma}^2_y\hat{\sigma}^2_x\hat{\sigma}^1_y$ is diagonal, b) measurement of $\hat{\sigma}^1_x\hat{\sigma}^2_x\hat{\sigma}^1_y\hat{\sigma}^2_y$ is diagonal}
\label{2vmermindiag}
\end{figure}
However, $\hat{\sigma}^1_z\hat{\sigma}^2_z$ is assigned different values in different pre-and-post-selections. 
It is precisely because of this connection between particular pre-and-post-selections and different values for  $\hat{\sigma}^1_z\hat{\sigma}^2_z$ that the issue of contextuality arises when we consider products of these observables.
In other words, the contextuality here is manifested by the fact that $\hat{\sigma}^1_x\hat{\sigma}^2_y\hat{\sigma}^2_x\hat{\sigma}^1_y=-1$  (given the pre-and-post-selection of fig. \ref{2vmermin2}.a) even though separately $\hat{\sigma}^1_x\hat{\sigma}^2_y=+1$ and $\hat{\sigma}^2_x\hat{\sigma}^1_y=+1$.  But these 3 outcomes can be measured weakly without contradiction because the product of weak-values is not equal to the weak-value of the product.  Therefore, instead of contextuality being an aspect of a hypothetical replacement for QM (the HVT), we have shown that contextuality is directly part of QM~\cite{jtcontextuality,jtcontextuality2,jt}.

\subsection{\bf Nonlocality}
\label{nonlocal}
Traditionally, it was believed that ``contextuality" was very closely related to ``kinematic-nonlocality." Typically, kinematic-nonlocality  refers to correlations, such as eq.~\ref{14.1}, that violate Bell's-inequality with the consequence that 
QM cannot be replaced with a {\bf local} realistic model.  Similarly, contextuality refers to the impossibility of 
replacing QM with a noncontextual realistic theory.
Applying this now to the relativistic-paradox (\S\ref{relparadox}), we see that Lorentz covariance in the state-description can be preserved in TSQM~\cite{ar}  
 because the post-selected vector $\sigma_z^A=+1$   propagates all the way back to the initial preparation of an EPR state, eq.~\ref{14.1}, $|\Psi_{EPR}\ra=\frac{1}{\sqrt{2}}\left\{\mid\uparrow\rangle_{\mathrm{\textcolor{BlueViolet}{A}}}\mid\downarrow\rangle_{\mathrm{\textcolor{Red}{B}}}
-\mid\downarrow\rangle_{\mathrm{\textcolor{BlueViolet}{A}}}\mid\uparrow\rangle_{\mathrm{\textcolor{Red}{B}}}\right\}$. E.g. if $A$ changes his mind and measures $\sigma_y^A$ instead of $\sigma_z^A$ or if we consider a different frame-of-reference, then this would change the post-selected vector all the way back to $|\Psi_{EPR}\ra$.   More explicitly, suppose the final post-selected-state is $\la\Psi_{fin}|=\frac{1}{\sqrt{2}}\la\uparrow_z|_A\left\{\la\downarrow |_{B}+\la\uparrow |_{B}\right\}=\frac{1}{\sqrt{2}}\left\{\la\uparrow_z |_{A}\la\uparrow |_{B}+
\la\uparrow_z |_{A}\la\downarrow |_{B}
\right\}$. The full state-description is the bra-ket combination (which is not just a scalar product):
\beq
\la\Psi_{fin}||\Psi_{EPR}\ra=\frac{1}{\sqrt{2}}\left\{\la\uparrow|_{A}\la\uparrow_{B}+
\la\uparrow|_{A}\la\downarrow|_{B}\right\}
\frac{1}{\sqrt{2}}\left\{\mid\uparrow\rangle_{A}\mid\downarrow\rangle_{B}
-\mid\downarrow\rangle_{A}\mid\uparrow\rangle_{B}\right\}
\eeq
There is no longer a need to specify a moment when a non-local collapse occurs, thereby removing the relativistic paradox.

Finally, TSQM and weak-measurements also provide insight into a very different kind of non-locality, namely  dynamical-nonlocality, e.g. that of the Aharonov-Bohm (AB) effect ~\cite{AB}.  We have shown how this novel kind of nonlocality can be measured with weak-measurements~\cite{at5}.

\section{\bf TSQM lead to new mathematics, simplifications in calculations,  and stimulated discoveries in other fields}
\label{simplify}
TSQM has influenced work in many areas of physics, e.g. in cosmology~\cite{gellman,hawking}, in black-holes~\cite{englert,thooft}, in superluminal tunneling~\cite{chiao,Stein}, in quantum information~\cite{adz,spie-entangle,spie-np}, etc.  We review two examples here.

\subsection{\bf Super-oscillations}
\label{superoscsec}
Superoscillations~\cite{berry} are functions which oscillate with an arbitrarily high frequency $\alpha$, but which, surprisingly, can be understood as  superpositions of low frequencies, $|k|<1$, seemingly a violation of the Fourier theorem: 
\beq
\sum_{|k|<1} c_k e^{ikx} \rightarrow e^{i\alpha x}
\eeq
Superoscillations were originally discovered through the study of weak-values. By way of example, consider again eq.~\ref{bigspina}:
\begin{eqnarray}
|\Phi_{fin}^{MD}\ra&=&\left\{\cos \frac{{\lambda} \hat{Q}_{\it md}}{N}-i\alpha_w\sin \frac{{\lambda} \hat{Q}_{\it md}}{N}\right\}^N|\Phi_{\mathrm{in}}^{\mathrm{MD}}\ra\nonumber\\
&=&\left\{\frac{e^{\frac{i{\lambda} \hat{Q}_{\it md}}{N}}+e^{-\frac{i{\lambda} \hat{Q}_{\it md}}{N}}}{2}+\alpha_w \frac{e^{\frac{i{\lambda} \hat{Q}_{\it md}}{N}}-e^{-\frac{i{\lambda} \hat{Q}_{\it md}}{N}}}{2}\right\}^N|\Phi_{\mathrm{in}}^{\mathrm{MD}}\ra\nonumber\\
&=&\underbrace{\left\{e^{\frac{i{\lambda} \hat{Q}_{\it md}}{N}}\frac{(1+\alpha_w )}{2}+e^{-\frac{i{\lambda} \hat{Q}_{\it md}}{N}}\frac{(1-\alpha_w )}{2}\right\}^N}_{\equiv \psi(x)}|\Phi_{\mathrm{in}}^{\mathrm{MD}}\ra
\label{bigspinc}
\end{eqnarray}
We already saw how this could be approximated as $ e^{i\lambda\alpha_w  \hat{Q}_{\it md} } |\Phi_{\mathrm{in}}^{\mathrm{MD}}\ra$ which produced a robust-shift in the measuring-device by the weak-value $\sqrt{2}$. 
However, we can also view $\psi (x)=\left\{e^{\frac{i{\lambda} \hat{Q}_{\it md}}{N}}\frac{(1+\alpha_w )}{2}+e^{-\frac{i{\lambda} \hat{Q}_{\it md}}{N}}\frac{(1-\alpha_w )}{2}\right\}^N$
in a very different way, by performing a binomial expansion:
\begin{eqnarray}
&\psi (x)=\sum_{n=0}^N \frac{(1+\alpha_w)^n(1-\alpha_w)^{N-n}}{2^N} \frac{N!}{n!(N-
n)!}\exp\left\{\frac{in\lambda\hat{Q}_{\mathrm{md}}}{N}\right\}\exp\left\{\frac{-i\lambda\hat{Q}_{\mathrm{md}} (N-n)}{ N}\right\} &
\nonumber\\
&=\sum_{n=0}^N c_n \exp\left\{\frac{i\lambda\hat{Q}_{\mathrm{md}} (2n-N)}{ N}\right\}=\sum_{n=0}^N c_n \exp\left\{\frac{i\lambda\hat{Q}_{\mathrm{md}}\lambda _n}{ N}\right\} 
\label{binom}
\end{eqnarray}
We see that this wavefunction is a 
superposition of waves with small wavenumbers $\vert k\vert\leq 1$ (because 
$-1<\frac{(2n-N)}{N}<1$).
For a small region (which can include several wavelengths $2\pi /\alpha_w$, depending on
how large one chooses $N$), $\psi (x)$ 
appears to have a very large momentum, since $\alpha_w$ can be arbitrarily large, i.e. a super-oscillation.  
Because these regions of superoscillations are created at the expense of having the
function grow exponentially in other regions, it would be natural to conclude that the superoscillations would be quickly ``over-taken" by tails coming from the exponential regions and would thus be short-lived.  However, it has been shown that superoscillations are remarkably robust~\cite{berry2006} and can last for a surprisingly long time.  This has therefore led to proposed/practical applications of superoscillations to situations which were previously probed by
evanescent waves (e.g. in the superresolution of very fine features with lasers).  
\footnote{In~\cite{jtsosc,spie-exttime}, we uncover several new relationships between the physical creation of the high-momenta associated with the superoscillations, eccentric weak-values, and modular variables which have been used to model the dynamical non-locality discussed in \S\ref{nonlocal}~\cite{app,app2,danny}.}

\begin{figure}[tbp] 
  \centering
  \includegraphics[width=4.14in,height=4.71in,keepaspectratio]{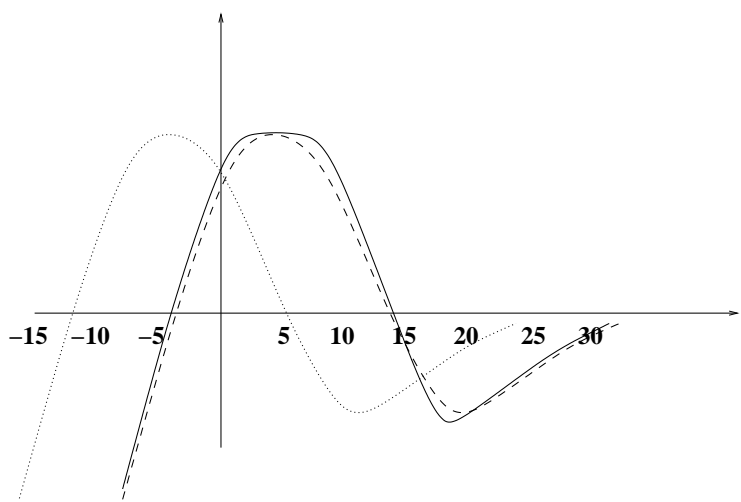}
\caption[]{\small { ``Demonstration of an approximate equality given by $\sum _{n=0}^{N}c_n f(t-a_n)\approx f(t-\alpha)$. The
sum of a function shifted by the 14 values cn between 0 and 1 and multiplied by the
coefficients, yields approximately
the same function shifted by the value 10. The dotted line shows $f(t)$; the dashed
line shows $f(t-10)$; and the solid line shows the sum." From \cite{av2v}}}
\label{lev5}
\end{figure}

As we mentioned in the introduction, TSQM is a {\bf re-formulation} of QM, and therefore it must be possible to view the novel effects from the traditional single-vector perspective.  This is precisely what super-oscillations teach us.  In summary,  
there are  2 ways to understand weak-values:
\begin{itemize}
\item the measuring-device is registering the weak-value as a property of the {\bf system} as characterized by TSQM
\item the weak-value is a result of a complex interference effect in the {\bf measuring-device}; the system continues to be described with a single-vector pursuant to QM
\end{itemize}
Oftentimes, calculations are either much simplified or can only be performed by utilizing the first approach (e.g. when the measuring-device is classical)~\cite{jt}.

\subsection{\bf Quantum Random Walk}
Another fundamental discovery arising out of TSQM is the Quantum-Random-Walk~\cite{adz} which has also stimulated discoveries in other areas of physics (for a review, see~\cite{kempe}).  
In the second bullet above, the measuring-device is shifted by the operator $\hat{\sigma}_\xi\upn$ with it's 
$N+1$ eigenvalues equally spaced between $-1$ and $+1$~\cite{jt}. 
How can a superposition of small shifts between $-1$ and $1$ give a shift that is arbitrarily far outside $\pm 1$?  The
answer is that states of the measuring-device interfere constructively for $\hat{P}_{\it md}^{(N)}=\alpha_w$ and destructively for all other
values of $\hat{P}_{\it md}^{(N)}$ such that
 $\Phi_{fin}^{MD}(P)\rightarrow\Phi_{in}^{MD}(P-A_w)$, the essence of quantum-random-walk\cite{adz}.  If the coefficients for a step to the left or right were probabilities, as would be the case in a classical random
walk, then $N$ steps of step size $1$ could generate an average
displacement of $\sqrt{N}$, but never a distance larger than $N$.  However when the steps are superposed with probability {\it amplitudes}, as with the quantum-random-walk, and when one considers probability amplitudes
that are determined by pre-and-post-selection, then the random walk can produce any
displacement.  In other words, instead of saying that a ``quantum step" is made up of
probabilities, we say that a quantum step is a superposition of the amplitude for a step ``to the
left" and the amplitude for a ``step to the right,"
then one can superpose small Fourier components and obtain a large shift.
This phenomenon is very general: if $f(t-a_n)$ is a function shifted by small numbers $a_n$, then a superposition can produce the same function but shifted by a value $\alpha$ well outside the range of $a_n$: $\sum _{n=0}^{N}c_n f(t-a_n)\approx f(t-\alpha)$.
The same values of $a_n$ and $c_n$ are appropriate for a wide 
class of functions and this relation can be made arbitrarily 
precise by increasing the number of terms in the sum, 
see figure \ref{lev5}.  The key to this phenomenon is the extremely rapid oscillations in the coefficients $c_n\equiv \frac{(1+\alpha_w)^n(1-\alpha_w)^{N-n}}{2^N} \frac{N!}{n!(N-
n)!}$ in $\sum_{n=0}^N c_n \exp\left\{i{\lambda} \hat{Q}_{\it md}k_n\right\}$.

\section{\bf TSQM suggests generalizations of QM}
\label{general}

\subsection{\bf Reformulation of Dynamics: each moment a new universe}
\label{eachmoment}
We review 
a generalization of QM suggested by TSQM~\cite{apt,jt} which addresses the ``artificial" separation in all areas of theoretical physics between the kinematic and dynamical  descriptions.
David Gross has predicted~\cite{dgross} that this distinction will be blurred as the  understanding of space and time is advanced and indeed we have developed a new way in which these traditionally distinct constructions can be united.
We note~\cite{apt,jt} that the description of the time evolution given by QM does not appropriately represent 
multi-time-correlations which are similar to  Einstein-Podolsky-Rosen/Bohm entanglement (eq. ~\ref{14.1}) but instead of being between two particles in space, they are correlations for a single particle between two different {\it times}.  
Multi-time-correlations, however, can be represented by using TSQM.
As a consequence, the general notion of time in QM is changed from the current conceptual framework which was inherited from CM, i.e.:
\begin{quote} {\bf 1): }the universe is viewed as unique, and the objects 
which inhabit it just change their state in time. In this view, time is ``empty," it just propagates a state forward; the operators of the theory create the time evolution;
\end{quote}
\noindent to a new conceptual framework in which:
\begin{quote} {\bf 2): } each  instant corresponds to a new pair of Hilbert spaces, (i.e., each instant is a new degree of freedom; in a sense, a new universe); instead of the operators creating the time evolution as in the previous approach, an entangled state (in time) ``creates" the propagation:  a whole new set of structures within time is able to ``propagate" a quantum state forward in time, 
\end{quote}

\noindent This new approach has a number of useful qualities, e.g.:  1) the dynamics and kinematics can both be represented simultaneously in  the same language, a single entangled vector (in many Hilbert spaces), and 2) a new, more fundamental complementarity between dynamics and kinematics is naturally introduced.
 This approach also leads to a new solution to the measurement problem which we model by uncertain Hamiltonians.  Finally, these considerations are also relevant to the problem of the ``Now," which was succinctly expressed by Davies~\cite{davies1995} as ``why is it 'now' {\it now}?"  
The kinematic-dynamics generalization~\cite{apt,jt}
 suggests a new fourth approach to time besides the traditional ``block universe," ``presentism," and ``possibilism" models.  

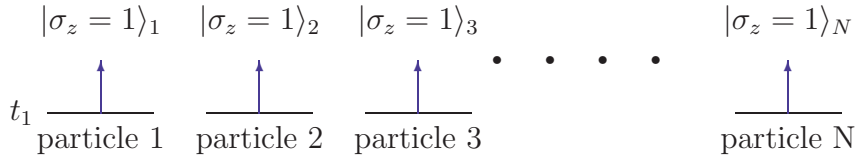
\begin{figure}[h] \epsfxsize=3.5truein
\begin{picture}(400,70)(0,0)
\put(40,10){\line(1,0){40}}
\color{BlueViolet}
\put(60,10){\vector(0,1){20}}

\color{Black}
\put(60,0){\makebox(0,0){particle 1}}
\put(60,45){\makebox(0,0){$|\sigma_z=1\rangle_1$}}
\put(100,10){\line(1,0){40}}
\color{BlueViolet}

\put(120,10){\vector(0,1){20}}
\color{RedViolet}

\color{Black}
\put(120,0){\makebox(0,0){particle 2}}
\put(120,45){\makebox(0,0){$|\sigma_z=1\rangle_2$}}
\put(160,10){\line(1,0){40}}
\color{BlueViolet}

\put(180,10){\vector(0,1){20}}
\color{Black}
\put(180,0){\makebox(0,0){particle 3}}
\put(180,45){\makebox(0,0){$|\sigma_z=1\rangle_3$}}

\put(210,30){\circle*{3}}
\put(230,30){\circle*{3}}
\put(250,30){\circle*{3}}
\put(270,30){\circle*{3}}
\put(300,10){\line(1,0){40}}
\color{BlueViolet}

\put(320,10){\vector(0,1){20}}
\color{Black}
\put(320,45){\makebox(0,0){$|\sigma_z=1\rangle_N$}}

\put(320,0){\makebox(0,0){particle N}}
\put(30,10){\makebox(0,0){$t_{1}$}}
\end{picture}

\caption[N spin-1/2 particles]
{\small N spin-1/2 particles all in the initial or pre-selected state of $\sigma_z =+1$.}
\label{nspins}
\end{figure}

While we leave all details to our other publications~\cite{apt,jt}), in brief,
consider a spin-1/2 particle, initially polarized ``up" 
along 
the $z$ axis, and having the Hamiltonian $H=0$. In this case the time evolution 
of the particle is trivial,
\begin{equation}
|\Psi(t)\rangle=constant=|\sigma_z=1\rangle.
\label{e2}
\end{equation}
To see the deficiency in representing multi-time-correlations, we will consider an isomorphism between the  correlations for a {\bf single} particle at {\bf multiple} instants of time and the correlations between {\bf multiple} particles at a {\bf single} instant of time.  Therefore, we ask if we could prepare $N$ spin-1/2 particles such that if we perform 
measurements
on them at some time $t_0$ we would obtain the same information as we would obtain by
measuring the state of the original particle at $N$ different time moments,
$t_1$, $t_2$...$t_N$?  Since the state of the original particle at all these
moments is $|\sigma_z=1\rangle$, one would suppose that this task can be 
accomplished by preparing the $N$ particles each polarized ``up" along 
the $z$ 
axis, that is eq. \ref{e3} (see also fig. \ref{nspins}):

\begin{equation}
|\sigma_z=1\rangle_1|\sigma_z=1\rangle_2...|\sigma_z=1\rangle_N\\
\label{e3}
\end{equation}

But this mapping is not appropriate for many reasons.  One reason 
is that the time evolution (\ref{e2}) contains subtle correlations (i.e. multi-time-correlations), which usually are 
not noticed, and which do not appear in the state \ref{e3} but which can actually be measured.  It is generally 
believed that since the particle is at every moment in a definite state of 
the 
$z$-spin component, the $z$-spin component is the only thing we know with 
certainty about the particle - all other spin components do not commute with 
$\sigma _z$ and cannot thus be well-defined. However, there are {\it multi-time} 
variables whose values are known with certainty, given the evolution (\ref{e2}). For 
example, although the $x$ spin component is not well defined when the spin is 
in 
the $|\sigma_z=1\rangle$ state, we know that it is constant in time, since the 
Hamiltonian is zero. Thus, for example, the two-time observable $\sigma_x(t_4)-\sigma_x(t_2)=0$
is definite ($t_2 < t_4$).
However, there is no 
state of $N$ spins such that
\beq
\hat{\sigma}_{\hat n}^1=\hat{\sigma}_{\hat n}^2=...=\hat{\sigma}_{\hat n}^N
\label{e10}
\eeq
for every direction $\hat n$ as would be required for all the multi-time-correlations. At best, one may find a {\bf two-particle state} eq.~\ref{14.1} for which the spins are {\it anti-correlated} instead of 
correlated i.e. $\hat{\sigma}_{\hat n}^1=-\hat{\sigma}_{\hat n}^2$.
However, e.g., for 3 particles, only 2 of them can be completely anti-correlated, thus it cannot be extended to $N$ particles.
\vskip .2cm
\begin{figure}[h] \epsfxsize=3.5truein
\begin{picture}(400,90)(0,0)
\put(40,10){\line(1,0){40}}
\put(40,90){\line(1,0){40}}
\color{BlueViolet}
\put(60,10){\vector(0,1){20}}
\color{RedViolet}
\put(55,90){\vector(0,-1){20}}
\put(65,90){\vector(0,-1){20}}
\put(35,85){\makebox(0,0){\tiny $\la\Psi_1|_1\!\equiv\!\la\d\!|_1$}}
\put(90,85){\makebox(0,0){\tiny $\la\Psi_2|_1\!\equiv\!\la\u\!|_1$}}

\color{OliveGreen}
\put(65,70){\oval(30,30)[tl]}
\put(50,70){\line(5,-4){69}}
\put(155,50){\makebox(0,0){\tiny correlated $\frac{1}{\sqrt{2}}\left\{\la\d\!|_1\!|\!\u\ra_2 -\la\!\u\!|_1|\!\d\ra_2\right\}$}}

\color{Black}
\put(100,10){\line(1,0){40}}
\put(100,90){\line(1,0){40}}
\color{BlueViolet}
\put(60,0){\makebox(0,0){$|\Phi\ra_1$}}

\put(115,10){\vector(0,1){20}}
\put(125,10){\vector(0,1){20}}
\put(95,15){\makebox(0,0){\tiny $|\Phi_1\ra_2\!\equiv\!|\!\u\ra_2$}}
\put(145,15){\makebox(0,0){\tiny $|\Phi_2\ra_2\!\equiv\!|\!\d\ra_2$}}

\color{RedViolet}

\put(120,100){\makebox(0,0){$\la\Psi|_2$}}

\put(120,90){\vector(0,-1){20}}
\color{OliveGreen}
\put(118,30){\oval(30,30)[br]}
\put(65,85){\line(5,-4){68}}

\color{Black}


\put(30,0){\makebox(0,0){$t_{1}$}}
\put(30,100){\makebox(0,0){$t_{3}$}}
\put(30,50){\makebox(0,0){$t_{2}$}}
\end{picture}

\caption[]
{\small  particle 1 is correlated to the pre-selected state of  particle 2.}
\label{2partcorr}
\end{figure}
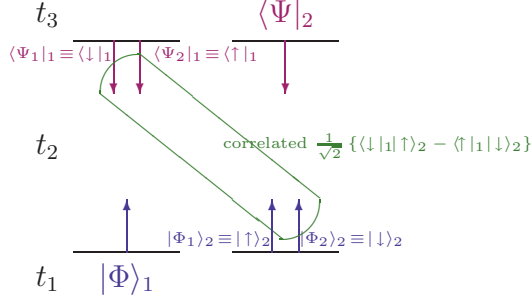
Although a state of $N$ spin 1/2 particles with complete correlations among all 
their spin components as required by eq. (\ref{e10}) doesn't exist in the usual sense,  there are pre-and-post-selected states with this property given by TSQM.
By way of example  (see figure \ref{2partcorr}), the post-selected state of particle 1 
can be completely correlated with the pre-selected state of particle 2 as described by the state $\Phi=\frac{1}{\sqrt{2}}\left\{\la\d\!|_1|\!\u\ra_2 -\la\u\!|_1|\!\d\ra_2\right\}$.
We are now able to preserve the single particle's multi-time-correlations by simply ``stacking" the $N$ spin-1/2 particles ``one 
on top of the other" along the time axis (fig. \ref{stacking}).  
As a result of the correlations between the pre-and-post-selected states, 
a verification measurement of $\hat{\sigma}_x(t_4)-\hat{\sigma}_x(t_2)$
(see left part of fig. \ref{stacking}), will yield $0$, i.e. perfect multi-time correlations because $\hat{\sigma}_x(t_2,particle \, 2)-\hat{\sigma}_x(t_2,particle \, 1)=0$ (see right part of fig. \ref{stacking}).
\begin{figure}[h] 
\vskip 3cm
\epsfxsize=6truein
\begin{picture}(500,200)(0,0)
\put(10,205){\vector(0,1){39}}
\put(10,195){\vector(0,-1){39}}
\put(10,200){\makebox(0,0){time}}
\put(70,5){\makebox(0,0){$t_{1}$}}
\put(70,90){\makebox(0,0){$t_{3}$}}
\put(70,47){\makebox(0,0){$t_{2}$}}
\put(70,137){\makebox(0,0){$t_{4}$}}

\put(40,47){\makebox(0,0){$\large \hat{\sigma}_x(t_2)$}}
\put(40,137){\makebox(0,0){$\large \hat{\sigma}_x(t_4)$}}
\put(15,47){\line(0,1){90}}
\put(15,47){\line(1,0){5}}
\put(15,137){\line(1,0){5}}

\put(210,47){\makebox(0,0){$\large \hat{\sigma}_x(t_2)$}}
\put(270,47){\makebox(0,0){$\large \hat{\sigma}_x(t_2)$}}

\put(90,10){\line(1,0){40}}
\put(90,90){\line(1,0){40}}
\color{BlueViolet}
\put(110,10){\vector(0,1){20}}
\color{RedViolet}
\put(105,90){\vector(0,-1){20}}
\put(115,90){\vector(0,-1){20}}
\color{Black}
\put(85,5){\dashbox{1}(50,90)}
\put(110,0){\makebox(0,0){single particle}}
\put(90,100){\line(1,0){40}}
\put(90,180){\line(1,0){40}}
\color{BlueViolet}
\put(105,100){\vector(0,1){20}}
\put(115,100){\vector(0,1){20}}
\color{RedViolet}
\put(105,180){\vector(0,-1){20}}
\put(115,180){\vector(0,-1){20}}
\color{Black}
\put(85,95){\dashbox{1}(50,90)}
\color{OliveGreen}
\put(135,95){\oval(80,25)}
\put(155,95){\makebox(0,0){\tiny correlated}}
\put(135,185){\oval(80,25)}
\put(155,185){\makebox(0,0){\tiny correlated}}
\color{Black}
\put(90,190){\line(1,0){40}}
\put(90,270){\line(1,0){40}}
\color{BlueViolet}
\put(105,190){\vector(0,1){20}}
\put(115,190){\vector(0,1){20}}
\color{RedViolet}
\put(110,270){\vector(0,-1){20}}
\color{Black}
\put(85,185){\dashbox{1}(50,90)}
\put(185,60){\vector(-1,0){39}}
\put(185,60){\vector(-1,0){39}}
\put(175,70){\makebox(0,0){stack}}
\put(270,140){\vector(-1,0){125}}
\put(270,140){\vector(-1,0){125}}
\put(270,95){\line(0,1){45}}
\put(175,150){\makebox(0,0){stack}}
\put(330,230){\vector(-1,0){195}}
\put(330,230){\vector(-1,0){195}}
\put(330,95){\line(0,1){135}}
\put(175,240){\makebox(0,0){stack}}
\put(185,5){\dashbox{1}(50,90)}
\put(190,10){\line(1,0){40}}
\put(190,90){\line(1,0){40}}
\color{BlueViolet}
\put(210,10){\vector(0,1){20}}
\color{RedViolet}
\put(205,90){\vector(0,-1){20}}
\put(215,90){\vector(0,-1){20}}
\color{OliveGreen}
\put(215,70){\oval(30,30)[tl]}
\put(200,70){\line(5,-4){69}}
\put(245,50){\makebox(0,0){\tiny correlated}}
\color{Black}
\put(210,0){\makebox(0,0){particle 1}}

\put(245,5){\dashbox{1}(50,90)}
\put(250,10){\line(1,0){40}}
\put(250,90){\line(1,0){40}}
\color{BlueViolet}
\put(265,10){\vector(0,1){20}}
\put(275,10){\vector(0,1){20}}
\color{RedViolet}
\put(265,90){\vector(0,-1){20}}
\put(275,90){\vector(0,-1){20}}
\color{OliveGreen}

\put(268,30){\oval(30,30)[br]}
\put(215,85){\line(5,-4){68}}
\put(305,50){\makebox(0,0){\tiny correlated}}
\color{Black}
\put(270,0){\makebox(0,0){particle 2}}

\put(305,5){\dashbox{1}(50,90)}
\put(310,10){\line(1,0){40}}
\put(310,90){\line(1,0){40}}
\color{BlueViolet}
\put(325,10){\vector(0,1){20}}
\put(335,10){\vector(0,1){20}}
\color{RedViolet}
\put(330,90){\vector(0,-1){20}}
\color{Black}
\put(275,70){\oval(30,30)[tl]}
\color{OliveGreen}

\put(260,70){\line(5,-4){69}}
\put(328,30){\oval(30,30)[br]}
\put(275,85){\line(5,-4){68}}
\color{Black}
\put(330,0){\makebox(0,0){particle 3}}
\end{picture}

\caption[Stacking the complete correlation between N particles]
{\small Measuring $\hat{\sigma}_x(t_4)-\hat{\sigma}_x(t_2)$ for the single spin-1/2 particle on the left ensures perfect multi-time-correlations because $\hat{\sigma}_x(t_2,particle \, 2)-\hat{\sigma}_x(t_2,particle \, 1)=0$.  }
\label{stacking}
\end{figure}
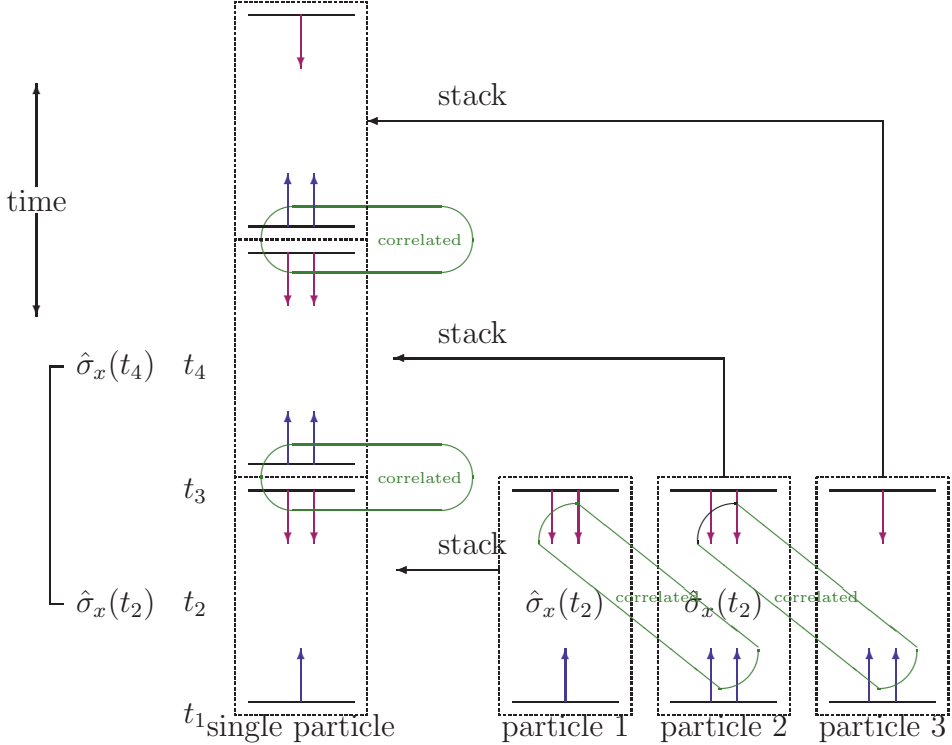
When ``stacked" onto the time axis, these correlations act like the identity operator and thus evolve the state forward,
handing-off or effectively propagating a  state 
from one moment to the next (although nothing is ``really" propagating in this picture). 

\subsection{\bf Destiny states: new solution to measurement problem}

\label{destiny}

Up until now we have limited ourselves to the possibility of 2 boundary
conditions which obtain their assignment due to selections made before and after a measurement. 
It is feasible and even suggestive to consider an extension of QM to include both a wavefunction
arriving from the past and a second ``destiny"  wavefunction coming from the future which are  determined
by 2 boundary conditions, rather than a measurement and selection.  This proposal could solve
the issue of the ``collapse'' of the wavefunction in a new and more natural way:
every time a measurement takes place and the possible measurement outcomes
decohere, then the future boundary condition simply selects one out of many possible outcomes~\cite{gross,jt}.
It also implies a kind of ``teleology" which might prove fruitful in addressing the anthropic and fine-tuning issues\cite{davies2007}
The possibility of a final boundary condition on the universe could be probed experimentally by searching for ``quantum miracles" on a cosmological scale.  
While a ``classical miracle"  is a rare event that
can be explained by a very unusual initial boundary-condition, ``Quantum Miracles" are those events which cannot naturally be explained through any special initial
boundary-condition, only through initial-and-final boundary-conditions.
By way of example, destiny-post-selection could be used to create the right dark energy or the right negative pressure (etc~\cite{hawking}).  

\section{Discussion of big questions and major unknowns concerning time-symmetry}
\label{bigquestions}

\subsection{\bf Why God Plays Dice}
\label{dice}

Why does uncertainty seem to play such a fundamental role in QM?  
First of all, uncertainty is necessary to obtain non-trivial pre-and-post-selections.  In addition, this uncertainty is needed in the measuring-device to preserve causality.  These two uncertainties work together perfectly~\cite{jt}.  Returning to the example discussed in \S's \ref{spinhalf}, \ref{counterfactual}, \ref{spin100}, since  
the weak-measurement result $\sqrt{2}$ was ``obtained" at a time  arbitrarily earlier than the post-selection time, couldn't we then ascertain that a future post-selection should produce $\sigma_y=+1$, seemingly in violation of causality?
While the weak-value depended on the post-selection, we now show that this cannot violate causality 
because  the uncertainty in the measuring-device forces us to interpret the outcomes of weak-measurements as errors.  If this were not true, then the outcome of a weak-measurement would force us to perform a particular post-selection (seemingly in violation of our free-will).
In summary, eccentric-weak-values like this cannot be discerned with certainty from the statistics of pre-selected-only-ensembles 
for two principle reasons:
\begin{itemize}
\item {\bf Analyticity of the measuring-device:}  Any measurement produces only bounded changes in the pointer variable\footnote{A bounded change in position, for example, occurs in the weak-measurements discussed in \S\ref{WM}.} which can produce an ``erroneous" value.  The disturbance to the wavefunction of the system being measured is bounded only if we prepare the measuring-device in an initial state with $Q$ bounded, i.e. $\tilde{\Phi}_{in}^{MD} (Q)$ has compact support. But this implies that the Fourier transform of $\tilde{\Phi}_{in}^{MD} (Q)$, i.e. $\Phi_{in}^{MD} (P)$ is analytic.  Therefore, there is a non-zero probability that the pointer produces ``erroneous" values even from the initial state $\Phi_{in}^{MD} (P)$.  
That is, it must be possible to constructively produce interference in the tails of $\Phi_{in}^{MD} (P)$ in order to reconstruct the initial wavefunction of the measuring-device in the ``forbidden" region, i.e.  $\Phi_{in}^{MD}(P-\la A\ra_w)$ centered around $A_w$, just as occurred with super-oscillations.
\item {\bf The probability to obtain the weak-value as an error of the measuring-device is greater than the probability of obtaining an actual weak-value.}
This follows from the requirement that the uncertainty in $P$ must be of the same order as the maximum separation between the eigenvalues (figure \ref{seqm1}.b), so that superposition of the measuring-device wavefunction can destructively interfere in the region where the normal spectrum is defined.\footnote{As shown in \S\ref{IM} and ~\cite{at3}, there are several regimes for valid weak-measurement and each rigorously preserves causality: 1) 
 we must have a small system-measuring-device interaction strength,  $\lambda$, as compared to strong-measurements in which the accuracy is increased by increasing $\lambda$; 2) a very rare pre-and-post-selection; 3) with robust-weak-measurement approach~\cite{at3}, causality is again preserved  because the corrections can only be made using the relative coordinates which can only be obtained after the particles go through the pre-and-post-selections.  If one attempted to utilize very eccentric-weak-values, then we note that as the weak-value goes further and further outside the operator spectrum, $|\langle \Psi _{\mathrm{fin}} \!\mid_j \!\Psi _{\mathrm{in}}
\rangle|^2$ from eq.~\ref{expweak} (the probability to see this particular weak-value) 
becomes smaller and smaller, and therefore the probability of obtaining these weak-values becomes smaller and smaller.  As the fluctuation in the system  increases, the probability of a rare or eccentric post-selection also increases.  
An attempt  to discern this fluctuation through the use of weak-measurements requires 
the spread in the measuring-device to be increased. This increases the probability of seeing
strange result as an error of the measuring-device.  This is a general condition which protects causality:  the probability of obtaining 
the weak-value as an error of the measuring-device must be greater than the probability of post-selection.  In other words, restricting 
$\hat{Q}$ to a finite interval forces $\Phi (P)$ to be analytic which means that $\Phi(P)$ has tails.  The tails allow 
the exponential to be expanded (eq. \ref{wv1}) and therefore, the measuring-device will register the weak-value-again without changing the shape of $\Phi (P)$.   The existence of these tails means that if the measuring-device registers $\sqrt{2}$, then it is more likely to be an error than a valid weak-value.  This prevents any ``acausal" indicator of the post-selection process.}
\end{itemize}
Therefore we conclude that the weak-value structure is completely hidden if we are looking at a pre-selected-only system because the measuring-device always hides the weak-value structure.  If the spread in $P$ did not hide the components of $A_w$, then we could obtain some information about the choice of the post-selection, which could violate causality.  
Nevertheless, usually one says that a causal connection between  events exists if the existence of a single event is ``followed" by many other events, i.e. that there is a one-way correlation.  
If we consider a number $N$ of weak-measurements during $t\in [t_{in},t_{fin}]$, then when the correct post-selection is obtained, then this post-selection forces all the weak-measurements to be centered on $A_w^1 = A_w^2=...=A_w^N=\weakv {\Psi_{fin}}{\hat{A}}{\Psi_{in} }$.  Therefore, the one-way correlation between  $\langle\!\Psi_{fin}\!\mid$ and $A_w$ is consistent with this ``causality" condition.  Finally, we have also used these considerations to probe the 
axiomatic structure of QM~\cite{jt,PopR2,PopR,danny}.  
Traditionally, the uncertainty of QM  meant that nature is capricious, i.e.``God playing dice." 
A different meaning for uncertainty can be obtained~\cite{jt}  
from two axioms: 1) the future is relevant to the present and  2) causality is maintained.  
In this program, uncertainty is derived as a consequence of the consistency between causality and weak-values;
in order to enrich 
nature with temporal non-locality,  
and yet preserve cause-effect 
relations, we must have uncertainty.

\bigskip

\subsection{\bf The Problem of Free-Will}
\label{timesymfw}

The ``destiny-generalization" of QM inspired by TSQM (\S\ref{destiny}) posits that what happens in the present is a superposition of effects, with equal contribution from past {\bf and} future events.
At first blush, it appears that perhaps we, at the present, are
   not free to decide in our own mind what our future steps may be\footnote{This was Bell's main concern with retrodictive solutions to Bell's theorem.}.
Nevertheless, we have shown~\cite{jt} that freedom-of-will and destiny can ``peacefully co-exist" in a way consistent with the aphorism ``All is foreseen, yet choice is given"~\cite{ha,price}.  
   
The concept of free-will is mainly that the past may define the future, yet after this future effect
   takes place, i.e. after it becomes past, then it cannot be changed:  we are free from the past, but, in this picture, we are not necessarily free from the future. 
Therefore, not knowing  the future is  a crucial requirement for the existence of free-will.  In other words, the destiny-vector cannot be used to inform us in the present of the result of our future free choices.

We have also shown~\cite{jt} that free-will does not necessarily mean that nobody can in principle know what the future will be because any attempt 
to communicate such knowledge will make 
the memory system 
unstable, 
thereby allowing the  freedom to change the
future.  
   Suppose there is a person who can see into the future, a prophet. Then while we, at the
   present are making a decision, and have not yet decided, the prophet knows exactly what this decision will
   be. At this point, as long as this prophet does not tell us what our decision
   will be, we are still free to make it, since we know that if the prophet had told us what our
   decision was going to be, then we would be free to change it and his prophecy would no longer be true.
Therefore, the prophet could be accurate as long as he doesn't tell us our
   future decision. I.e., we are still free to make decisions based
   on nothing but the past and our own mind. Our decisions stand alone and the prophet's knowledge does not affect our free-will.

   From TSQM and the destiny-generalization, we may say that this prophet is the information of future measurements propagating back from the future to affect
   the results of measurements conducted at the present.
Since a measurement of a weak-value is dependent upon a certain type of post-selection (which is only one of a few possible post-selections), but
   we at the present do not know whether the weak-value measured is due to an experimental error, or
   due to the post-selection. 
In addition, because a weak-measurement could be an error, there is nothing that forces us to perform a particular post-selection in the future.
Only in the future,
 when all the measurements are finished and we actually make the post-selection, can we
 retrospectively conclude whether the eccentric-weak-value shown by the measuring-device was either an error, or a
   real result due to the concrete post-selection.  Again, 
 the conditions for a weak-measurements require a high probability of experimental error. 

   From this we conclude that our prophet, the post-selected vector coming from the future, does
   not tell us the information we need to violate our free-will, and we are still free to decide what
   kind of future measurements to conduct. Therefore, {\bf free-will survives.}

\subsection{Emergence and Origin of Laws}
TSQM also provides novel perspectives on several other themes explored in this volume, e.g. on the question of emergence~\cite{ellis}:
\begin{itemize}  
\item Contextuality (\S\ref{contextuality}) suggests that the measuring-device determines the sets of possible micro-states~\cite{jtcontextuality,jtcontextuality2,jt}.   
\item A crucial component of contextuality, namely the failure of the product rule\footnote{The weak-value of a product of observables is not equal to the product of their weak-values.} suggests other novel forms of emergence~\cite{jt}.  
 By way of example, another surprising pre-and-post-selection effect is the ability to separate a system from it's properties~\cite{jt,danny}, as suggested by the Cheshire cat story: ``Well! I've often seen a cat without a grin," thought Alice; ``but a grin without a cat! It's the most curious thing I ever saw in all my life!"~\cite{alice}.  We approximate the cat by a single particle with grin states given by $|\sigma_z=+1\ra$ (grinning) and $|\sigma_z=-1\ra$ (frowning). Besides spin, we also specify the particle's location as either in a box on the left $|\psi_L\ra$, or a box on the right $|\psi_R\ra$ (figure \ref{cat}). Consider the pre-selection: $|\Psi_{in}\ra=|\psi_L\ra\left\{|\sigma_z=+1\ra+|\sigma_z=-1\ra\right\} +|\psi_R\ra|\sigma_z=+1\ra$ and the post-selected state: $|\Psi_{fin}\ra=\left\{|\psi_L\ra-|\psi_R\ra\ra\right\}\left\{|\sigma_z=+1\ra-|\sigma_z=-1\ra\right\}$. Using the isomorphism between spin states and boxes, if $N_L(+1)$ is the number of $\sigma_z=+1$ particles in the left box (etc.), then the total number of particles in the left box is: $N_L(+1)+N_L(-1)=0$. But the magnetic moment in the left box is: $N_L(+1)-N_L(-1)=2N$. Thus, there are no particles in the left box, yet there is twice the magnetic field there! Alice would say ``Curiouser and curiouser":  the particles are all in the right box, but there is no field there, thereby challenging the notion that all properties ``sit" on the particle.   
\begin{figure}[tbp] 
  \centering
  \includegraphics[width=2.14in,height=3.21in,keepaspectratio]{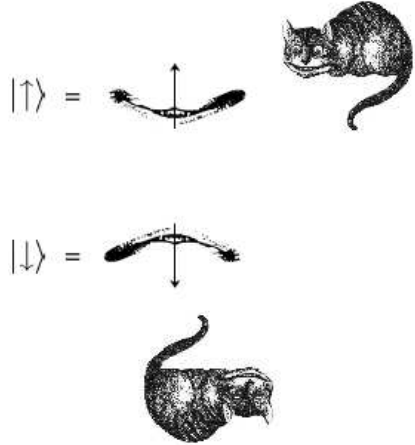}
\caption[]{\small {\bf Chesire Cat grin states.  From \cite{danny}}}
\label{cat}
\end{figure}
\item Finally, the ``destiny-vector" (\S\ref{destiny}) suggests a form of 
top-down causality which is stable to fluctuations because post-selections are performed on the entire Universe and by definition no fluctuation exists outside the Universe.
\end{itemize}
These are examples of  emergence with respect to {\em properties}.  
As Barrows and Davies~\cite{davies2007} have emphasized, the questions of fine-tuning, the origin of the physical laws, and the anthropic principle are significant outstanding problems in physics.
What novel perspectives can be gleaned from TSQM on these questions?
E.g. the dynamics-kinematics generalization (\S\ref{eachmoment}~\cite{apt,jt}) suggests a novel way to think about dynamical laws.  One implication is the fact that although we may know the dynamics on a particular time-scale $T$, this doesn't mean that we know anything about the dynamics on a smaller time-scale: consider a superposition of unitary evolutions (using $e^{-iH T}=\lbrace e^{\frac{-iHT}{N}} \rbrace ^N$):
\begin{equation}
\int g(\nu) e^{-iH(\nu) t}d\nu\rightarrow \int g(\nu) \lbrace 1+ iH(\nu) t\rbrace d\nu\underbrace{\longrightarrow}_{if \,\int g(\nu)d\nu=1}1+ i\int g(\nu)H(\nu) t d\nu
\end{equation}
This theory is the same as the usual theory but with an effective Hamiltonian
\beq
H_{eff}=\int g(\nu)H(\nu) t d\nu
\eeq
The finer grained Hamiltonian can be expressed as a superposition of evolutions $e^{\frac{-iHT}{N}}=\sum_n\alpha_n e^{\frac{-i\beta_n HT}{N}}$, i.e. the Hamiltonian can be represented as a superposition of different laws given by pre-and-post-selection~\cite{jt}.  

\bigskip

\noindent {\bf Acknowledgments}   We thank everybody involved in the development of TSQM.  To name a few: David Albert, Alonso Botero, Sandu Popescu, Benni Reznik, Daniel Rohrlich, Lev Vaidman. We also thank Lev Vaidman for providing several figures and J.E. Gray for help with editing.  JT thanks  the Templeton foundation for support.


\begin{thebibliography}{99}
\bibitem{zeilinger} Zeilinger, A., ``Quantum randomness, entanglement, and the free will of the experimentalist," 
this volume.
\bibitem{avg}

{Theorem:} for every observable $A$ and a normalized state $|\psi\rangle$, we have:
 $A|\psi\rangle=\langle  A\rangle|\psi\rangle+\Delta A|\psi_\perp\rangle$
   for some state $|\psi_\perp\rangle$ which is orthogonal to $|\psi\rangle$.
To prove this, we begin with:
$A|\psi\rangle=\langle A\rangle|\psi\rangle+A|\psi\rangle-\langle A\rangle|\psi\rangle$
   now, we set:
$|\widetilde{\psi}_\perp\rangle =A|\psi\rangle-\langle A\rangle|\psi\rangle$,   so: $\langle \widetilde{\psi}_\perp |\psi\rangle =(\langle\psi|A-\langle\psi|\langle
      A\rangle)|\psi\rangle=\langle\psi| A|\psi\rangle - \langle A\rangle\langle\psi|\psi\rangle = 0$
   now we set: $|\psi_\perp\rangle=b|\widetilde{\psi}_\perp\rangle$, where $|\psi_\perp\rangle$ is normalized and $b$
   real (note that $\langle\psi|\psi_\perp\rangle=0$).
   so:
$A|\psi\rangle=\langle A\rangle|\psi\rangle+b|\psi_\perp\rangle$.
   Now we multiply from the left by $\langle\psi_\perp|$, and we get:
   $\langle\psi_\perp|A|\psi\rangle=b$.   Now we can see that:
$\langle\psi|A^2|\psi\rangle=\langle\psi|A(\langle A\rangle|\psi\rangle+b|\psi_\perp\rangle)=
      \langle\psi|({\langle A\rangle}^2|\psi\rangle+b\langle A\rangle|\psi_\perp\rangle+bA|\psi_\perp\rangle)=
      {\langle A\rangle}^2+b\langle\psi|A|\psi_\perp\rangle$  so: $\langle A^2\rangle-{\langle A\rangle}^2=b\langle\psi|A|\psi_\perp\rangle=b^2$
   which means that: $b=\sqrt{\langle A^2\rangle-{\langle A\rangle}^2}=\Delta A$
   and the result: 
   $A|\psi\rangle=\langle  A\rangle|\psi\rangle+\Delta
   A|\psi_\perp\rangle$  is proved.
\bibitem{proof} The identity $\exp\left\{{i\alpha\hat{\sigma}_{\vec{n}}}\right\}=\cos\alpha+i\hat{\sigma}_{\vec{n}}\sin\alpha$ is easily proven using the fact that for any integer $k$:
$\sigma_{\hat{n}}^{2k}=I$ and $\sigma_{\hat{n}}^{2k+1}=\sigma_{\hat{n}}$
   and now it follows that:
$      e^{i\alpha\sigma_{\hat{n}}}=\sum_{k=0}^\infty\frac{(i\alpha)^k\sigma_{\hat{n}}^k}{k!}=\sum_{k=0}^\infty\frac{(i\alpha)^{2k}}{(2k)!}+\sigma_{\hat{n}}\sum_{k=0}^\infty\frac{(i\alpha)^{2k+1}}{(2k+1)!}
      =e^{i\alpha\sigma_{\hat{n}}}=\cos\alpha+i\sigma_{\hat{n}}\sin\alpha$  and the identity is proven.
\bibitem{proofthm2}
\noindent Proof: Given that $\mathrm{\bf \hat{P}}_{A}=\sum_n a_n\ket{\alpha_n}\bra{\alpha_n}$, if an eigenvalue, e.g. $\mathrm{\bf \hat{P}}_{A}=a_n$,  is obtained with certainty, then for $n\neq m$, $\mathrm{\bf \hat{P}}_{A}\equiv \ket{\alpha_m}\bra{\alpha_m}=0$ because the probability to obtain another eigenvalue by ABL is $\propto\la\Psi_{\mathrm{fin}}\ket{\alpha_m}\bra{\alpha_m}\Psi_{\mathrm{in}}\ra=0$.  In this case, the weak-value $(\mathrm{\bf \hat{P}}_{A})_{\mathrm{w}}=(\ket{\alpha_m}\bra{\alpha_m})_{\mathrm{w}}=\weakv{\Psi_{\mathrm{fin}}}{\, {\alpha_m}\ra\la{\alpha_m} \, }{\Psi_{\mathrm{in}}}=0$. In addition,
$\sum_m\weakv{\Psi_{\mathrm{fin}}}{\, {\alpha_m}\ra\la{\alpha_m} \, }{\Psi_{\mathrm{in}}}=1$ 
because $\sum_m \ket{\alpha_m}\bra{\alpha_m}=1$.  But since $\la\Psi_{\mathrm{fin}}\ket{\alpha_m}\bra{\alpha_m}\Psi_{\mathrm{in}}\ra=0$ for $n\neq m$, the only term left is $n$. Therefore, the weak-value is $1$, the same as the ideal value.
\end{thebibliography}
\end{document}